\documentclass[twocolumn,american,floatfix, table, dvipsnames,prl]{revtex4-2}
\usepackage[T1]{fontenc}
\usepackage[latin9]{inputenc}
\usepackage{color}
\usepackage{babel}
\usepackage{amsmath}
\usepackage{amssymb}
\usepackage{graphicx}
\usepackage{esint}
\usepackage[unicode=true,
 bookmarks=false,
 breaklinks=true,pdfborder={0 0 1},backref=false,colorlinks=true]
 {hyperref}
\hypersetup{
 pdfcreator={},pdfproducer={LaTeX with hyperref},linkcolor=blue,anchorcolor=blue,citecolor=blue,filecolor=red,menucolor=red,pagecolor=red,urlcolor=blue,pdfstartview=FitV,pdfhighlight=/I,pdfpagelayout=TwoColumns,hypertexnames=true}

\makeatletter
\usepackage{lmodern}
\usepackage{mathptmx, newtxtext, newtxmath, xspace}
\usepackage{bbm}
\usepackage{graphicx}

\usepackage{floatrow}

\usepackage{accents}
\newlength{\dhatheight}

\usepackage{natbib}
\setcitestyle{numbers}

\makeatother

\begin{document}
\title{A critical nematic phase with pseudogap-like behavior in twisted bilayers}
\author{Virginia Gali}
\affiliation{School of Physics and Astronomy, University of Minnesota, Minneapolis
55455 MN, USA}
\author{Matthias Hecker }
\affiliation{School of Physics and Astronomy, University of Minnesota, Minneapolis
55455 MN, USA}
\author{Rafael M. Fernandes}
\affiliation{School of Physics and Astronomy, University of Minnesota, Minneapolis
55455 MN, USA}
\date{\today }
\begin{abstract}
The crystallographic restriction theorem constrains two-dimensional
nematicity to display either Ising ($Z_{2}$) or three-state-Potts
($Z_{3}$) critical behaviors, both of which are dominated by amplitude
fluctuations. Here, we use group theory and microscopic modeling to
show that this constraint is circumvented in a $30^{\circ}$-twisted
hexagonal bilayer due to its emergent quasicrystalline symmetries.
We find a critical phase dominated by phase fluctuations of a $Z_{6}$
nematic order parameter and bounded by two Berezinskii-Kosterlitz-Thouless
(BKT) transitions, which displays only quasi-long-range nematic order.
The electronic spectrum in the critical phase displays a thermal pseudogap-like
behavior, whose properties depend on the anomalous critical exponent.
We also show that an out-of-plane magnetic field induces nematic phase
fluctuations that suppress the two BKT transitions via a mechanism
analogous to the Hall viscoelastic response of the lattice, giving
rise to a putative nematic quantum critical point with emergent continuous
symmetry. Finally, we demonstrate that even in the case of an untwisted
bilayer, a critical phase emerges when the nematic order parameter
changes sign between the two layers, establishing an odd-parity nematic
state.
\end{abstract}
\maketitle
The discovery of magic-angle twisted bilayer graphene \citep{Cao2018_a,Cao2018_b,Yankowitz2019,Sharpe2019,Lu2019}
heralded the field of twistronics, enabled by the remarkable precision
with which twist angles can be tuned \citep{Kennes2021,Mak2022,Shen2022,Zhang2022,Uri2023superconductivity}.
By imposing an underlying superlattice potential on the charge carriers
\citep{Cano2023}, twisting affects the electronic properties of 2D
systems in various ways. Besides the emergence of flat bands from
the band folding reconstruction at magic twist angles \citep{Bistritzer2011},
the superlattice, being an incommensurate array of registered sites,
displays features with no counterpart on crystalline lattices that
significantly affect the electronic degrees of freedom. Indeed, for
small twist angles, the elastic excitations of the resulting moiré
superlattice behave very differently from standard acoustic phonons
\citep{Koshino2019,Ochoa2019,Ochoa2022,Gao2022,Samajdar2022}, thus
influencing transport properties \citep{Ochoa2023,Ishizuka2021} and
possibly superconductivity \citep{DasSarma2019,Bernevig2019}. Conversely,
for certain large twist angles, the twisted superlattice acquires
symmetries forbidden by the crystallographic restriction theorem,
which enables new electronically ordered states \citep{Yang2023_b,Yang2023_c}.
For instance, the superlattice formed by twisting two tetragonal layers
by $45^{\circ}$ has an eight-fold improper rotational symmetry \citep{Haenel2022}.
In the presence of $d$-wave pairing interactions, this symmetry enforces
the superconducting state to be the exotic $d+id$ \citep{Can2021},
as recently proposed to be realized in twisted cuprates \citep{Can2021,Tummuru2022,Vishwanath2022,Yang2023_a,Volkov2023_a,Volkov2023_b,Yuan2023,Wang2023,Kim2023}.

\begin{figure}[t]
\includegraphics[width=1\columnwidth]{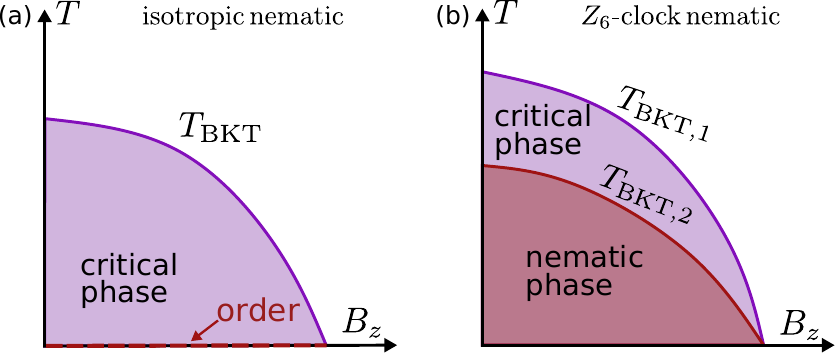}

\caption{Schematic phase diagrams in 2D of (a) the isotropic nematic model
and (b) the $Z_{6}$-clock nematic model realized on hexagonal layers
twisted by $30^{\circ}$. While both feature a BKT transition towards
a critical phase with quasi-long-range nematic order (purple), long-range
order (red) is established at nonzero temperatures in the $Z_{6}$
case via another BKT transition and only at $T=0$ in the isotropic
case. $B_{z}$ is an out-of-plane magnetic field. \label{fig:Z6_clock}}
\end{figure}

Besides superconductivity, another electronic state strongly affected
by the symmetries of the underlying lattice is the electronic nematic
\citep{Kivelson1998}, in which electron-electron interactions lead
to the spontaneous breaking of rotational symmetry \citep{Fradkin2010,Fernandes2014}.
In 2D isotropic space, this is a continuous $U(1)$ (or XY) symmetry
that, when broken, triggers a nematic Goldstone mode that couples
directly to the electronic charge density \textendash{} unlike magnon
or phonon Goldstone modes \textendash{} and promotes non-Fermi-liquid
behavior at zero temperature \citep{Oganesyan2001,Watanabe2014}.
At nonzero temperatures, only quasi-long-range nematic order is allowed
across a critical phase {[}Fig. \ref{fig:Z6_clock}(a){]}, whose impact
on the electronic spectrum remains unexplored. These appealing features
of the isotropic nematic, however, are not realized in 2D crystals,
since the underlying lattice lowers the continuous rotational symmetry
to $Z_{2}$ (Ising, in tetragonal lattices) \citep{Fradkin2010,Fernandes2014}
or $Z_{3}$ (3-state Potts/clock, in hexagonal lattices) \citep{Hecker2018,Fernandes2020,Chakraborty2023}. 

Here, we show that these crystallographic restrictions on nematic
phenomena are circumvented in hexagonal bilayers twisted by $30^{\circ}$,
as the resulting twisted superlattice is a quasicrystal with a non-crystallographic
twelve-fold improper rotation axis \citep{Stampfli1986,Ahn2018}.
Using group-theory and a microscopic model, we find that the nematic
order parameter has the $Z_{6}$ symmetry of a six-state clock model
\citep{Jose1977}. As a result, before the onset of long-range nematic
order, the system displays a critical nematic phase with quasi-long-range
order (like the isotropic nematic) bounded by two Berezinskii-Kosterlitz-Thouless
(BKT) transitions, see Fig. \ref{fig:Z6_clock}(b). 

Upon computing the electronic self-energy, we find that the nematic
phase fluctuations inside the critical phase suppress the density
of states (DOS) at the Fermi level and promote a pronounced peak in
the spectral function at a frequency set by the anomalous exponent,
a behavior reminiscent of a pseudogap. We also demonstrate that an
out-of-plane magnetic field triggers fluctuations of the nematic phase
via a mechanism analogous to the viscoelastic Hall response of the
lattice \citep{Maciejko2013,Fradkin2013,Fradkin2014}. Consequently,
a magnetic field acts as an effective transverse nematic field, driving
the two BKT transitions towards a nematic quantum critical point (QCP)
{[}Fig. \ref{fig:Z6_clock}(b){]}. This QCP is expected to belong
to the XY universality class and to trigger a pseudo-Goldstone mode,
which can promote a non-Fermi-liquid to Fermi-liquid crossover at
$T=0$ \textendash{} similarly to the recently studied valley-polarized
$T=0$ nematic state \citep{Mandal2023}.

We start by considering two identical hexagonal layers, each with
point group $D_{6}$ and described by the non-interacting Hamiltonian
$\mathcal{H}_{0,\mu}=\sum_{\boldsymbol{k}}\boldsymbol{c}_{\boldsymbol{k},\mu}^{\dagger}\left(\varepsilon_{\boldsymbol{k}}+\delta_{0}f_{\boldsymbol{k}}^{A_{1}}\right)\sigma^{0}\boldsymbol{c}_{\boldsymbol{k},\mu}^{\phantom{\dagger}}$,
with electronic operators $\boldsymbol{c}_{\boldsymbol{k},\mu}\equiv\left(c_{\mathbf{k}\uparrow,\mu},c_{\mathbf{k}\downarrow,\mu}\right)$,
momentum $\boldsymbol{k}=\left|\boldsymbol{k}\right|\left(\cos\theta,\sin\theta\right)$,
spin-space Pauli matrices $\sigma^{i}$ , and layer index $\mu=t,b$
for top and bottom layers, respectively. To keep the analysis general,
the electronic dispersion consists of an isotropic term $\varepsilon_{\boldsymbol{k}}=\frac{\boldsymbol{k}^{2}}{2m}-\mu_{0}$
and a hexagonal warping term with coefficient $\delta_{0}$ and form
factor $f_{\boldsymbol{k}}^{A_{1}}=\cos\left(6\theta\right)$. The
electronic nematic degrees of freedom are described by the two-component
collective field $\boldsymbol{\phi}_{\mu}=\left|\boldsymbol{\phi}_{\mu}\right|\left(\cos\alpha_{\mu},\sin\alpha_{\mu}\right)$
that couples to the $d_{x^{2}-y^{2}}$ and $d_{xy}$ quadrupolar charge
densities of each layer, $\mathcal{H}_{\mathrm{nem},\mu}=-g\sum_{\mathbf{k}}\boldsymbol{c}_{\boldsymbol{k},\mu}^{\dagger}\left(\boldsymbol{\phi}_{\mu}\cdot\boldsymbol{f}_{\boldsymbol{k}}^{E_{2}}\right)\sigma^{0}\boldsymbol{c}_{\boldsymbol{k},\mu}^{\phantom{\dagger}}$,
with form-factor $\boldsymbol{f}_{\boldsymbol{k}}^{E_{2}}=\left(\cos2\theta,\,-\sin2\theta\right)$
and coupling constant $g$ \citep{Fernandes2020}. The nematic properties
of an isolated layer can be obtained from the Landau free-energy $\mathcal{F}\left[\boldsymbol{\phi}_{\mu}\right]$,
which we derive directly from the microscopic Hamiltonian $\mathcal{H}_{\mu}\equiv\mathcal{H}_{0,\mu}+\mathcal{H}_{\mathrm{nem},\mu}$
(details in the Supplementary Material (SM)):

\begin{equation}
\mathcal{F}\left[\boldsymbol{\phi}\right]=\mathcal{F}_{0}\left[\left|\boldsymbol{\phi}\right|^{2}\right]+\lambda_{q}\left|\boldsymbol{\phi}\right|^{q}\cos\left(q\alpha\right),\label{eq:F_nem}
\end{equation}
where $\mathcal{F}_{0}$ depends only on the amplitude and we omitted
the layer subscript $\mu$. In our case, $q=3$, i.e. the nematic
transition belongs to the 2D three-state Potts/clock universality
class \textendash{} recall that the $Z_{q}$-Potts and $Z_{q}$-clock
models are equivalent for $q=3$, but not $q>3$ \citep{Wu1982}.
This is a well-established result \citep{Hecker2018,Fernandes2020,Xu2020}
that can also be derived from group-theory. Since $\boldsymbol{\phi}_{\mu}$
transforms as the $E_{2}$ irreducible representation (irrep) of the
$D_{6}$ group, the free energy must have a cubic invariant because
the decomposition of the product $E_{2}\otimes E_{2}\otimes E_{2}=A_{1}\oplus A_{2}\oplus E_{2}$
contains a term that transforms as the trivial irrep $A_{1}$. More
broadly, for any of the ten 2D crystallographic point groups that
admit a non-trivial nematic order parameter, the nematic free energy
must have the form of Eq. (\ref{eq:F_nem}) with $q=2$ ($Z_{2}$
Ising) or $q=3$ ($Z_{3}$ Potts/clock). In either case, the phase
$\alpha_{\mu}$ is strongly constrained to discrete values and phase
fluctuations do not play an important role, unlike the isotropic nematic
case.

The situation changes when the two layers are coupled and twisted
by an angle $\theta_{\mathrm{tw}}$, since the system can acquire
crystallographically-forbidden symmetries, thus enabling other $q$
values in Eq. (\ref{eq:F_nem}). Instead of $\boldsymbol{\phi}_{t}$
and $\boldsymbol{\phi}_{b}$, we consider their symmetric and antisymmetric
rotated combinations, $\boldsymbol{\phi}_{\pm}=\frac{1}{2}\left[\mathcal{R}_{z}\left(-\theta_{\mathrm{tw}}\right)\boldsymbol{\phi}_{b}\pm\mathcal{R}_{z}\left(\theta_{\mathrm{tw}}\right)\boldsymbol{\phi}_{t}\right]$,
where $\mathcal{R}_{z}\left(\theta\right)$ is the rotation matrix
with respect to the $z$-axis. This change of basis is convenient
because the nematic directors in the two layers are rotated against
each other by the horizontal mirror reflection $\sigma_{h}$ (corresponding
to switching the layer indices), $\boldsymbol{\phi}_{b/t}\overset{\sigma_{h}}{\longrightarrow}\mathcal{R}_{z}\left(\pm2\theta_{\mathrm{tw}}\right)\boldsymbol{\phi}_{t/b}$,
implying that the combinations $\boldsymbol{\phi}_{\pm}$ are eigenstates
of $\sigma_{h}$, $\boldsymbol{\phi}_{\pm}\overset{\sigma_{h}}{\longrightarrow}\pm\boldsymbol{\phi}_{\pm}$.
Note that $\sigma_{h}$ does not necessarily leave the twisted bilayer
invariant. 

\begin{figure}[t]
\includegraphics[width=1\columnwidth]{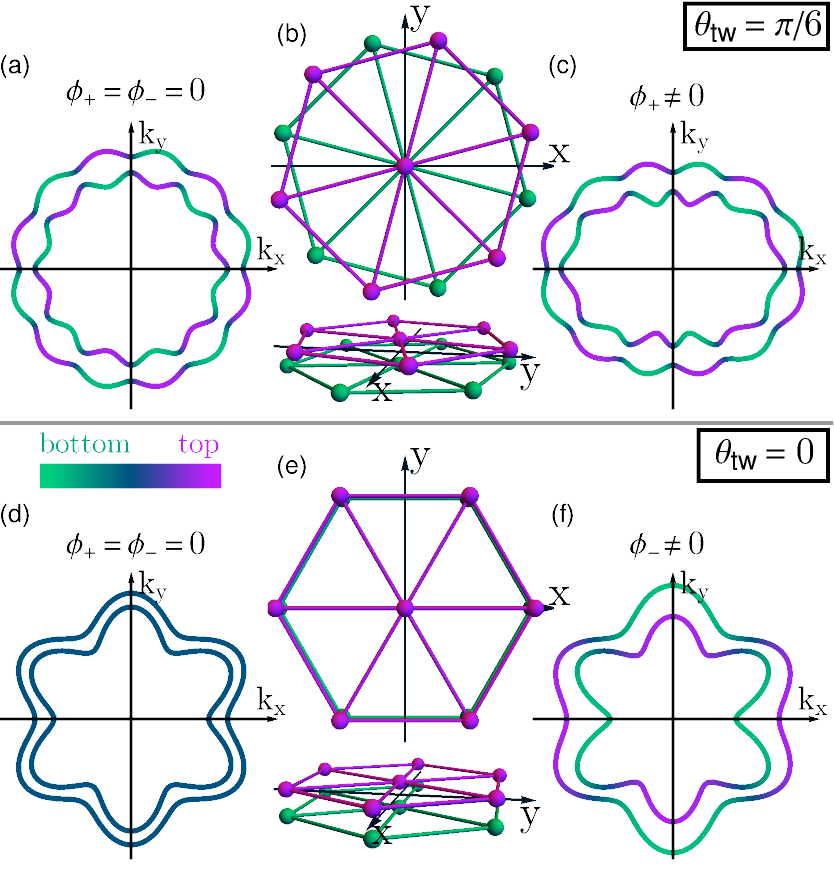}

\caption{$30^{\circ}$-twisted {[}(a)-(c){]} and untwisted {[}(d)-(f){]} hexagonal
bilayer system. In each case we show the lattice structure {[}(b),
(e){]}, the FS without nematic order {[}(a), (d){]}, and the FS in
the presence of pure nematic (c) or odd-parity nematic order (f).
The color code denotes the spectral weight from the top and bottom
layers. \label{fig:FS}}
\end{figure}

Consider first the untwisted case, $\theta_{\mathrm{tw}}=0$ {[}Fig.
\ref{fig:FS}(e){]}. In this case, the reflection $\sigma_{h}$ is
a symmetry of the bilayer, such that its point group becomes $D_{6h}=D_{6}\times\left\{ E,\,\sigma_{h}\right\} $
rather than $D_{6}$ (here $E$ denotes the identity operator). With
$\boldsymbol{\phi}_{\pm}$ being even/odd under $\sigma_{h}$, they
transform respectively as the irreps $E_{2g}$ and $E_{2u}$ of $D_{6h}$.
Thus, $\boldsymbol{\phi}_{+}$ still behaves as a $Z_{3}$ nematic
order parameter with the free energy given by Eq. (\ref{eq:F_nem})
with $q=3$. In contrast, the threefold rotational-symmetry-breaking
pattern due to $\boldsymbol{\phi}_{-}$ changes sign between the two
layers \textendash{} hence we dub $\boldsymbol{\phi}_{-}$ an odd-parity
nematic order. Importantly, because $\boldsymbol{\phi}_{-}$ is odd
under $\sigma_{h}$, the cubic term in Eq. (\ref{eq:F_nem}) is no
longer allowed, and the leading anisotropic term is the $q=6$ one.

We now twist the layers by $\theta_{\mathrm{tw}}=\pi/6$. The resulting
twisted lattice, shown in Fig. \ref{fig:FS}(b), is not invariant
under $\sigma_{h}$ but it is symmetric under an improper rotation
$S_{12}=\sigma_{h}C_{12z}$, i.e. a twelve-fold rotation followed
by a reflection, an operation that is forbidden in periodic crystals.
Thus, the $30^{\circ}$-twisted bilayer is actually an aperiodic quasicrystal
described by the non-crystallographic point group $D_{6d}=D_{6}\times\left\{ E,\,S_{12}\right\} $
\citep{Stampfli1986,Ahn2018}, see also Fig. S1 in the SM. This construction,
which was previously proposed to realize $f+if$ and $g+ig$ superconductivity
\citep{Yang2023_b,Yang2023_c}, is analogous to the $45^{\circ}$-twisted
tetragonal bilayer, characterized by a non-crystallographic $D_{4d}$
point group that enables $d+id$ superconductivity \citep{Can2021}.

For the $30^{\circ}$-twisted bilayer, $\boldsymbol{\phi}_{+}$, corresponding
to pure nematic order, transforms as the $E_{2}$ irrep of $D_{6d}$,
whereas the odd-parity nematic $\boldsymbol{\phi}_{-}$ transforms
as $E_{4}$. Importantly, the combinations  $\boldsymbol{\phi}_{+}$,
$\boldsymbol{\phi}_{-}$ are odd/even under another symmetry operation
of the twisted bilayer: the four-fold improper rotation $S_{4}=\sigma_{h}C_{4z}$,
$\boldsymbol{\phi}_{\pm}\overset{S_{4}}{\longrightarrow}\mp\boldsymbol{\phi}_{\pm}$.
Consequently, the cubic term $q=3$ in the free energy (\ref{eq:F_nem})
is forbidden for $\boldsymbol{\phi}_{+}$, and the leading-order anisotropy
term is the $q=6$ one, which corresponds to the 2D six-state ($Z_{6}$)
clock model \citep{Jose1977}. The resulting phase diagram, shown
schematically in Fig. \ref{fig:Z6_clock}(b), displays two BKT phase
transitions \citep{Jose1977,Xu2020,Podolsky2016,Arnold2022}. Below
$T_{\mathrm{BKT},1}$, vortices and anti-vortices associated with
the nematic phase $\alpha_{+}$ bind into pairs, like in the XY model.
Below $T_{\mathrm{BKT},2}$, the discrete nature of $\alpha_{+}$
becomes relevant and long-range nematic order emerges. Thus, for $T_{\mathrm{BKT},2}\leq T\leq T_{\mathrm{BKT},1}$
the system displays a critical phase with quasi-long-range nematic
order. Remarkably, the quasicrystalline symmetry of the $30^{\circ}$-twisted
bilayer enables the system to display the same behavior as that of
the isotropic (XY) nematic phase over a wide temperature range.

We now show that the same results obtained using group-theory follow
from the microscopic model. Upon coupling the two layers via a simple
tunneling Hamiltonian \citep{Can2021}, $\mathcal{H}_{b-t}=t\sum_{\boldsymbol{k}}\boldsymbol{c}_{\boldsymbol{k},t}^{\dagger}\sigma^{0}\boldsymbol{c}_{\boldsymbol{k},b}^{\phantom{\dagger}}+\mathrm{h.c.}$,
and rotating the top/bottom layer by $\pm\theta_{\mathrm{tw}}/2$,
respectively, we obtain $\mathcal{H}_{\mathrm{tot}}=\mathcal{H}_{b}+\mathcal{H}_{t}+\mathcal{H}_{b-t}$:

\begin{align}
\mathcal{H}_{\mathrm{tot}} & =\sum_{\boldsymbol{k}}\boldsymbol{c}_{\boldsymbol{k}}^{\dagger}\Big\{\varepsilon_{\boldsymbol{k}}\tau^{0}+\delta_{0}f_{\boldsymbol{k}}^{A_{1}}\cos\left(3\theta_{\mathrm{tw}}\right)\tau^{0}+\delta_{0}f_{\boldsymbol{k}}^{A_{2}}\sin\left(3\theta_{\mathrm{tw}}\right)\tau^{z}\nonumber \\
 & \quad-g\left(\boldsymbol{\phi}_{+}\cdot\boldsymbol{f}_{\boldsymbol{k}}^{E_{2}}\tau^{0}+\boldsymbol{\phi}_{-}\cdot\boldsymbol{f}_{\boldsymbol{k}}^{E_{2}}\tau^{z}\right)+t\,\tau^{x}\Big\}\sigma^{0}\boldsymbol{c}_{\boldsymbol{k}}^{\phantom{\dagger}},\label{eq:H_tot}
\end{align}
where $\boldsymbol{c}_{\boldsymbol{k}}=\left(\boldsymbol{c}_{\boldsymbol{k},b},\,\boldsymbol{c}_{\boldsymbol{k},t}\right)$,
$f_{\boldsymbol{k}}^{A_{2}}=\sin\left(6\theta\right)$, and $\tau^{i}$
are layer-space Pauli matrices. The free energy of this model, $\mathcal{F}\left[\boldsymbol{\phi}_{+},\,\boldsymbol{\phi}_{-}\right]$,
is derived in the SM. Regardless of $\theta_{\mathrm{tw}}$, the ``pure''
nematic order parameter $\boldsymbol{\phi}_{+}$ and $\boldsymbol{\phi}_{-}$
decouple to quadratic order, although $\theta_{\mathrm{tw}}$ affects
their higher-order couplings (see SM). Most importantly, there is
always a $q=3$ cubic term for both $\boldsymbol{\phi}_{+}$ and $\boldsymbol{\phi}_{-}$,
except when $\theta_{\mathrm{tw}}=\pi/6$ ($\theta_{\mathrm{tw}}=0$),
in which case the cubic term vanishes for the pure nematic $\boldsymbol{\phi}_{+}$
(odd-parity nematic $\boldsymbol{\phi}_{-}$) and the corresponding
free energy becomes that of Eq. (\ref{eq:F_nem}) with $q=6$. 

We also diagonalize $\mathcal{H}_{\mathrm{tot}}$ to obtain the reconstructed
Fermi surfaces (FS). Figs. \ref{fig:FS}(a),(d) show the FS in the
disordered phase for the $30^{\circ}$-twisted ($\theta_{\mathrm{tw}}=\pi/6$)
and untwisted ($\theta_{\mathrm{tw}}=0$) cases. Despite the incommensurate
nature of the twisted system for arbitrary $\theta_{\mathrm{tw}}$,
its FS can be represented in the untwisted Brillouin zone provided
the tunneling $t$ is small, in which case the main effect of the
aperiodic potential is to split the crossings between the top- and
bottom-layer FS \citep{Yang2023_b}, as highlighted by the color coding
in the figure. A similar approach is often employed for incommensurate
charge density-waves \citep{Norman2007} and magic-angle TBG, whose
moiré superlattice is also incommensurate \citep{Uri2023superconductivity}.
In Figs. \ref{fig:FS}(c),(f) we show the impact on the FS of long-range
nematic order $\boldsymbol{\phi}_{+}=\phi_{0}\left(1,0\right)$ in
the twisted case and odd-parity nematic order $\boldsymbol{\phi}_{-}=\phi_{0}\left(1,0\right)$
in the untwisted case, highlighting the threefold rotational-symmetry-breaking. 

While below $T_{\mathrm{BKT},2}$ the electronic spectrum is determined
by $\boldsymbol{\phi}_{\pm}\neq0$, as shown in Figs. \ref{fig:FS}(c),(f),
the spectrum inside the critical phase $T_{\mathrm{BKT},2}\leq T\leq T_{\mathrm{BKT},1}$
is governed by phase fluctuations. To capture this effect, we compute
the one-loop electronic self-energy $\Sigma\left(\omega,\boldsymbol{k}\right)=g^{2}A_{0}T\sum_{ij}\int_{q}\chi_{\mathrm{nem}}^{ij}\left(\Omega,\boldsymbol{q}\right)f_{\boldsymbol{k}-\frac{\boldsymbol{q}}{2},i}^{E_{2}}f_{\boldsymbol{k}-\frac{\boldsymbol{q}}{2},j}^{E_{2}}\mathcal{G}\left(\omega-\Omega,\boldsymbol{k}-\boldsymbol{q}\right)$,
where $\chi_{\mathrm{nem}}^{ij}$ is the nematic susceptibility, $A_{0}$
is the area, and $\mathcal{G}$ is the non-interacting Green's function.
While for the specific model considered above $\mathcal{G}$ can be
read off from Eq. (\ref{eq:H_tot}), here we consider a generic linearized
electronic dispersion. This enables us to extend the results to other
systems that display a critical phase associated with a $Z_{6}$ nematic-like
order parameter, such as the odd-parity nematic order for $\theta_{\mathrm{tw}}=0$
and the recently discussed spin-polarized \citep{Wu2007,Classen2020,Chichinadze2020}
and valley-polarized \citep{Xu2020,Mandal2023} nematic orders.

\begin{figure}[t]
\includegraphics[width=1\columnwidth]{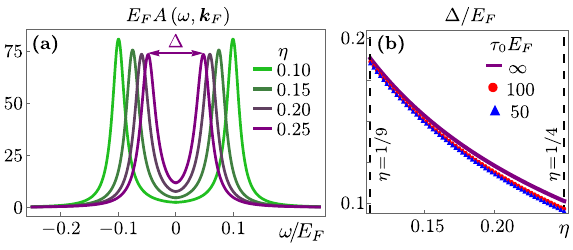} \caption{(a) The electronic spectral function $A\left(\omega,\boldsymbol{k}_{F}\right)$
in the critical nematic phase for $\eta$ within the range $1/9\protect\leq\eta\protect\leq1/4$;
here, $T=0.01E_{F}$ and $\tau_{0}=50E_{F}^{-1}$. (b) The pseudogap-like
energy scale $\Delta$ as a function of $\eta$ for different lifetime
values $\tau_{0}$. The solid line is the analytical result (\ref{eq:Delta}).
\label{fig:peaks}}
\end{figure}

Inside the critical phase, the nematic susceptibility $\chi_{\mathrm{nem}}^{ij}\left(0,\boldsymbol{q}\right)=\left(\left|\boldsymbol{q}\right|\big/k_{F}\right)^{\eta-2}\chi_{0}\delta_{ij}$
is characterized by the anomalous exponent $\eta$. Related to the
phase-fluctuation stiffness $\rho_{s}$ by $\eta(T)=T/\left(2\pi\rho_{s}\right)$,
$\eta$ acquires the universal value $\eta\left(T_{\mathrm{BKT},1}\right)=1/4$
at the upper BKT transition and decreases continuously until $\eta\left(T_{\mathrm{BKT},2}\right)=1/9$
is reached at the lower BKT transition \citep{Jose1977}, where long-range
nematic order onsets. Note that, since the critical phase occurs at
finite temperatures, we set the frequency of the bosonic propagator
to zero in our calculation of the self-energy; such a quasistatic
approximation is commonly employed to describe the effects of fluctuating
order on the electronic spectrum at non-zero temperatures \citep{Schmalian1998,Chubukov2010,Lin_Millis}.
Moreover, we also introduce a quasiparticle lifetime $\tau_{0}^{-1}$
to model the thermal broadening of the spectral function arising from
correlations not associated with nematicity, such that $\mathcal{G}^{-1}=\omega-\boldsymbol{v}_{F}\cdot\boldsymbol{q}+i\tau_{0}^{-1}/2$,
with Fermi velocity $\boldsymbol{v}_{F}$. Computing the one-loop
self-energy at the Fermi momentum $k_{F}$, we find (see SM):

\begin{equation}
\Sigma\left(\omega,\boldsymbol{k}_{F}\right)=\kappa Tc_{\eta}e^{-i\frac{\eta\pi}{2}}\left(\frac{2\omega\tau_{0}+i}{2E_{F}\tau_{0}}\right)^{\eta-1},\label{eq:self_energy}
\end{equation}
where $c_{\eta}=\frac{\pi^{\frac{3}{2}}\Gamma\left(\frac{1}{2}-\frac{\eta}{2}\right)}{2^{\eta}\Gamma\left(1-\frac{\eta}{2}\right)}\frac{1}{\sin\left(\frac{\eta\pi}{2}\right)}$,
$E_{F}=\frac{v_{F}k_{F}}{2}$ is the Fermi energy, and $\kappa=\frac{g^{2}k_{F}^{2}A\chi_{0}}{(2\pi)^{2}E_{F}}$
is a dimensionless parameter. Fig. \ref{fig:peaks}(a) shows the corresponding
electronic spectral function $A\left(\omega,\boldsymbol{k}_{F}\right)=-2\mathrm{Im}\left[\tilde{\mathcal{G}}\left(\omega,\boldsymbol{k}_{F}\right)\right]$,
with $\tilde{\mathcal{G}}^{-1}=\mathcal{G}^{-1}-\Sigma$, for the
values $\eta$ assumes inside the critical phase, $1/9\leq\eta\leq1/4$.
The spectral weight is strongly suppressed at the Fermi level and
transferred to peaks at higher energies $\pm\omega_{\mathrm{peak}}$,
a behavior reminiscent of a pseudogap. As shown in Fig. \ref{fig:peaks}(b),
the pseudogap-like energy scale $\Delta=2\omega_{\mathrm{peak}}$
increases with decreasing $\eta$ and depends very weakly on the lifetime
$\tau_{0}$, being well approximated by the $\tau_{0}\rightarrow\infty$
analytical expression:

\begin{equation}
\Delta\approx2E_{F}\left[\frac{\kappa T}{E_{F}}\frac{a_{\eta}\left(1-2\eta\right)}{\eta+1}\left(\sqrt{1+\frac{3c_{\eta}^{2}\left(1-\eta^{2}\right)}{a_{\eta}^{2}\left(1-2\eta\right)}}-1\right)\right]^{\frac{1}{2-\eta}}\label{eq:Delta}
\end{equation}
where $a_{\eta}\equiv c_{\eta}\cos\left(\frac{\eta\pi}{2}\right)$.
We emphasize that $\Delta\ll E_{F}$ within our weak-coupling approach.
Moreover, as expected within a quasistatic approximation, $\Delta$
is a purely thermal effect that would vanish as $T\rightarrow0$,
which is not reached here since long-range nematic order onsets at
$T_{\mathrm{BKT},2}$. The pseudogap-like behavior found here is qualitatively
different from previously studied mechanisms \citep{Vilk1996,Schmalian1998,Norman1998,Millis1998,Chubukov2010,Sachdev2019},
as it arises entirely from the nematic phase fluctuations in the critical
phase. Nevertheless, it is interesting to note that finite-energy
peaks in the spectral function were also obtained in Ref. \citep{Dasgupta2011},
which included the anomalous exponent in their model of pairing phase
fluctuations.

The behaviors discussed so far occur at finite temperatures. This
motivates us to ask if a tuning parameter can suppress the BKT transitions
and promote quantum critical fluctuations. Using group-theory, we
find that an out-of-plane magnetic field $B_{z}$ induces nematic
phase fluctuations, via the following coupling in the nematic action
$\mathcal{S}\left[\boldsymbol{\phi}_{+}\right]=\mathcal{F}\left[\boldsymbol{\phi}_{+}\right]/T$:

\begin{equation}
\mathcal{S}\left[\boldsymbol{\phi}_{+}\right]=\lambda B_{z}\int_{r}\left(\boldsymbol{\phi}_{+}i\rho^{y}\dot{\boldsymbol{\phi}}_{+}\right)=\lambda B_{z}\int_{r}\left|\boldsymbol{\phi}_{+}\left(r\right)\right|^{2}\,\frac{d\alpha_{+}\left(r\right)}{dt}\label{eq:Bz}
\end{equation}
where $r=\left(\mathbf{r},t\right)$, $\lambda$ is a coupling constant,
and $\rho^{i}$ are nematic-space Pauli matrices. Thus, $B_{z}$ acts
similarly to a transverse field to the emergent XY nematic order parameter,
which is expected to suppress nematic order. This term is the nematic
analogue of the (non-dissipative) viscoelastic response of a hexagonal
lattice, by which external shear stress $\sigma_{xy}$ induces a time-changing
transverse strain $\epsilon_{x^{2}-y^{2}}$ via $\sigma_{xy}=\eta_{H}\,\frac{d\epsilon_{x^{2}-y^{2}}}{dt}$
\citep{Bradlyn2012,Barkeshli2012,Link2018}. Therefore, $\lambda B_{z}$
plays a similar role as the Hall viscosity coefficient $\eta_{H}$.
Interestingly, a term analogous to Eq. (\ref{eq:Bz}) was shown to
emerge in nematic quantum Hall states as a result of the explicitly
broken time-reversal symmetry \citep{Maciejko2013,Fradkin2013,Fradkin2014}. 

The $\left(B_{z},\,T\right)$ nematic phase diagram of the $30^{\circ}$-twisted
bilayer should then be similar to the phase diagram of the quantum
six-state clock model previously investigated in other contexts \citep{Xu2020,Podolsky2016,Arnold2022},
as schematically shown in Fig. \ref{fig:Z6_clock}(b). Because at
$T=0$ the system is above the upper critical dimension $d_{u}\lesssim3$
of the $Z_{6}$-clock model, the coefficient of the $q=6$ term in
the free energy (\ref{eq:F_nem}) is dangerously irrelevant, resulting
in an XY nematic QCP and in a pseudo-Goldstone mode inside the $T=0$
nematically ordered state \citep{Oshikawa2000,Sandvik2021}. As discussed
in Ref. \citep{Mandal2023}, its coupling to the low-energy electronic
states results in a crossover from non-Fermi-liquid to Fermi-liquid
behavior. An interesting question is how the system interpolates between
the pseudogap-like behavior in the critical phase ($T_{\mathrm{BKT},2}\leq T\leq T_{\mathrm{BKT},1}$)
and the non-Fermi-liquid to Fermi-liquid crossover at $T=0$ as $T_{\mathrm{BKT},i}\rightarrow0$.
Addressing this question will require incorporating nematic temporal
fluctuations, which is beyond the scope of this work.

In summary, we established the nematic phase diagram of the $30^{\circ}$-twisted
hexagonal bilayer. Its non-crystallographic symmetries endow nematicity
with an enhanced $Z_{6}$-clock symmetry that is forbidden in any
2D crystalline system, leading to a critical phase displaying quasi-long-range
nematic order over an extended temperature range. Phase fluctuations
in this critical phase have a pronounced impact on the electronic
spectrum, promoting a thermal pseudogap-like behavior. The two BKT
transitions bounding the critical phase can be suppressed by a perpendicular
magnetic field, which promotes quantum nematic phase fluctuations
via a mechanism analogous to the Hall viscoelastic response of the
lattice. An interesting direction is whether superconductivity emerges
at this putative QCP. While it is well-established that Ising-nematic
quantum critical fluctuations lead to enhanced pairing \citep{Lederer2017,Klein2018},
the case of an emergent XY nematic demands further investigations.

Experimentally, $30^{\circ}$-twisted bilayer graphene has been realized
\citep{Ahn2018}. However, nematic order has not yet been observed
in graphene, although it is theoretically predicted to emerge at the
van Hove filling \citep{Valenzuela2008,Kiesel2013}; furthermore,
nematic order has been experimentally reported in Bernal bilayer graphene
\citep{Novoselov2011}. Importantly, recent advances in twisting non
van-der-Waals materials \citep{Shen2022} suggest the potential feasibility
of $30^{\circ}$-twisted bilayers of materials beyond graphene. In
this regard, nematic order is observed in several compounds whose
lattices have threefold rotational symmetry, such as doped Bi$_{2}$Se$_{3}$
\citep{Tamegai2019,Cho2020}, bismuth \citep{Feldman2016}, intercalated
Fe$_{1/3}$NbS$_{2}$ \citep{Little2020,Analytis2020} and the transition
metal phosphorous trichalcogenides $M$P$X_{3}$ which, by virtue
of their van der Waals bonding, can be grown in few-layer form \citep{LiangWu2023,Hwangbo2023,Sun2023,Tan2023}.
Besides crystals, $Z_{3}$-Potts nematicity has been observed in optical
lattices \citep{Jin2021}, which are also amenable to twisting \citep{Cirac2019}.
We emphasize that our analysis shows that the critical $Z_{6}$ nematic
phase emerges not only when the twist angle of the hexagonal bilayer
is $30^{\circ}$, but even in the untwisted case, provided that the
nematic order parameter changes sign between the two layers. Such
an odd-parity nematic order is the counterpart of the time-reversal-odd
spin- and valley-polarized nematic orders \citep{Wu2007,Classen2020,Chichinadze2020,Xu2020,Mandal2023}.
The latter, proposed to emerge in threefold symmetric systems near
van Hove fillings, including twisted bilayer graphene, should also
support a critical nematic phase that displays pseudogap-like behavior.
\begin{acknowledgments}
We thank A. Chubukov, I. Mandal, J. Schmalian, C. Xu for fruitful
discussions. This work was supported by the U.S. Department of Energy,
Office of Science, Basic Energy Sciences, Materials Sciences and Engineering
Division, under Award No. DE-SC0020045.
\end{acknowledgments}

\bibliographystyle{apsrev4-1-titles}
\bibliography{refs_twisted_nematics}

\begin{thebibliography}{85}%
\makeatletter
\providecommand \@ifxundefined [1]{%
 \@ifx{#1\undefined}
}%
\providecommand \@ifnum [1]{%
 \ifnum #1\expandafter \@firstoftwo
 \else \expandafter \@secondoftwo
 \fi
}%
\providecommand \@ifx [1]{%
 \ifx #1\expandafter \@firstoftwo
 \else \expandafter \@secondoftwo
 \fi
}%
\providecommand \natexlab [1]{#1}%
\providecommand \enquote  [1]{``#1''}%
\providecommand \bibnamefont  [1]{#1}%
\providecommand \bibfnamefont [1]{#1}%
\providecommand \citenamefont [1]{#1}%
\providecommand \href@noop [0]{\@secondoftwo}%
\providecommand \href [0]{\begingroup \@sanitize@url \@href}%
\providecommand \@href[1]{\@@startlink{#1}\@@href}%
\providecommand \@@href[1]{\endgroup#1\@@endlink}%
\providecommand \@sanitize@url [0]{\catcode `\\12\catcode `\$12\catcode
  `\&12\catcode `\#12\catcode `\^12\catcode `\_12\catcode `\%12\relax}%
\providecommand \@@startlink[1]{}%
\providecommand \@@endlink[0]{}%
\providecommand \url  [0]{\begingroup\@sanitize@url \@url }%
\providecommand \@url [1]{\endgroup\@href {#1}{\urlprefix }}%
\providecommand \urlprefix  [0]{URL }%
\providecommand \Eprint [0]{\href }%
\providecommand \doibase [0]{http://doi.org/}%
\providecommand \selectlanguage [0]{\@gobble}%
\providecommand \bibinfo  [0]{\@secondoftwo}%
\providecommand \bibfield  [0]{\@secondoftwo}%
\providecommand \translation [1]{[#1]}%
\providecommand \BibitemOpen [0]{}%
\providecommand \bibitemStop [0]{}%
\providecommand \bibitemNoStop [0]{.\EOS\space}%
\providecommand \EOS [0]{\spacefactor3000\relax}%
\providecommand \BibitemShut  [1]{\csname bibitem#1\endcsname}%
\let\auto@bib@innerbib\@empty
%</preamble>
\bibitem [{\citenamefont {Cao}\ \emph {et~al.}(2018{\natexlab{a}})\citenamefont
  {Cao}, \citenamefont {Fatemi}, \citenamefont {Fang}, \citenamefont
  {Watanabe}, \citenamefont {Taniguchi}, \citenamefont {Kaxiras},\ and\
  \citenamefont {Jarillo-Herrero}}]{Cao2018_a}%
  \BibitemOpen
  \bibfield  {author} {\bibinfo {author} {\bibfnamefont {Y.}~\bibnamefont
  {Cao}}, \bibinfo {author} {\bibfnamefont {V.}~\bibnamefont {Fatemi}},
  \bibinfo {author} {\bibfnamefont {S.}~\bibnamefont {Fang}}, \bibinfo {author}
  {\bibfnamefont {K.}~\bibnamefont {Watanabe}}, \bibinfo {author}
  {\bibfnamefont {T.}~\bibnamefont {Taniguchi}}, \bibinfo {author}
  {\bibfnamefont {E.}~\bibnamefont {Kaxiras}}, \ and\ \bibinfo {author}
  {\bibfnamefont {P.}~\bibnamefont {Jarillo-Herrero}},\ }\bibfield  {title}
  {\emph {\bibinfo {title} {Unconventional superconductivity in magic-angle
  graphene superlattices}},\ }\href
  {https://www.nature.com/articles/nature26160} {\bibfield  {journal} {\bibinfo
   {journal} {Nature}\ }\textbf {\bibinfo {volume} {556}},\ \bibinfo {pages}
  {43} (\bibinfo {year} {2018}{\natexlab{a}})}\BibitemShut {NoStop}%
\bibitem [{\citenamefont {Cao}\ \emph {et~al.}(2018{\natexlab{b}})\citenamefont
  {Cao}, \citenamefont {Fatemi}, \citenamefont {Demir}, \citenamefont {Fang},
  \citenamefont {Tomarken}, \citenamefont {Luo}, \citenamefont
  {Sanchez-Yamagishi}, \citenamefont {Watanabe}, \citenamefont {Taniguchi},
  \citenamefont {Kaxiras} \emph {et~al.}}]{Cao2018_b}%
  \BibitemOpen
  \bibfield  {author} {\bibinfo {author} {\bibfnamefont {Y.}~\bibnamefont
  {Cao}}, \bibinfo {author} {\bibfnamefont {V.}~\bibnamefont {Fatemi}},
  \bibinfo {author} {\bibfnamefont {A.}~\bibnamefont {Demir}}, \bibinfo
  {author} {\bibfnamefont {S.}~\bibnamefont {Fang}}, \bibinfo {author}
  {\bibfnamefont {S.~L.}\ \bibnamefont {Tomarken}}, \bibinfo {author}
  {\bibfnamefont {J.~Y.}\ \bibnamefont {Luo}}, \bibinfo {author} {\bibfnamefont
  {J.~D.}\ \bibnamefont {Sanchez-Yamagishi}}, \bibinfo {author} {\bibfnamefont
  {K.}~\bibnamefont {Watanabe}}, \bibinfo {author} {\bibfnamefont
  {T.}~\bibnamefont {Taniguchi}}, \bibinfo {author} {\bibfnamefont
  {E.}~\bibnamefont {Kaxiras}},  \emph {et~al.},\ }\bibfield  {title} {\emph
  {\bibinfo {title} {Correlated insulator behaviour at half-filling in
  magic-angle graphene superlattices}},\ }\href
  {https://www.nature.com/articles/nature26154} {\bibfield  {journal} {\bibinfo
   {journal} {Nature}\ }\textbf {\bibinfo {volume} {556}},\ \bibinfo {pages}
  {80} (\bibinfo {year} {2018}{\natexlab{b}})}\BibitemShut {NoStop}%
\bibitem [{\citenamefont {Yankowitz}\ \emph {et~al.}(2019)\citenamefont
  {Yankowitz}, \citenamefont {Chen}, \citenamefont {Polshyn}, \citenamefont
  {Zhang}, \citenamefont {Watanabe}, \citenamefont {Taniguchi}, \citenamefont
  {Graf}, \citenamefont {Young},\ and\ \citenamefont {Dean}}]{Yankowitz2019}%
  \BibitemOpen
  \bibfield  {author} {\bibinfo {author} {\bibfnamefont {M.}~\bibnamefont
  {Yankowitz}}, \bibinfo {author} {\bibfnamefont {S.}~\bibnamefont {Chen}},
  \bibinfo {author} {\bibfnamefont {H.}~\bibnamefont {Polshyn}}, \bibinfo
  {author} {\bibfnamefont {Y.}~\bibnamefont {Zhang}}, \bibinfo {author}
  {\bibfnamefont {K.}~\bibnamefont {Watanabe}}, \bibinfo {author}
  {\bibfnamefont {T.}~\bibnamefont {Taniguchi}}, \bibinfo {author}
  {\bibfnamefont {D.}~\bibnamefont {Graf}}, \bibinfo {author} {\bibfnamefont
  {A.~F.}\ \bibnamefont {Young}}, \ and\ \bibinfo {author} {\bibfnamefont
  {C.~R.}\ \bibnamefont {Dean}},\ }\bibfield  {title} {\emph {\bibinfo {title}
  {Tuning superconductivity in twisted bilayer graphene}},\ }\href
  {https://www.science.org/doi/10.1126/science.aav1910} {\bibfield  {journal}
  {\bibinfo  {journal} {Science}\ }\textbf {\bibinfo {volume} {363}},\ \bibinfo
  {pages} {1059} (\bibinfo {year} {2019})}\BibitemShut {NoStop}%
\bibitem [{\citenamefont {Sharpe}\ \emph {et~al.}(2019)\citenamefont {Sharpe},
  \citenamefont {Fox}, \citenamefont {Barnard}, \citenamefont {Finney},
  \citenamefont {Watanabe}, \citenamefont {Taniguchi}, \citenamefont
  {Kastner},\ and\ \citenamefont {Goldhaber-Gordon}}]{Sharpe2019}%
  \BibitemOpen
  \bibfield  {author} {\bibinfo {author} {\bibfnamefont {A.~L.}\ \bibnamefont
  {Sharpe}}, \bibinfo {author} {\bibfnamefont {E.~J.}\ \bibnamefont {Fox}},
  \bibinfo {author} {\bibfnamefont {A.~W.}\ \bibnamefont {Barnard}}, \bibinfo
  {author} {\bibfnamefont {J.}~\bibnamefont {Finney}}, \bibinfo {author}
  {\bibfnamefont {K.}~\bibnamefont {Watanabe}}, \bibinfo {author}
  {\bibfnamefont {T.}~\bibnamefont {Taniguchi}}, \bibinfo {author}
  {\bibfnamefont {M.}~\bibnamefont {Kastner}}, \ and\ \bibinfo {author}
  {\bibfnamefont {D.}~\bibnamefont {Goldhaber-Gordon}},\ }\bibfield  {title}
  {\emph {\bibinfo {title} {Emergent ferromagnetism near three-quarters filling
  in twisted bilayer graphene}},\ }\href
  {https://www.science.org/doi/10.1126/science.aaw3780} {\bibfield  {journal}
  {\bibinfo  {journal} {Science}\ }\textbf {\bibinfo {volume} {365}},\ \bibinfo
  {pages} {605} (\bibinfo {year} {2019})}\BibitemShut {NoStop}%
\bibitem [{\citenamefont {Lu}\ \emph {et~al.}(2019)\citenamefont {Lu},
  \citenamefont {Stepanov}, \citenamefont {Yang}, \citenamefont {Xie},
  \citenamefont {Aamir}, \citenamefont {Das}, \citenamefont {Urgell},
  \citenamefont {Watanabe}, \citenamefont {Taniguchi}, \citenamefont {Zhang}
  \emph {et~al.}}]{Lu2019}%
  \BibitemOpen
  \bibfield  {author} {\bibinfo {author} {\bibfnamefont {X.}~\bibnamefont
  {Lu}}, \bibinfo {author} {\bibfnamefont {P.}~\bibnamefont {Stepanov}},
  \bibinfo {author} {\bibfnamefont {W.}~\bibnamefont {Yang}}, \bibinfo {author}
  {\bibfnamefont {M.}~\bibnamefont {Xie}}, \bibinfo {author} {\bibfnamefont
  {M.~A.}\ \bibnamefont {Aamir}}, \bibinfo {author} {\bibfnamefont
  {I.}~\bibnamefont {Das}}, \bibinfo {author} {\bibfnamefont {C.}~\bibnamefont
  {Urgell}}, \bibinfo {author} {\bibfnamefont {K.}~\bibnamefont {Watanabe}},
  \bibinfo {author} {\bibfnamefont {T.}~\bibnamefont {Taniguchi}}, \bibinfo
  {author} {\bibfnamefont {G.}~\bibnamefont {Zhang}},  \emph {et~al.},\
  }\bibfield  {title} {\emph {\bibinfo {title} {Superconductors, orbital
  magnets and correlated states in magic-angle bilayer graphene}},\ }\href
  {https://www.nature.com/articles/s41586-019-1695-0} {\bibfield  {journal}
  {\bibinfo  {journal} {Nature}\ }\textbf {\bibinfo {volume} {574}},\ \bibinfo
  {pages} {653} (\bibinfo {year} {2019})}\BibitemShut {NoStop}%
\bibitem [{\citenamefont {Kennes}\ \emph {et~al.}(2021)\citenamefont {Kennes},
  \citenamefont {Claassen}, \citenamefont {Xian}, \citenamefont {Georges},
  \citenamefont {Millis}, \citenamefont {Hone}, \citenamefont {Dean},
  \citenamefont {Basov}, \citenamefont {Pasupathy},\ and\ \citenamefont
  {Rubio}}]{Kennes2021}%
  \BibitemOpen
  \bibfield  {author} {\bibinfo {author} {\bibfnamefont {D.~M.}\ \bibnamefont
  {Kennes}}, \bibinfo {author} {\bibfnamefont {M.}~\bibnamefont {Claassen}},
  \bibinfo {author} {\bibfnamefont {L.}~\bibnamefont {Xian}}, \bibinfo {author}
  {\bibfnamefont {A.}~\bibnamefont {Georges}}, \bibinfo {author} {\bibfnamefont
  {A.~J.}\ \bibnamefont {Millis}}, \bibinfo {author} {\bibfnamefont
  {J.}~\bibnamefont {Hone}}, \bibinfo {author} {\bibfnamefont {C.~R.}\
  \bibnamefont {Dean}}, \bibinfo {author} {\bibfnamefont {D.}~\bibnamefont
  {Basov}}, \bibinfo {author} {\bibfnamefont {A.~N.}\ \bibnamefont
  {Pasupathy}}, \ and\ \bibinfo {author} {\bibfnamefont {A.}~\bibnamefont
  {Rubio}},\ }\bibfield  {title} {\emph {\bibinfo {title} {Moir{\'e}
  heterostructures as a condensed-matter quantum simulator}},\ }\href
  {https://www.nature.com/articles/s41567-020-01154-3} {\bibfield  {journal}
  {\bibinfo  {journal} {Nature Physics}\ }\textbf {\bibinfo {volume} {17}},\
  \bibinfo {pages} {155} (\bibinfo {year} {2021})}\BibitemShut {NoStop}%
\bibitem [{\citenamefont {Mak}\ and\ \citenamefont {Shan}(2022)}]{Mak2022}%
  \BibitemOpen
  \bibfield  {author} {\bibinfo {author} {\bibfnamefont {K.~F.}\ \bibnamefont
  {Mak}}\ and\ \bibinfo {author} {\bibfnamefont {J.}~\bibnamefont {Shan}},\
  }\bibfield  {title} {\emph {\bibinfo {title} {Semiconductor moir{\'e}
  materials}},\ }\href {https://www.nature.com/articles/s41565-022-01165-6}
  {\bibfield  {journal} {\bibinfo  {journal} {Nature Nanotechnology}\ }\textbf
  {\bibinfo {volume} {17}},\ \bibinfo {pages} {686} (\bibinfo {year}
  {2022})}\BibitemShut {NoStop}%
\bibitem [{\citenamefont {Shen}\ \emph {et~al.}(2022)\citenamefont {Shen},
  \citenamefont {Dong}, \citenamefont {Qi}, \citenamefont {Zhang},
  \citenamefont {Zhu}, \citenamefont {Wu},\ and\ \citenamefont
  {Li}}]{Shen2022}%
  \BibitemOpen
  \bibfield  {author} {\bibinfo {author} {\bibfnamefont {J.}~\bibnamefont
  {Shen}}, \bibinfo {author} {\bibfnamefont {Z.}~\bibnamefont {Dong}}, \bibinfo
  {author} {\bibfnamefont {M.}~\bibnamefont {Qi}}, \bibinfo {author}
  {\bibfnamefont {Y.}~\bibnamefont {Zhang}}, \bibinfo {author} {\bibfnamefont
  {C.}~\bibnamefont {Zhu}}, \bibinfo {author} {\bibfnamefont {Z.}~\bibnamefont
  {Wu}}, \ and\ \bibinfo {author} {\bibfnamefont {D.}~\bibnamefont {Li}},\
  }\bibfield  {title} {\emph {\bibinfo {title} {{Observation of Moir{\'e}
  Patterns in Twisted Stacks of Bilayer Perovskite Oxide Nanomembranes with
  Various Lattice Symmetries}}},\ }\href
  {https://pubs.acs.org/doi/10.1021/acsami.2c14746} {\bibfield  {journal}
  {\bibinfo  {journal} {ACS Applied Materials \& Interfaces}\ }\textbf
  {\bibinfo {volume} {14}},\ \bibinfo {pages} {50386} (\bibinfo {year}
  {2022})}\BibitemShut {NoStop}%
\bibitem [{\citenamefont {Zhang}\ \emph {et~al.}(2022)\citenamefont {Zhang},
  \citenamefont {Polski}, \citenamefont {Lewandowski}, \citenamefont {Thomson},
  \citenamefont {Peng}, \citenamefont {Choi}, \citenamefont {Kim},
  \citenamefont {Watanabe}, \citenamefont {Taniguchi}, \citenamefont {Alicea}
  \emph {et~al.}}]{Zhang2022}%
  \BibitemOpen
  \bibfield  {author} {\bibinfo {author} {\bibfnamefont {Y.}~\bibnamefont
  {Zhang}}, \bibinfo {author} {\bibfnamefont {R.}~\bibnamefont {Polski}},
  \bibinfo {author} {\bibfnamefont {C.}~\bibnamefont {Lewandowski}}, \bibinfo
  {author} {\bibfnamefont {A.}~\bibnamefont {Thomson}}, \bibinfo {author}
  {\bibfnamefont {Y.}~\bibnamefont {Peng}}, \bibinfo {author} {\bibfnamefont
  {Y.}~\bibnamefont {Choi}}, \bibinfo {author} {\bibfnamefont {H.}~\bibnamefont
  {Kim}}, \bibinfo {author} {\bibfnamefont {K.}~\bibnamefont {Watanabe}},
  \bibinfo {author} {\bibfnamefont {T.}~\bibnamefont {Taniguchi}}, \bibinfo
  {author} {\bibfnamefont {J.}~\bibnamefont {Alicea}},  \emph {et~al.},\
  }\bibfield  {title} {\emph {\bibinfo {title} {Promotion of superconductivity
  in magic-angle graphene multilayers}},\ }\href
  {https://www.science.org/doi/10.1126/science.abn8585} {\bibfield  {journal}
  {\bibinfo  {journal} {Science}\ }\textbf {\bibinfo {volume} {377}},\ \bibinfo
  {pages} {1538} (\bibinfo {year} {2022})}\BibitemShut {NoStop}%
\bibitem [{\citenamefont {Uri}\ \emph {et~al.}(2023)\citenamefont {Uri},
  \citenamefont {de~la Barrera}, \citenamefont {Randeria}, \citenamefont
  {Rodan-Legrain}, \citenamefont {Devakul}, \citenamefont {Crowley},
  \citenamefont {Paul}, \citenamefont {Watanabe}, \citenamefont {Taniguchi},
  \citenamefont {Lifshitz}, \citenamefont {Fu}, \citenamefont {Ashoori},\ and\
  \citenamefont {Jarillo-Herrero}}]{Uri2023superconductivity}%
  \BibitemOpen
  \bibfield  {author} {\bibinfo {author} {\bibfnamefont {A.}~\bibnamefont
  {Uri}}, \bibinfo {author} {\bibfnamefont {S.~C.}\ \bibnamefont {de~la
  Barrera}}, \bibinfo {author} {\bibfnamefont {M.~T.}\ \bibnamefont
  {Randeria}}, \bibinfo {author} {\bibfnamefont {D.}~\bibnamefont
  {Rodan-Legrain}}, \bibinfo {author} {\bibfnamefont {T.}~\bibnamefont
  {Devakul}}, \bibinfo {author} {\bibfnamefont {P.~J.~D.}\ \bibnamefont
  {Crowley}}, \bibinfo {author} {\bibfnamefont {N.}~\bibnamefont {Paul}},
  \bibinfo {author} {\bibfnamefont {K.}~\bibnamefont {Watanabe}}, \bibinfo
  {author} {\bibfnamefont {T.}~\bibnamefont {Taniguchi}}, \bibinfo {author}
  {\bibfnamefont {R.}~\bibnamefont {Lifshitz}}, \bibinfo {author}
  {\bibfnamefont {L.}~\bibnamefont {Fu}}, \bibinfo {author} {\bibfnamefont
  {R.~C.}\ \bibnamefont {Ashoori}}, \ and\ \bibinfo {author} {\bibfnamefont
  {P.}~\bibnamefont {Jarillo-Herrero}},\ }\bibfield  {title} {\emph {\bibinfo
  {title} {Superconductivity and strong interactions in a tunable moir{\'e}
  quasicrystal}},\ }\href {https://www.nature.com/articles/s41586-023-06294-z}
  {\bibfield  {journal} {\bibinfo  {journal} {Nature (London)}\ }\textbf
  {\bibinfo {volume} {620}},\ \bibinfo {pages} {762} (\bibinfo {year}
  {2023})}\BibitemShut {NoStop}%
\bibitem [{\citenamefont {Ghorashi}\ \emph {et~al.}(2023)\citenamefont
  {Ghorashi}, \citenamefont {Dunbrack}, \citenamefont {Abouelkomsan},
  \citenamefont {Sun}, \citenamefont {Du},\ and\ \citenamefont
  {Cano}}]{Cano2023}%
  \BibitemOpen
  \bibfield  {author} {\bibinfo {author} {\bibfnamefont {S.~A.~A.}\
  \bibnamefont {Ghorashi}}, \bibinfo {author} {\bibfnamefont {A.}~\bibnamefont
  {Dunbrack}}, \bibinfo {author} {\bibfnamefont {A.}~\bibnamefont
  {Abouelkomsan}}, \bibinfo {author} {\bibfnamefont {J.}~\bibnamefont {Sun}},
  \bibinfo {author} {\bibfnamefont {X.}~\bibnamefont {Du}}, \ and\ \bibinfo
  {author} {\bibfnamefont {J.}~\bibnamefont {Cano}},\ }\bibfield  {title}
  {\emph {\bibinfo {title} {{Topological and Stacked Flat Bands in Bilayer
  Graphene with a Superlattice Potential}}},\ }\href
  {https://journals.aps.org/prl/abstract/10.1103/PhysRevLett.130.196201}
  {\bibfield  {journal} {\bibinfo  {journal} {Phys. Rev. Lett.}\ }\textbf
  {\bibinfo {volume} {130}},\ \bibinfo {pages} {196201} (\bibinfo {year}
  {2023})}\BibitemShut {NoStop}%
\bibitem [{\citenamefont {Bistritzer}\ and\ \citenamefont
  {MacDonald}(2011)}]{Bistritzer2011}%
  \BibitemOpen
  \bibfield  {author} {\bibinfo {author} {\bibfnamefont {R.}~\bibnamefont
  {Bistritzer}}\ and\ \bibinfo {author} {\bibfnamefont {A.~H.}\ \bibnamefont
  {MacDonald}},\ }\bibfield  {title} {\emph {\bibinfo {title} {Moir{\'e} bands
  in twisted double-layer graphene}},\ }\href
  {https://www.pnas.org/doi/10.1073/pnas.1108174108} {\bibfield  {journal}
  {\bibinfo  {journal} {Proceedings of the National Academy of Sciences}\
  }\textbf {\bibinfo {volume} {108}},\ \bibinfo {pages} {12233} (\bibinfo
  {year} {2011})}\BibitemShut {NoStop}%
\bibitem [{\citenamefont {Koshino}\ and\ \citenamefont
  {Son}(2019)}]{Koshino2019}%
  \BibitemOpen
  \bibfield  {author} {\bibinfo {author} {\bibfnamefont {M.}~\bibnamefont
  {Koshino}}\ and\ \bibinfo {author} {\bibfnamefont {Y.-W.}\ \bibnamefont
  {Son}},\ }\bibfield  {title} {\emph {\bibinfo {title} {Moir{\'e} phonons in
  twisted bilayer graphene}},\ }\href {\doibase 10.1103/PhysRevB.100.075416}
  {\bibfield  {journal} {\bibinfo  {journal} {Phys. Rev. B}\ }\textbf {\bibinfo
  {volume} {100}},\ \bibinfo {pages} {075416} (\bibinfo {year}
  {2019})}\BibitemShut {NoStop}%
\bibitem [{\citenamefont {Ochoa}(2019)}]{Ochoa2019}%
  \BibitemOpen
  \bibfield  {author} {\bibinfo {author} {\bibfnamefont {H.}~\bibnamefont
  {Ochoa}},\ }\bibfield  {title} {\emph {\bibinfo {title} {Moir{\'e}-pattern
  fluctuations and electron-phason coupling in twisted bilayer graphene}},\
  }\href {\doibase 10.1103/PhysRevB.100.155426} {\bibfield  {journal} {\bibinfo
   {journal} {Phys. Rev. B}\ }\textbf {\bibinfo {volume} {100}},\ \bibinfo
  {pages} {155426} (\bibinfo {year} {2019})}\BibitemShut {NoStop}%
\bibitem [{\citenamefont {Ochoa}\ and\ \citenamefont
  {Fernandes}(2022)}]{Ochoa2022}%
  \BibitemOpen
  \bibfield  {author} {\bibinfo {author} {\bibfnamefont {H.}~\bibnamefont
  {Ochoa}}\ and\ \bibinfo {author} {\bibfnamefont {R.~M.}\ \bibnamefont
  {Fernandes}},\ }\bibfield  {title} {\emph {\bibinfo {title} {{Degradation of
  Phonons in Disordered Moir{\'e} Superlattices}}},\ }\href
  {https://link.aps.org/doi/10.1103/PhysRevLett.128.065901} {\bibfield
  {journal} {\bibinfo  {journal} {Phys. Rev. Lett.}\ }\textbf {\bibinfo
  {volume} {128}},\ \bibinfo {pages} {065901} (\bibinfo {year}
  {2022})}\BibitemShut {NoStop}%
\bibitem [{\citenamefont {Gao}\ and\ \citenamefont {Khalaf}(2022)}]{Gao2022}%
  \BibitemOpen
  \bibfield  {author} {\bibinfo {author} {\bibfnamefont {Q.}~\bibnamefont
  {Gao}}\ and\ \bibinfo {author} {\bibfnamefont {E.}~\bibnamefont {Khalaf}},\
  }\bibfield  {title} {\emph {\bibinfo {title} {{Symmetry origin of lattice
  vibration modes in twisted multilayer graphene: Phasons versus moir{\'e}
  phonons}}},\ }\href {\doibase 10.1103/PhysRevB.106.075420} {\bibfield
  {journal} {\bibinfo  {journal} {Phys. Rev. B}\ }\textbf {\bibinfo {volume}
  {106}},\ \bibinfo {pages} {075420} (\bibinfo {year} {2022})}\BibitemShut
  {NoStop}%
\bibitem [{\citenamefont {Samajdar}\ \emph {et~al.}(2022)\citenamefont
  {Samajdar}, \citenamefont {Teng},\ and\ \citenamefont
  {Scheurer}}]{Samajdar2022}%
  \BibitemOpen
  \bibfield  {author} {\bibinfo {author} {\bibfnamefont {R.}~\bibnamefont
  {Samajdar}}, \bibinfo {author} {\bibfnamefont {Y.}~\bibnamefont {Teng}}, \
  and\ \bibinfo {author} {\bibfnamefont {M.~S.}\ \bibnamefont {Scheurer}},\
  }\bibfield  {title} {\emph {\bibinfo {title} {Moir{\'e} phonons and impact of
  electronic symmetry breaking in twisted trilayer graphene}},\ }\href
  {\doibase 10.1103/PhysRevB.106.L201403} {\bibfield  {journal} {\bibinfo
  {journal} {Phys. Rev. B}\ }\textbf {\bibinfo {volume} {106}},\ \bibinfo
  {pages} {L201403} (\bibinfo {year} {2022})}\BibitemShut {NoStop}%
\bibitem [{\citenamefont {Ochoa}\ and\ \citenamefont
  {Fernandes}(2023)}]{Ochoa2023}%
  \BibitemOpen
  \bibfield  {author} {\bibinfo {author} {\bibfnamefont {H.}~\bibnamefont
  {Ochoa}}\ and\ \bibinfo {author} {\bibfnamefont {R.~M.}\ \bibnamefont
  {Fernandes}},\ }\bibfield  {title} {\emph {\bibinfo {title} {{Extended
  linear-in-$T$ resistivity due to electron-phason scattering in moir{\'e}
  superlattices}}},\ }\href {\doibase 10.1103/PhysRevB.108.075168} {\bibfield
  {journal} {\bibinfo  {journal} {Phys. Rev. B}\ }\textbf {\bibinfo {volume}
  {108}},\ \bibinfo {pages} {075168} (\bibinfo {year} {2023})}\BibitemShut
  {NoStop}%
\bibitem [{\citenamefont {Ishizuka}\ \emph {et~al.}(2021)\citenamefont
  {Ishizuka}, \citenamefont {Fahimniya}, \citenamefont {Guinea},\ and\
  \citenamefont {Levitov}}]{Ishizuka2021}%
  \BibitemOpen
  \bibfield  {author} {\bibinfo {author} {\bibfnamefont {H.}~\bibnamefont
  {Ishizuka}}, \bibinfo {author} {\bibfnamefont {A.}~\bibnamefont {Fahimniya}},
  \bibinfo {author} {\bibfnamefont {F.}~\bibnamefont {Guinea}}, \ and\ \bibinfo
  {author} {\bibfnamefont {L.}~\bibnamefont {Levitov}},\ }\bibfield  {title}
  {\emph {\bibinfo {title} {{Purcell-like Enhancement of Electron--Phonon
  Interactions in Long-Period Superlattices: Linear-Temperature Resistivity and
  Cooling Power}}},\ }\href {https://doi.org/10.1021/acs.nanolett.1c00565}
  {\bibfield  {journal} {\bibinfo  {journal} {Nano Letters}\ }\textbf {\bibinfo
  {volume} {21}},\ \bibinfo {pages} {7465} (\bibinfo {year}
  {2021})}\BibitemShut {NoStop}%
\bibitem [{\citenamefont {Wu}\ \emph {et~al.}(2019)\citenamefont {Wu},
  \citenamefont {Hwang},\ and\ \citenamefont {Das~Sarma}}]{DasSarma2019}%
  \BibitemOpen
  \bibfield  {author} {\bibinfo {author} {\bibfnamefont {F.}~\bibnamefont
  {Wu}}, \bibinfo {author} {\bibfnamefont {E.}~\bibnamefont {Hwang}}, \ and\
  \bibinfo {author} {\bibfnamefont {S.}~\bibnamefont {Das~Sarma}},\ }\bibfield
  {title} {\emph {\bibinfo {title} {{Phonon-induced giant linear-in-$T$
  resistivity in magic angle twisted bilayer graphene: Ordinary strangeness and
  exotic superconductivity}}},\ }\href {\doibase 10.1103/PhysRevB.99.165112}
  {\bibfield  {journal} {\bibinfo  {journal} {Phys. Rev. B}\ }\textbf {\bibinfo
  {volume} {99}},\ \bibinfo {pages} {165112} (\bibinfo {year}
  {2019})}\BibitemShut {NoStop}%
\bibitem [{\citenamefont {Lian}\ \emph {et~al.}(2019)\citenamefont {Lian},
  \citenamefont {Wang},\ and\ \citenamefont {Bernevig}}]{Bernevig2019}%
  \BibitemOpen
  \bibfield  {author} {\bibinfo {author} {\bibfnamefont {B.}~\bibnamefont
  {Lian}}, \bibinfo {author} {\bibfnamefont {Z.}~\bibnamefont {Wang}}, \ and\
  \bibinfo {author} {\bibfnamefont {B.~A.}\ \bibnamefont {Bernevig}},\
  }\bibfield  {title} {\emph {\bibinfo {title} {{Twisted Bilayer Graphene: A
  Phonon-Driven Superconductor}}},\ }\href
  {https://link.aps.org/doi/10.1103/PhysRevLett.122.257002} {\bibfield
  {journal} {\bibinfo  {journal} {Phys. Rev. Lett.}\ }\textbf {\bibinfo
  {volume} {122}},\ \bibinfo {pages} {257002} (\bibinfo {year}
  {2019})}\BibitemShut {NoStop}%
\bibitem [{\citenamefont {Liu}\ \emph {et~al.}(2023{\natexlab{a}})\citenamefont
  {Liu}, \citenamefont {Zhou}, \citenamefont {Zhang}, \citenamefont {Chen},\
  and\ \citenamefont {Yang}}]{Yang2023_b}%
  \BibitemOpen
  \bibfield  {author} {\bibinfo {author} {\bibfnamefont {Y.-B.}\ \bibnamefont
  {Liu}}, \bibinfo {author} {\bibfnamefont {J.}~\bibnamefont {Zhou}}, \bibinfo
  {author} {\bibfnamefont {Y.}~\bibnamefont {Zhang}}, \bibinfo {author}
  {\bibfnamefont {W.-Q.}\ \bibnamefont {Chen}}, \ and\ \bibinfo {author}
  {\bibfnamefont {F.}~\bibnamefont {Yang}},\ }\bibfield  {title} {\emph
  {\bibinfo {title} {Making chiral topological superconductors from
  nontopological superconductors through large angle twists}},\ }\href
  {\doibase 10.1103/PhysRevB.108.064508} {\bibfield  {journal} {\bibinfo
  {journal} {Phys. Rev. B}\ }\textbf {\bibinfo {volume} {108}},\ \bibinfo
  {pages} {064508} (\bibinfo {year} {2023}{\natexlab{a}})}\BibitemShut
  {NoStop}%
\bibitem [{\citenamefont {Liu}\ \emph {et~al.}(2023{\natexlab{b}})\citenamefont
  {Liu}, \citenamefont {Zhang}, \citenamefont {Chen},\ and\ \citenamefont
  {Yang}}]{Yang2023_c}%
  \BibitemOpen
  \bibfield  {author} {\bibinfo {author} {\bibfnamefont {Y.-B.}\ \bibnamefont
  {Liu}}, \bibinfo {author} {\bibfnamefont {Y.}~\bibnamefont {Zhang}}, \bibinfo
  {author} {\bibfnamefont {W.-Q.}\ \bibnamefont {Chen}}, \ and\ \bibinfo
  {author} {\bibfnamefont {F.}~\bibnamefont {Yang}},\ }\bibfield  {title}
  {\emph {\bibinfo {title} {High-angular-momentum topological
  superconductivities in twisted bilayer quasicrystal systems}},\ }\href
  {\doibase 10.1103/PhysRevB.107.014501} {\bibfield  {journal} {\bibinfo
  {journal} {Phys. Rev. B}\ }\textbf {\bibinfo {volume} {107}},\ \bibinfo
  {pages} {014501} (\bibinfo {year} {2023}{\natexlab{b}})}\BibitemShut
  {NoStop}%
\bibitem [{\citenamefont {Haenel}\ \emph {et~al.}(2022)\citenamefont {Haenel},
  \citenamefont {Tummuru},\ and\ \citenamefont {Franz}}]{Haenel2022}%
  \BibitemOpen
  \bibfield  {author} {\bibinfo {author} {\bibfnamefont {R.}~\bibnamefont
  {Haenel}}, \bibinfo {author} {\bibfnamefont {T.}~\bibnamefont {Tummuru}}, \
  and\ \bibinfo {author} {\bibfnamefont {M.}~\bibnamefont {Franz}},\ }\bibfield
   {title} {\emph {\bibinfo {title} {Incoherent tunneling and topological
  superconductivity in twisted cuprate bilayers}},\ }\href
  {https://journals.aps.org/prb/abstract/10.1103/PhysRevB.106.104505}
  {\bibfield  {journal} {\bibinfo  {journal} {Phys. Rev. B}\ }\textbf {\bibinfo
  {volume} {106}},\ \bibinfo {pages} {104505} (\bibinfo {year}
  {2022})}\BibitemShut {NoStop}%
\bibitem [{\citenamefont {Can}\ \emph {et~al.}(2021)\citenamefont {Can},
  \citenamefont {Tummuru}, \citenamefont {Day}, \citenamefont {Elfimov},
  \citenamefont {Damascelli},\ and\ \citenamefont {Franz}}]{Can2021}%
  \BibitemOpen
  \bibfield  {author} {\bibinfo {author} {\bibfnamefont {O.}~\bibnamefont
  {Can}}, \bibinfo {author} {\bibfnamefont {T.}~\bibnamefont {Tummuru}},
  \bibinfo {author} {\bibfnamefont {R.~P.}\ \bibnamefont {Day}}, \bibinfo
  {author} {\bibfnamefont {I.}~\bibnamefont {Elfimov}}, \bibinfo {author}
  {\bibfnamefont {A.}~\bibnamefont {Damascelli}}, \ and\ \bibinfo {author}
  {\bibfnamefont {M.}~\bibnamefont {Franz}},\ }\bibfield  {title} {\emph
  {\bibinfo {title} {High-temperature topological superconductivity in twisted
  double-layer copper oxides}},\ }\href
  {https://www.nature.com/articles/s41567-020-01142-7} {\bibfield  {journal}
  {\bibinfo  {journal} {Nature Physics}\ }\textbf {\bibinfo {volume} {17}},\
  \bibinfo {pages} {519} (\bibinfo {year} {2021})}\BibitemShut {NoStop}%
\bibitem [{\citenamefont {Tummuru}\ \emph {et~al.}(2022)\citenamefont
  {Tummuru}, \citenamefont {Lantagne-Hurtubise},\ and\ \citenamefont
  {Franz}}]{Tummuru2022}%
  \BibitemOpen
  \bibfield  {author} {\bibinfo {author} {\bibfnamefont {T.}~\bibnamefont
  {Tummuru}}, \bibinfo {author} {\bibfnamefont {E.}~\bibnamefont
  {Lantagne-Hurtubise}}, \ and\ \bibinfo {author} {\bibfnamefont
  {M.}~\bibnamefont {Franz}},\ }\bibfield  {title} {\emph {\bibinfo {title}
  {Twisted multilayer nodal superconductors}},\ }\href
  {https://link.aps.org/doi/10.1103/PhysRevB.106.014520} {\bibfield  {journal}
  {\bibinfo  {journal} {Phys. Rev. B}\ }\textbf {\bibinfo {volume} {106}},\
  \bibinfo {pages} {014520} (\bibinfo {year} {2022})}\BibitemShut {NoStop}%
\bibitem [{\citenamefont {Song}\ \emph {et~al.}(2022)\citenamefont {Song},
  \citenamefont {Zhang},\ and\ \citenamefont {Vishwanath}}]{Vishwanath2022}%
  \BibitemOpen
  \bibfield  {author} {\bibinfo {author} {\bibfnamefont {X.-Y.}\ \bibnamefont
  {Song}}, \bibinfo {author} {\bibfnamefont {Y.-H.}\ \bibnamefont {Zhang}}, \
  and\ \bibinfo {author} {\bibfnamefont {A.}~\bibnamefont {Vishwanath}},\
  }\bibfield  {title} {\emph {\bibinfo {title} {{Doping a moir{\'e} Mott
  insulator: A $t\ensuremath{-}J$ model study of twisted cuprates}}},\ }\href
  {\doibase 10.1103/PhysRevB.105.L201102} {\bibfield  {journal} {\bibinfo
  {journal} {Phys. Rev. B}\ }\textbf {\bibinfo {volume} {105}},\ \bibinfo
  {pages} {L201102} (\bibinfo {year} {2022})}\BibitemShut {NoStop}%
\bibitem [{\citenamefont {Liu}\ \emph {et~al.}(2023{\natexlab{c}})\citenamefont
  {Liu}, \citenamefont {Zhou}, \citenamefont {Wu},\ and\ \citenamefont
  {Yang}}]{Yang2023_a}%
  \BibitemOpen
  \bibfield  {author} {\bibinfo {author} {\bibfnamefont {Y.-B.}\ \bibnamefont
  {Liu}}, \bibinfo {author} {\bibfnamefont {J.}~\bibnamefont {Zhou}}, \bibinfo
  {author} {\bibfnamefont {C.}~\bibnamefont {Wu}}, \ and\ \bibinfo {author}
  {\bibfnamefont {F.}~\bibnamefont {Yang}},\ }\bibfield  {title} {\emph
  {\bibinfo {title} {{Charge-4e superconductivity and chiral metal in
  45$^\circ$-twisted bilayer cuprates and related bilayers}}},\ }\href
  {https://www.nature.com/articles/s41467-023-43782-2} {\bibfield  {journal}
  {\bibinfo  {journal} {Nature Communications}\ }\textbf {\bibinfo {volume}
  {14}},\ \bibinfo {pages} {7926} (\bibinfo {year}
  {2023}{\natexlab{c}})}\BibitemShut {NoStop}%
\bibitem [{\citenamefont {Volkov}\ \emph
  {et~al.}(2023{\natexlab{a}})\citenamefont {Volkov}, \citenamefont {Wilson},
  \citenamefont {Lucht},\ and\ \citenamefont {Pixley}}]{Volkov2023_a}%
  \BibitemOpen
  \bibfield  {author} {\bibinfo {author} {\bibfnamefont {P.~A.}\ \bibnamefont
  {Volkov}}, \bibinfo {author} {\bibfnamefont {J.~H.}\ \bibnamefont {Wilson}},
  \bibinfo {author} {\bibfnamefont {K.~P.}\ \bibnamefont {Lucht}}, \ and\
  \bibinfo {author} {\bibfnamefont {J.~H.}\ \bibnamefont {Pixley}},\ }\bibfield
   {title} {\emph {\bibinfo {title} {{Current- and Field-Induced Topology in
  Twisted Nodal Superconductors}}},\ }\href
  {https://link.aps.org/doi/10.1103/PhysRevLett.130.186001} {\bibfield
  {journal} {\bibinfo  {journal} {Phys. Rev. Lett.}\ }\textbf {\bibinfo
  {volume} {130}},\ \bibinfo {pages} {186001} (\bibinfo {year}
  {2023}{\natexlab{a}})}\BibitemShut {NoStop}%
\bibitem [{\citenamefont {Volkov}\ \emph
  {et~al.}(2023{\natexlab{b}})\citenamefont {Volkov}, \citenamefont {Wilson},
  \citenamefont {Lucht},\ and\ \citenamefont {Pixley}}]{Volkov2023_b}%
  \BibitemOpen
  \bibfield  {author} {\bibinfo {author} {\bibfnamefont {P.~A.}\ \bibnamefont
  {Volkov}}, \bibinfo {author} {\bibfnamefont {J.~H.}\ \bibnamefont {Wilson}},
  \bibinfo {author} {\bibfnamefont {K.~P.}\ \bibnamefont {Lucht}}, \ and\
  \bibinfo {author} {\bibfnamefont {J.~H.}\ \bibnamefont {Pixley}},\ }\bibfield
   {title} {\emph {\bibinfo {title} {Magic angles and correlations in twisted
  nodal superconductors}},\ }\href {\doibase 10.1103/PhysRevB.107.174506}
  {\bibfield  {journal} {\bibinfo  {journal} {Phys. Rev. B}\ }\textbf {\bibinfo
  {volume} {107}},\ \bibinfo {pages} {174506} (\bibinfo {year}
  {2023}{\natexlab{b}})}\BibitemShut {NoStop}%
\bibitem [{\citenamefont {Yuan}\ \emph {et~al.}(2023)\citenamefont {Yuan},
  \citenamefont {Vituri}, \citenamefont {Berg}, \citenamefont {Spivak},\ and\
  \citenamefont {Kivelson}}]{Yuan2023}%
  \BibitemOpen
  \bibfield  {author} {\bibinfo {author} {\bibfnamefont {A.~C.}\ \bibnamefont
  {Yuan}}, \bibinfo {author} {\bibfnamefont {Y.}~\bibnamefont {Vituri}},
  \bibinfo {author} {\bibfnamefont {E.}~\bibnamefont {Berg}}, \bibinfo {author}
  {\bibfnamefont {B.}~\bibnamefont {Spivak}}, \ and\ \bibinfo {author}
  {\bibfnamefont {S.~A.}\ \bibnamefont {Kivelson}},\ }\bibfield  {title} {\emph
  {\bibinfo {title} {Inhomogeneity-induced time-reversal symmetry breaking in
  cuprate twist junctions}},\ }\href {\doibase 10.1103/PhysRevB.108.L100505}
  {\bibfield  {journal} {\bibinfo  {journal} {Phys. Rev. B}\ }\textbf {\bibinfo
  {volume} {108}},\ \bibinfo {pages} {L100505} (\bibinfo {year}
  {2023})}\BibitemShut {NoStop}%
\bibitem [{\citenamefont {Wang}\ \emph {et~al.}(2023)\citenamefont {Wang},
  \citenamefont {Zhu}, \citenamefont {Bai}, \citenamefont {Wang}, \citenamefont
  {Hu}, \citenamefont {Xie}, \citenamefont {Hu}, \citenamefont {Cui},
  \citenamefont {Huang}, \citenamefont {Chen} \emph {et~al.}}]{Wang2023}%
  \BibitemOpen
  \bibfield  {author} {\bibinfo {author} {\bibfnamefont {H.}~\bibnamefont
  {Wang}}, \bibinfo {author} {\bibfnamefont {Y.}~\bibnamefont {Zhu}}, \bibinfo
  {author} {\bibfnamefont {Z.}~\bibnamefont {Bai}}, \bibinfo {author}
  {\bibfnamefont {Z.}~\bibnamefont {Wang}}, \bibinfo {author} {\bibfnamefont
  {S.}~\bibnamefont {Hu}}, \bibinfo {author} {\bibfnamefont {H.-Y.}\
  \bibnamefont {Xie}}, \bibinfo {author} {\bibfnamefont {X.}~\bibnamefont
  {Hu}}, \bibinfo {author} {\bibfnamefont {J.}~\bibnamefont {Cui}}, \bibinfo
  {author} {\bibfnamefont {M.}~\bibnamefont {Huang}}, \bibinfo {author}
  {\bibfnamefont {J.}~\bibnamefont {Chen}},  \emph {et~al.},\ }\bibfield
  {title} {\emph {\bibinfo {title} {{Prominent Josephson tunneling between
  twisted single copper oxide planes of
  ${\mathrm{Bi}}_{2}{\mathrm{Sr}}_{2-x}{\mathrm{La}}_{x}{\mathrm{Cu}}{\mathrm{O}}_{6+y}$}}},\
  }\href {https://www.nature.com/articles/s41467-023-40525-1} {\bibfield
  {journal} {\bibinfo  {journal} {Nature Communications}\ }\textbf {\bibinfo
  {volume} {14}},\ \bibinfo {pages} {5201} (\bibinfo {year}
  {2023})}\BibitemShut {NoStop}%
\bibitem [{\citenamefont {Zhao}\ \emph {et~al.}(2023)\citenamefont {Zhao},
  \citenamefont {Cui}, \citenamefont {Volkov}, \citenamefont {Yoo},
  \citenamefont {Lee}, \citenamefont {Gardener}, \citenamefont {Akey},
  \citenamefont {Engelke}, \citenamefont {Ronen}, \citenamefont {Zhong},
  \citenamefont {Gu}, \citenamefont {Plugge}, \citenamefont {Tummuru},
  \citenamefont {Kim}, \citenamefont {Franz}, \citenamefont {Pixley},
  \citenamefont {Poccia},\ and\ \citenamefont {Kim}}]{Kim2023}%
  \BibitemOpen
  \bibfield  {author} {\bibinfo {author} {\bibfnamefont {S.~Y.~F.}\
  \bibnamefont {Zhao}}, \bibinfo {author} {\bibfnamefont {X.}~\bibnamefont
  {Cui}}, \bibinfo {author} {\bibfnamefont {P.~A.}\ \bibnamefont {Volkov}},
  \bibinfo {author} {\bibfnamefont {H.}~\bibnamefont {Yoo}}, \bibinfo {author}
  {\bibfnamefont {S.}~\bibnamefont {Lee}}, \bibinfo {author} {\bibfnamefont
  {J.~A.}\ \bibnamefont {Gardener}}, \bibinfo {author} {\bibfnamefont {A.~J.}\
  \bibnamefont {Akey}}, \bibinfo {author} {\bibfnamefont {R.}~\bibnamefont
  {Engelke}}, \bibinfo {author} {\bibfnamefont {Y.}~\bibnamefont {Ronen}},
  \bibinfo {author} {\bibfnamefont {R.}~\bibnamefont {Zhong}}, \bibinfo
  {author} {\bibfnamefont {G.}~\bibnamefont {Gu}}, \bibinfo {author}
  {\bibfnamefont {S.}~\bibnamefont {Plugge}}, \bibinfo {author} {\bibfnamefont
  {T.}~\bibnamefont {Tummuru}}, \bibinfo {author} {\bibfnamefont
  {M.}~\bibnamefont {Kim}}, \bibinfo {author} {\bibfnamefont {M.}~\bibnamefont
  {Franz}}, \bibinfo {author} {\bibfnamefont {J.~H.}\ \bibnamefont {Pixley}},
  \bibinfo {author} {\bibfnamefont {N.}~\bibnamefont {Poccia}}, \ and\ \bibinfo
  {author} {\bibfnamefont {P.}~\bibnamefont {Kim}},\ }\bibfield  {title} {\emph
  {\bibinfo {title} {Time-reversal symmetry breaking superconductivity between
  twisted cuprate superconductors}},\ }\href {\doibase 10.1126/science.abl8371}
  {\bibfield  {journal} {\bibinfo  {journal} {Science}\ ,\ \bibinfo {pages}
  {eabl8371}} (\bibinfo {year} {2023})}\BibitemShut {NoStop}%
\bibitem [{\citenamefont {Kivelson}\ \emph {et~al.}(1998)\citenamefont
  {Kivelson}, \citenamefont {Fradkin},\ and\ \citenamefont
  {Emery}}]{Kivelson1998}%
  \BibitemOpen
  \bibfield  {author} {\bibinfo {author} {\bibfnamefont {S.~A.}\ \bibnamefont
  {Kivelson}}, \bibinfo {author} {\bibfnamefont {E.}~\bibnamefont {Fradkin}}, \
  and\ \bibinfo {author} {\bibfnamefont {V.~J.}\ \bibnamefont {Emery}},\
  }\bibfield  {title} {\emph {\bibinfo {title} {{Electronic liquid-crystal
  phases of a doped Mott insulator}}},\ }\href
  {https://www.nature.com/articles/31177} {\bibfield  {journal} {\bibinfo
  {journal} {Nature}\ }\textbf {\bibinfo {volume} {393}},\ \bibinfo {pages}
  {550} (\bibinfo {year} {1998})}\BibitemShut {NoStop}%
\bibitem [{\citenamefont {Fradkin}\ \emph {et~al.}(2010)\citenamefont
  {Fradkin}, \citenamefont {Kivelson}, \citenamefont {Lawler}, \citenamefont
  {Eisenstein},\ and\ \citenamefont {Mackenzie}}]{Fradkin2010}%
  \BibitemOpen
  \bibfield  {author} {\bibinfo {author} {\bibfnamefont {E.}~\bibnamefont
  {Fradkin}}, \bibinfo {author} {\bibfnamefont {S.~A.}\ \bibnamefont
  {Kivelson}}, \bibinfo {author} {\bibfnamefont {M.~J.}\ \bibnamefont
  {Lawler}}, \bibinfo {author} {\bibfnamefont {J.~P.}\ \bibnamefont
  {Eisenstein}}, \ and\ \bibinfo {author} {\bibfnamefont {A.~P.}\ \bibnamefont
  {Mackenzie}},\ }\bibfield  {title} {\emph {\bibinfo {title} {{Nematic Fermi
  Fluids in Condensed Matter Physics}}},\ }\href
  {https://www.annualreviews.org/doi/abs/10.1146/annurev-conmatphys-070909-103925}
  {\bibfield  {journal} {\bibinfo  {journal} {Annu. Rev. Condens. Matter
  Phys.}\ }\textbf {\bibinfo {volume} {1}},\ \bibinfo {pages} {153} (\bibinfo
  {year} {2010})}\BibitemShut {NoStop}%
\bibitem [{\citenamefont {Fernandes}\ \emph {et~al.}(2014)\citenamefont
  {Fernandes}, \citenamefont {Chubukov},\ and\ \citenamefont
  {Schmalian}}]{Fernandes2014}%
  \BibitemOpen
  \bibfield  {author} {\bibinfo {author} {\bibfnamefont {R.~M.}\ \bibnamefont
  {Fernandes}}, \bibinfo {author} {\bibfnamefont {A.~V.}\ \bibnamefont
  {Chubukov}}, \ and\ \bibinfo {author} {\bibfnamefont {J.}~\bibnamefont
  {Schmalian}},\ }\bibfield  {title} {\emph {\bibinfo {title} {What drives
  nematic order in iron-based superconductors?}},\ }\href
  {https://www.nature.com/articles/nphys2877} {\bibfield  {journal} {\bibinfo
  {journal} {Nature Physics}\ }\textbf {\bibinfo {volume} {10}},\ \bibinfo
  {pages} {97} (\bibinfo {year} {2014})}\BibitemShut {NoStop}%
\bibitem [{\citenamefont {Oganesyan}\ \emph {et~al.}(2001)\citenamefont
  {Oganesyan}, \citenamefont {Kivelson},\ and\ \citenamefont
  {Fradkin}}]{Oganesyan2001}%
  \BibitemOpen
  \bibfield  {author} {\bibinfo {author} {\bibfnamefont {V.}~\bibnamefont
  {Oganesyan}}, \bibinfo {author} {\bibfnamefont {S.~A.}\ \bibnamefont
  {Kivelson}}, \ and\ \bibinfo {author} {\bibfnamefont {E.}~\bibnamefont
  {Fradkin}},\ }\bibfield  {title} {\emph {\bibinfo {title} {{Quantum theory of
  a nematic Fermi fluid}}},\ }\href {\doibase 10.1103/PhysRevB.64.195109}
  {\bibfield  {journal} {\bibinfo  {journal} {Phys. Rev. B}\ }\textbf {\bibinfo
  {volume} {64}},\ \bibinfo {pages} {195109} (\bibinfo {year}
  {2001})}\BibitemShut {NoStop}%
\bibitem [{\citenamefont {Watanabe}\ and\ \citenamefont
  {Vishwanath}(2014)}]{Watanabe2014}%
  \BibitemOpen
  \bibfield  {author} {\bibinfo {author} {\bibfnamefont {H.}~\bibnamefont
  {Watanabe}}\ and\ \bibinfo {author} {\bibfnamefont {A.}~\bibnamefont
  {Vishwanath}},\ }\bibfield  {title} {\emph {\bibinfo {title} {{Criterion for
  stability of Goldstone modes and Fermi liquid behavior in a metal with broken
  symmetry}}},\ }\href {https://www.pnas.org/doi/10.1073/pnas.1415592111}
  {\bibfield  {journal} {\bibinfo  {journal} {Proceedings of the National
  Academy of Sciences}\ }\textbf {\bibinfo {volume} {111}},\ \bibinfo {pages}
  {16314} (\bibinfo {year} {2014})}\BibitemShut {NoStop}%
\bibitem [{\citenamefont {Hecker}\ and\ \citenamefont
  {Schmalian}(2018)}]{Hecker2018}%
  \BibitemOpen
  \bibfield  {author} {\bibinfo {author} {\bibfnamefont {M.}~\bibnamefont
  {Hecker}}\ and\ \bibinfo {author} {\bibfnamefont {J.}~\bibnamefont
  {Schmalian}},\ }\bibfield  {title} {\emph {\bibinfo {title} {{Vestigial
  nematic order and superconductivity in the doped topological insulator
  Cu$_x$Bi$_2$Se$_3$}}},\ }\href
  {https://www.nature.com/articles/s41535-018-0098-z} {\bibfield  {journal}
  {\bibinfo  {journal} {npj Quantum Materials}\ }\textbf {\bibinfo {volume}
  {3}},\ \bibinfo {pages} {26} (\bibinfo {year} {2018})}\BibitemShut {NoStop}%
\bibitem [{\citenamefont {Fernandes}\ and\ \citenamefont
  {Venderbos}(2020)}]{Fernandes2020}%
  \BibitemOpen
  \bibfield  {author} {\bibinfo {author} {\bibfnamefont {R.~M.}\ \bibnamefont
  {Fernandes}}\ and\ \bibinfo {author} {\bibfnamefont {J.~W.}\ \bibnamefont
  {Venderbos}},\ }\bibfield  {title} {\emph {\bibinfo {title} {{Nematicity with
  a twist: Rotational symmetry breaking in a moir{\'e} superlattice}}},\ }\href
  {https://www.science.org/doi/10.1126/sciadv.aba8834} {\bibfield  {journal}
  {\bibinfo  {journal} {Science Advances}\ }\textbf {\bibinfo {volume} {6}},\
  \bibinfo {pages} {eaba8834} (\bibinfo {year} {2020})}\BibitemShut {NoStop}%
\bibitem [{\citenamefont {Chakraborty}\ and\ \citenamefont
  {Fernandes}(2023)}]{Chakraborty2023}%
  \BibitemOpen
  \bibfield  {author} {\bibinfo {author} {\bibfnamefont {A.~R.}\ \bibnamefont
  {Chakraborty}}\ and\ \bibinfo {author} {\bibfnamefont {R.~M.}\ \bibnamefont
  {Fernandes}},\ }\bibfield  {title} {\emph {\bibinfo {title} {{Strain-tuned
  quantum criticality in electronic Potts-nematic systems}}},\ }\href
  {https://journals.aps.org/prb/abstract/10.1103/PhysRevB.107.195136}
  {\bibfield  {journal} {\bibinfo  {journal} {Phys. Rev. B}\ }\textbf {\bibinfo
  {volume} {107}},\ \bibinfo {pages} {195136} (\bibinfo {year}
  {2023})}\BibitemShut {NoStop}%
\bibitem [{\citenamefont {Stampfli}(1986)}]{Stampfli1986}%
  \BibitemOpen
  \bibfield  {author} {\bibinfo {author} {\bibfnamefont {P.}~\bibnamefont
  {Stampfli}},\ }\bibfield  {title} {\emph {\bibinfo {title} {A dodecagonal
  quasiperiodic lattice in two dimensions}},\ }\href@noop {} {\bibfield
  {journal} {\bibinfo  {journal} {Helv. Phys. Acta}\ }\textbf {\bibinfo
  {volume} {59}},\ \bibinfo {pages} {1260} (\bibinfo {year}
  {1986})}\BibitemShut {NoStop}%
\bibitem [{\citenamefont {Ahn}\ \emph {et~al.}(2018)\citenamefont {Ahn},
  \citenamefont {Moon}, \citenamefont {Kim}, \citenamefont {Kim}, \citenamefont
  {Shin}, \citenamefont {Kim}, \citenamefont {Cha}, \citenamefont {Kahng},
  \citenamefont {Kim}, \citenamefont {Koshino} \emph {et~al.}}]{Ahn2018}%
  \BibitemOpen
  \bibfield  {author} {\bibinfo {author} {\bibfnamefont {S.~J.}\ \bibnamefont
  {Ahn}}, \bibinfo {author} {\bibfnamefont {P.}~\bibnamefont {Moon}}, \bibinfo
  {author} {\bibfnamefont {T.-H.}\ \bibnamefont {Kim}}, \bibinfo {author}
  {\bibfnamefont {H.-W.}\ \bibnamefont {Kim}}, \bibinfo {author} {\bibfnamefont
  {H.-C.}\ \bibnamefont {Shin}}, \bibinfo {author} {\bibfnamefont {E.~H.}\
  \bibnamefont {Kim}}, \bibinfo {author} {\bibfnamefont {H.~W.}\ \bibnamefont
  {Cha}}, \bibinfo {author} {\bibfnamefont {S.-J.}\ \bibnamefont {Kahng}},
  \bibinfo {author} {\bibfnamefont {P.}~\bibnamefont {Kim}}, \bibinfo {author}
  {\bibfnamefont {M.}~\bibnamefont {Koshino}},  \emph {et~al.},\ }\bibfield
  {title} {\emph {\bibinfo {title} {Dirac electrons in a dodecagonal graphene
  quasicrystal}},\ }\href {https://www.science.org/doi/10.1126/science.aar8412}
  {\bibfield  {journal} {\bibinfo  {journal} {Science}\ }\textbf {\bibinfo
  {volume} {361}},\ \bibinfo {pages} {782} (\bibinfo {year}
  {2018})}\BibitemShut {NoStop}%
\bibitem [{\citenamefont {Jos\'e}\ \emph {et~al.}(1977)\citenamefont {Jos\'e},
  \citenamefont {Kadanoff}, \citenamefont {Kirkpatrick},\ and\ \citenamefont
  {Nelson}}]{Jose1977}%
  \BibitemOpen
  \bibfield  {author} {\bibinfo {author} {\bibfnamefont {J.~V.}\ \bibnamefont
  {Jos\'e}}, \bibinfo {author} {\bibfnamefont {L.~P.}\ \bibnamefont
  {Kadanoff}}, \bibinfo {author} {\bibfnamefont {S.}~\bibnamefont
  {Kirkpatrick}}, \ and\ \bibinfo {author} {\bibfnamefont {D.~R.}\ \bibnamefont
  {Nelson}},\ }\bibfield  {title} {\emph {\bibinfo {title} {Renormalization,
  vortices, and symmetry-breaking perturbations in the two-dimensional planar
  model}},\ }\href {\doibase 10.1103/PhysRevB.16.1217} {\bibfield  {journal}
  {\bibinfo  {journal} {Phys. Rev. B}\ }\textbf {\bibinfo {volume} {16}},\
  \bibinfo {pages} {1217} (\bibinfo {year} {1977})}\BibitemShut {NoStop}%
\bibitem [{\citenamefont {Maciejko}\ \emph {et~al.}(2013)\citenamefont
  {Maciejko}, \citenamefont {Hsu}, \citenamefont {Kivelson}, \citenamefont
  {Park},\ and\ \citenamefont {Sondhi}}]{Maciejko2013}%
  \BibitemOpen
  \bibfield  {author} {\bibinfo {author} {\bibfnamefont {J.}~\bibnamefont
  {Maciejko}}, \bibinfo {author} {\bibfnamefont {B.}~\bibnamefont {Hsu}},
  \bibinfo {author} {\bibfnamefont {S.~A.}\ \bibnamefont {Kivelson}}, \bibinfo
  {author} {\bibfnamefont {Y.}~\bibnamefont {Park}}, \ and\ \bibinfo {author}
  {\bibfnamefont {S.~L.}\ \bibnamefont {Sondhi}},\ }\bibfield  {title} {\emph
  {\bibinfo {title} {{Field theory of the quantum Hall nematic transition}}},\
  }\href {\doibase 10.1103/PhysRevB.88.125137} {\bibfield  {journal} {\bibinfo
  {journal} {Phys. Rev. B}\ }\textbf {\bibinfo {volume} {88}},\ \bibinfo
  {pages} {125137} (\bibinfo {year} {2013})}\BibitemShut {NoStop}%
\bibitem [{\citenamefont {You}\ and\ \citenamefont
  {Fradkin}(2013)}]{Fradkin2013}%
  \BibitemOpen
  \bibfield  {author} {\bibinfo {author} {\bibfnamefont {Y.}~\bibnamefont
  {You}}\ and\ \bibinfo {author} {\bibfnamefont {E.}~\bibnamefont {Fradkin}},\
  }\bibfield  {title} {\emph {\bibinfo {title} {{Field theory of nematicity in
  the spontaneous quantum anomalous Hall effect}}},\ }\href
  {https://link.aps.org/doi/10.1103/PhysRevB.88.235124} {\bibfield  {journal}
  {\bibinfo  {journal} {Phys. Rev. B}\ }\textbf {\bibinfo {volume} {88}},\
  \bibinfo {pages} {235124} (\bibinfo {year} {2013})}\BibitemShut {NoStop}%
\bibitem [{\citenamefont {You}\ \emph {et~al.}(2014)\citenamefont {You},
  \citenamefont {Cho},\ and\ \citenamefont {Fradkin}}]{Fradkin2014}%
  \BibitemOpen
  \bibfield  {author} {\bibinfo {author} {\bibfnamefont {Y.}~\bibnamefont
  {You}}, \bibinfo {author} {\bibfnamefont {G.~Y.}\ \bibnamefont {Cho}}, \ and\
  \bibinfo {author} {\bibfnamefont {E.}~\bibnamefont {Fradkin}},\ }\bibfield
  {title} {\emph {\bibinfo {title} {{Theory of Nematic Fractional Quantum Hall
  States}}},\ }\href {\doibase 10.1103/PhysRevX.4.041050} {\bibfield  {journal}
  {\bibinfo  {journal} {Phys. Rev. X}\ }\textbf {\bibinfo {volume} {4}},\
  \bibinfo {pages} {041050} (\bibinfo {year} {2014})}\BibitemShut {NoStop}%
\bibitem [{\citenamefont {Mandal}\ and\ \citenamefont
  {Fernandes}(2023)}]{Mandal2023}%
  \BibitemOpen
  \bibfield  {author} {\bibinfo {author} {\bibfnamefont {I.}~\bibnamefont
  {Mandal}}\ and\ \bibinfo {author} {\bibfnamefont {R.~M.}\ \bibnamefont
  {Fernandes}},\ }\bibfield  {title} {\emph {\bibinfo {title}
  {{Valley-polarized nematic order in twisted moir{\'e} systems: In-plane
  orbital magnetism and crossover from non-Fermi liquid to Fermi liquid}}},\
  }\href {\doibase 10.1103/PhysRevB.107.125142} {\bibfield  {journal} {\bibinfo
   {journal} {Phys. Rev. B}\ }\textbf {\bibinfo {volume} {107}},\ \bibinfo
  {pages} {125142} (\bibinfo {year} {2023})}\BibitemShut {NoStop}%
\bibitem [{\citenamefont {Wu}(1982)}]{Wu1982}%
  \BibitemOpen
  \bibfield  {author} {\bibinfo {author} {\bibfnamefont {F.~Y.}\ \bibnamefont
  {Wu}},\ }\bibfield  {title} {\emph {\bibinfo {title} {{The Potts model}}},\
  }\href {\doibase 10.1103/RevModPhys.54.235} {\bibfield  {journal} {\bibinfo
  {journal} {Rev. Mod. Phys.}\ }\textbf {\bibinfo {volume} {54}},\ \bibinfo
  {pages} {235} (\bibinfo {year} {1982})}\BibitemShut {NoStop}%
\bibitem [{\citenamefont {Xu}\ \emph {et~al.}(2020)\citenamefont {Xu},
  \citenamefont {Wu}, \citenamefont {Jian},\ and\ \citenamefont {Xu}}]{Xu2020}%
  \BibitemOpen
  \bibfield  {author} {\bibinfo {author} {\bibfnamefont {Y.}~\bibnamefont
  {Xu}}, \bibinfo {author} {\bibfnamefont {X.-C.}\ \bibnamefont {Wu}}, \bibinfo
  {author} {\bibfnamefont {C.-M.}\ \bibnamefont {Jian}}, \ and\ \bibinfo
  {author} {\bibfnamefont {C.}~\bibnamefont {Xu}},\ }\bibfield  {title} {\emph
  {\bibinfo {title} {{Orbital order and possible non-Fermi liquid in moir{\'e}
  systems}}},\ }\href {\doibase 10.1103/PhysRevB.101.205426} {\bibfield
  {journal} {\bibinfo  {journal} {Phys. Rev. B}\ }\textbf {\bibinfo {volume}
  {101}},\ \bibinfo {pages} {205426} (\bibinfo {year} {2020})}\BibitemShut
  {NoStop}%
\bibitem [{\citenamefont {Podolsky}\ \emph {et~al.}(2016)\citenamefont
  {Podolsky}, \citenamefont {Shimshoni}, \citenamefont {Morigi},\ and\
  \citenamefont {Fishman}}]{Podolsky2016}%
  \BibitemOpen
  \bibfield  {author} {\bibinfo {author} {\bibfnamefont {D.}~\bibnamefont
  {Podolsky}}, \bibinfo {author} {\bibfnamefont {E.}~\bibnamefont {Shimshoni}},
  \bibinfo {author} {\bibfnamefont {G.}~\bibnamefont {Morigi}}, \ and\ \bibinfo
  {author} {\bibfnamefont {S.}~\bibnamefont {Fishman}},\ }\bibfield  {title}
  {\emph {\bibinfo {title} {{Buckling Transitions and Clock Order of
  Two-Dimensional Coulomb Crystals}}},\ }\href
  {https://journals.aps.org/prx/abstract/10.1103/PhysRevX.6.031025} {\bibfield
  {journal} {\bibinfo  {journal} {Phys. Rev. X}\ }\textbf {\bibinfo {volume}
  {6}},\ \bibinfo {pages} {031025} (\bibinfo {year} {2016})}\BibitemShut
  {NoStop}%
\bibitem [{\citenamefont {Arnold}\ and\ \citenamefont
  {Nigmatullin}(2022)}]{Arnold2022}%
  \BibitemOpen
  \bibfield  {author} {\bibinfo {author} {\bibfnamefont {M.}~\bibnamefont
  {Arnold}}\ and\ \bibinfo {author} {\bibfnamefont {R.}~\bibnamefont
  {Nigmatullin}},\ }\bibfield  {title} {\emph {\bibinfo {title} {{Dynamics of
  vortex defect formation in two-dimensional Coulomb crystals}}},\ }\href
  {\doibase 10.1103/PhysRevB.106.104106} {\bibfield  {journal} {\bibinfo
  {journal} {Phys. Rev. B}\ }\textbf {\bibinfo {volume} {106}},\ \bibinfo
  {pages} {104106} (\bibinfo {year} {2022})}\BibitemShut {NoStop}%
\bibitem [{\citenamefont {Norman}\ \emph {et~al.}(2007)\citenamefont {Norman},
  \citenamefont {Kanigel}, \citenamefont {Randeria}, \citenamefont
  {Chatterjee},\ and\ \citenamefont {Campuzano}}]{Norman2007}%
  \BibitemOpen
  \bibfield  {author} {\bibinfo {author} {\bibfnamefont {M.~R.}\ \bibnamefont
  {Norman}}, \bibinfo {author} {\bibfnamefont {A.}~\bibnamefont {Kanigel}},
  \bibinfo {author} {\bibfnamefont {M.}~\bibnamefont {Randeria}}, \bibinfo
  {author} {\bibfnamefont {U.}~\bibnamefont {Chatterjee}}, \ and\ \bibinfo
  {author} {\bibfnamefont {J.~C.}\ \bibnamefont {Campuzano}},\ }\bibfield
  {title} {\emph {\bibinfo {title} {{Modeling the Fermi arc in underdoped
  cuprates}}},\ }\href {\doibase 10.1103/PhysRevB.76.174501} {\bibfield
  {journal} {\bibinfo  {journal} {Phys. Rev. B}\ }\textbf {\bibinfo {volume}
  {76}},\ \bibinfo {pages} {174501} (\bibinfo {year} {2007})}\BibitemShut
  {NoStop}%
\bibitem [{\citenamefont {Wu}\ \emph {et~al.}(2007)\citenamefont {Wu},
  \citenamefont {Sun}, \citenamefont {Fradkin},\ and\ \citenamefont
  {Zhang}}]{Wu2007}%
  \BibitemOpen
  \bibfield  {author} {\bibinfo {author} {\bibfnamefont {C.}~\bibnamefont
  {Wu}}, \bibinfo {author} {\bibfnamefont {K.}~\bibnamefont {Sun}}, \bibinfo
  {author} {\bibfnamefont {E.}~\bibnamefont {Fradkin}}, \ and\ \bibinfo
  {author} {\bibfnamefont {S.-C.}\ \bibnamefont {Zhang}},\ }\bibfield  {title}
  {\emph {\bibinfo {title} {Fermi liquid instabilities in the spin channel}},\
  }\href {\doibase 10.1103/PhysRevB.75.115103} {\bibfield  {journal} {\bibinfo
  {journal} {Phys. Rev. B}\ }\textbf {\bibinfo {volume} {75}},\ \bibinfo
  {pages} {115103} (\bibinfo {year} {2007})}\BibitemShut {NoStop}%
\bibitem [{\citenamefont {Classen}\ \emph {et~al.}(2020)\citenamefont
  {Classen}, \citenamefont {Chubukov}, \citenamefont {Honerkamp},\ and\
  \citenamefont {Scherer}}]{Classen2020}%
  \BibitemOpen
  \bibfield  {author} {\bibinfo {author} {\bibfnamefont {L.}~\bibnamefont
  {Classen}}, \bibinfo {author} {\bibfnamefont {A.~V.}\ \bibnamefont
  {Chubukov}}, \bibinfo {author} {\bibfnamefont {C.}~\bibnamefont {Honerkamp}},
  \ and\ \bibinfo {author} {\bibfnamefont {M.~M.}\ \bibnamefont {Scherer}},\
  }\bibfield  {title} {\emph {\bibinfo {title} {{Competing orders at
  higher-order Van Hove points}}},\ }\href
  {https://link.aps.org/doi/10.1103/PhysRevB.102.125141} {\bibfield  {journal}
  {\bibinfo  {journal} {Phys. Rev. B}\ }\textbf {\bibinfo {volume} {102}},\
  \bibinfo {pages} {125141} (\bibinfo {year} {2020})}\BibitemShut {NoStop}%
\bibitem [{\citenamefont {Chichinadze}\ \emph {et~al.}(2020)\citenamefont
  {Chichinadze}, \citenamefont {Classen},\ and\ \citenamefont
  {Chubukov}}]{Chichinadze2020}%
  \BibitemOpen
  \bibfield  {author} {\bibinfo {author} {\bibfnamefont {D.~V.}\ \bibnamefont
  {Chichinadze}}, \bibinfo {author} {\bibfnamefont {L.}~\bibnamefont
  {Classen}}, \ and\ \bibinfo {author} {\bibfnamefont {A.~V.}\ \bibnamefont
  {Chubukov}},\ }\bibfield  {title} {\emph {\bibinfo {title} {{Valley
  magnetism, nematicity, and density wave orders in twisted bilayer
  graphene}}},\ }\href {\doibase 10.1103/PhysRevB.102.125120} {\bibfield
  {journal} {\bibinfo  {journal} {Phys. Rev. B}\ }\textbf {\bibinfo {volume}
  {102}},\ \bibinfo {pages} {125120} (\bibinfo {year} {2020})}\BibitemShut
  {NoStop}%
\bibitem [{\citenamefont {Schmalian}\ \emph {et~al.}(1998)\citenamefont
  {Schmalian}, \citenamefont {Pines},\ and\ \citenamefont
  {Stojkovi\ifmmode~\acute{c}\else \'{c}\fi{}}}]{Schmalian1998}%
  \BibitemOpen
  \bibfield  {author} {\bibinfo {author} {\bibfnamefont {J.}~\bibnamefont
  {Schmalian}}, \bibinfo {author} {\bibfnamefont {D.}~\bibnamefont {Pines}}, \
  and\ \bibinfo {author} {\bibfnamefont {B.}~\bibnamefont
  {Stojkovi\ifmmode~\acute{c}\else \'{c}\fi{}}},\ }\bibfield  {title} {\emph
  {\bibinfo {title} {Weak pseudogap behavior in the underdoped cuprate
  superconductors}},\ }\href {\doibase 10.1103/PhysRevLett.80.3839} {\bibfield
  {journal} {\bibinfo  {journal} {Phys. Rev. Lett.}\ }\textbf {\bibinfo
  {volume} {80}},\ \bibinfo {pages} {3839} (\bibinfo {year}
  {1998})}\BibitemShut {NoStop}%
\bibitem [{\citenamefont {Sedrakyan}\ and\ \citenamefont
  {Chubukov}(2010)}]{Chubukov2010}%
  \BibitemOpen
  \bibfield  {author} {\bibinfo {author} {\bibfnamefont {T.~A.}\ \bibnamefont
  {Sedrakyan}}\ and\ \bibinfo {author} {\bibfnamefont {A.~V.}\ \bibnamefont
  {Chubukov}},\ }\bibfield  {title} {\emph {\bibinfo {title} {Pseudogap in
  underdoped cuprates and spin-density-wave fluctuations}},\ }\href
  {https://link.aps.org/doi/10.1103/PhysRevB.81.174536} {\bibfield  {journal}
  {\bibinfo  {journal} {Phys. Rev. B}\ }\textbf {\bibinfo {volume} {81}},\
  \bibinfo {pages} {174536} (\bibinfo {year} {2010})}\BibitemShut {NoStop}%
\bibitem [{\citenamefont {Lin}\ and\ \citenamefont
  {Millis}(2011)}]{Lin_Millis}%
  \BibitemOpen
  \bibfield  {author} {\bibinfo {author} {\bibfnamefont {J.}~\bibnamefont
  {Lin}}\ and\ \bibinfo {author} {\bibfnamefont {A.~J.}\ \bibnamefont
  {Millis}},\ }\bibfield  {title} {\emph {\bibinfo {title} {Optical and hall
  conductivities of a thermally disordered two-dimensional spin-density wave:
  Two-particle response in the pseudogap regime of electron-doped
  high-${T}_{c}$ superconductors}},\ }\href {\doibase
  10.1103/PhysRevB.83.125108} {\bibfield  {journal} {\bibinfo  {journal} {Phys.
  Rev. B}\ }\textbf {\bibinfo {volume} {83}},\ \bibinfo {pages} {125108}
  (\bibinfo {year} {2011})}\BibitemShut {NoStop}%
\bibitem [{\citenamefont {Vilk}\ and\ \citenamefont
  {Tremblay}(1996)}]{Vilk1996}%
  \BibitemOpen
  \bibfield  {author} {\bibinfo {author} {\bibfnamefont {Y.~M.}\ \bibnamefont
  {Vilk}}\ and\ \bibinfo {author} {\bibfnamefont {A.-M.~S.}\ \bibnamefont
  {Tremblay}},\ }\bibfield  {title} {\emph {\bibinfo {title} {{Destruction of
  Fermi-liquid quasiparticles in two dimensions by critical fluctuations}}},\
  }\href {\doibase 10.1209/epl/i1996-00315-2} {\bibfield  {journal} {\bibinfo
  {journal} {Europhysics Letters}\ }\textbf {\bibinfo {volume} {33}},\ \bibinfo
  {pages} {159} (\bibinfo {year} {1996})}\BibitemShut {NoStop}%
\bibitem [{\citenamefont {Norman}\ \emph {et~al.}(1998)\citenamefont {Norman},
  \citenamefont {Randeria}, \citenamefont {Ding},\ and\ \citenamefont
  {Campuzano}}]{Norman1998}%
  \BibitemOpen
  \bibfield  {author} {\bibinfo {author} {\bibfnamefont {M.~R.}\ \bibnamefont
  {Norman}}, \bibinfo {author} {\bibfnamefont {M.}~\bibnamefont {Randeria}},
  \bibinfo {author} {\bibfnamefont {H.}~\bibnamefont {Ding}}, \ and\ \bibinfo
  {author} {\bibfnamefont {J.~C.}\ \bibnamefont {Campuzano}},\ }\bibfield
  {title} {\emph {\bibinfo {title} {Phenomenology of the low-energy spectral
  function in high-${T}_{c}$ superconductors}},\ }\href
  {https://link.aps.org/doi/10.1103/PhysRevB.57.R11093} {\bibfield  {journal}
  {\bibinfo  {journal} {Phys. Rev. B}\ }\textbf {\bibinfo {volume} {57}},\
  \bibinfo {pages} {R11093} (\bibinfo {year} {1998})}\BibitemShut {NoStop}%
\bibitem [{\citenamefont {Franz}\ and\ \citenamefont
  {Millis}(1998)}]{Millis1998}%
  \BibitemOpen
  \bibfield  {author} {\bibinfo {author} {\bibfnamefont {M.}~\bibnamefont
  {Franz}}\ and\ \bibinfo {author} {\bibfnamefont {A.~J.}\ \bibnamefont
  {Millis}},\ }\bibfield  {title} {\emph {\bibinfo {title} {Phase fluctuations
  and spectral properties of underdoped cuprates}},\ }\href
  {https://journals.aps.org/prb/abstract/10.1103/PhysRevB.58.14572} {\bibfield
  {journal} {\bibinfo  {journal} {Phys. Rev. B}\ }\textbf {\bibinfo {volume}
  {58}},\ \bibinfo {pages} {14572} (\bibinfo {year} {1998})}\BibitemShut
  {NoStop}%
\bibitem [{\citenamefont {Sachdev}\ \emph {et~al.}(2019)\citenamefont
  {Sachdev}, \citenamefont {Scammell}, \citenamefont {Scheurer},\ and\
  \citenamefont {Tarnopolsky}}]{Sachdev2019}%
  \BibitemOpen
  \bibfield  {author} {\bibinfo {author} {\bibfnamefont {S.}~\bibnamefont
  {Sachdev}}, \bibinfo {author} {\bibfnamefont {H.~D.}\ \bibnamefont
  {Scammell}}, \bibinfo {author} {\bibfnamefont {M.~S.}\ \bibnamefont
  {Scheurer}}, \ and\ \bibinfo {author} {\bibfnamefont {G.}~\bibnamefont
  {Tarnopolsky}},\ }\bibfield  {title} {\emph {\bibinfo {title} {Gauge theory
  for the cuprates near optimal doping}},\ }\href
  {https://link.aps.org/doi/10.1103/PhysRevB.99.054516} {\bibfield  {journal}
  {\bibinfo  {journal} {Phys. Rev. B}\ }\textbf {\bibinfo {volume} {99}},\
  \bibinfo {pages} {054516} (\bibinfo {year} {2019})}\BibitemShut {NoStop}%
\bibitem [{\citenamefont {Banerjee}\ \emph {et~al.}(2011)\citenamefont
  {Banerjee}, \citenamefont {Ramakrishnan},\ and\ \citenamefont
  {Dasgupta}}]{Dasgupta2011}%
  \BibitemOpen
  \bibfield  {author} {\bibinfo {author} {\bibfnamefont {S.}~\bibnamefont
  {Banerjee}}, \bibinfo {author} {\bibfnamefont {T.~V.}\ \bibnamefont
  {Ramakrishnan}}, \ and\ \bibinfo {author} {\bibfnamefont {C.}~\bibnamefont
  {Dasgupta}},\ }\bibfield  {title} {\emph {\bibinfo {title} {{Effect of
  pairing fluctuations on low-energy electronic spectra in cuprate
  superconductors}}},\ }\href {\doibase 10.1103/PhysRevB.84.144525} {\bibfield
  {journal} {\bibinfo  {journal} {Phys. Rev. B}\ }\textbf {\bibinfo {volume}
  {84}},\ \bibinfo {pages} {144525} (\bibinfo {year} {2011})}\BibitemShut
  {NoStop}%
\bibitem [{\citenamefont {Bradlyn}\ \emph {et~al.}(2012)\citenamefont
  {Bradlyn}, \citenamefont {Goldstein},\ and\ \citenamefont
  {Read}}]{Bradlyn2012}%
  \BibitemOpen
  \bibfield  {author} {\bibinfo {author} {\bibfnamefont {B.}~\bibnamefont
  {Bradlyn}}, \bibinfo {author} {\bibfnamefont {M.}~\bibnamefont {Goldstein}},
  \ and\ \bibinfo {author} {\bibfnamefont {N.}~\bibnamefont {Read}},\
  }\bibfield  {title} {\emph {\bibinfo {title} {{Kubo formulas for viscosity:
  Hall viscosity, Ward identities, and the relation with conductivity}}},\
  }\href {\doibase 10.1103/PhysRevB.86.245309} {\bibfield  {journal} {\bibinfo
  {journal} {Phys. Rev. B}\ }\textbf {\bibinfo {volume} {86}},\ \bibinfo
  {pages} {245309} (\bibinfo {year} {2012})}\BibitemShut {NoStop}%
\bibitem [{\citenamefont {Barkeshli}\ \emph {et~al.}(2012)\citenamefont
  {Barkeshli}, \citenamefont {Chung},\ and\ \citenamefont
  {Qi}}]{Barkeshli2012}%
  \BibitemOpen
  \bibfield  {author} {\bibinfo {author} {\bibfnamefont {M.}~\bibnamefont
  {Barkeshli}}, \bibinfo {author} {\bibfnamefont {S.~B.}\ \bibnamefont
  {Chung}}, \ and\ \bibinfo {author} {\bibfnamefont {X.-L.}\ \bibnamefont
  {Qi}},\ }\bibfield  {title} {\emph {\bibinfo {title} {{Dissipationless phonon
  Hall viscosity}}},\ }\href {\doibase 10.1103/PhysRevB.85.245107} {\bibfield
  {journal} {\bibinfo  {journal} {Phys. Rev. B}\ }\textbf {\bibinfo {volume}
  {85}},\ \bibinfo {pages} {245107} (\bibinfo {year} {2012})}\BibitemShut
  {NoStop}%
\bibitem [{\citenamefont {Link}\ \emph {et~al.}(2018)\citenamefont {Link},
  \citenamefont {Sheehy}, \citenamefont {Narozhny},\ and\ \citenamefont
  {Schmalian}}]{Link2018}%
  \BibitemOpen
  \bibfield  {author} {\bibinfo {author} {\bibfnamefont {J.~M.}\ \bibnamefont
  {Link}}, \bibinfo {author} {\bibfnamefont {D.~E.}\ \bibnamefont {Sheehy}},
  \bibinfo {author} {\bibfnamefont {B.~N.}\ \bibnamefont {Narozhny}}, \ and\
  \bibinfo {author} {\bibfnamefont {J.}~\bibnamefont {Schmalian}},\ }\bibfield
  {title} {\emph {\bibinfo {title} {{Elastic response of the electron fluid in
  intrinsic graphene: The collisionless regime}}},\ }\href
  {https://link.aps.org/doi/10.1103/PhysRevB.98.195103} {\bibfield  {journal}
  {\bibinfo  {journal} {Phys. Rev. B}\ }\textbf {\bibinfo {volume} {98}},\
  \bibinfo {pages} {195103} (\bibinfo {year} {2018})}\BibitemShut {NoStop}%
\bibitem [{\citenamefont {Oshikawa}(2000)}]{Oshikawa2000}%
  \BibitemOpen
  \bibfield  {author} {\bibinfo {author} {\bibfnamefont {M.}~\bibnamefont
  {Oshikawa}},\ }\bibfield  {title} {\emph {\bibinfo {title} {{Ordered phase
  and scaling in ${Z}_{n}$ models and the three-state antiferromagnetic Potts
  model in three dimensions}}},\ }\href {\doibase 10.1103/PhysRevB.61.3430}
  {\bibfield  {journal} {\bibinfo  {journal} {Phys. Rev. B}\ }\textbf {\bibinfo
  {volume} {61}},\ \bibinfo {pages} {3430} (\bibinfo {year}
  {2000})}\BibitemShut {NoStop}%
\bibitem [{\citenamefont {Patil}\ \emph {et~al.}(2021)\citenamefont {Patil},
  \citenamefont {Shao},\ and\ \citenamefont {Sandvik}}]{Sandvik2021}%
  \BibitemOpen
  \bibfield  {author} {\bibinfo {author} {\bibfnamefont {P.}~\bibnamefont
  {Patil}}, \bibinfo {author} {\bibfnamefont {H.}~\bibnamefont {Shao}}, \ and\
  \bibinfo {author} {\bibfnamefont {A.~W.}\ \bibnamefont {Sandvik}},\
  }\bibfield  {title} {\emph {\bibinfo {title} {{Unconventional U(1) to
  ${Z}_{q}$ crossover in quantum and classical $q$-state clock models}}},\
  }\href {\doibase 10.1103/PhysRevB.103.054418} {\bibfield  {journal} {\bibinfo
   {journal} {Phys. Rev. B}\ }\textbf {\bibinfo {volume} {103}},\ \bibinfo
  {pages} {054418} (\bibinfo {year} {2021})}\BibitemShut {NoStop}%
\bibitem [{\citenamefont {Lederer}\ \emph {et~al.}(2017)\citenamefont
  {Lederer}, \citenamefont {Schattner}, \citenamefont {Berg},\ and\
  \citenamefont {Kivelson}}]{Lederer2017}%
  \BibitemOpen
  \bibfield  {author} {\bibinfo {author} {\bibfnamefont {S.}~\bibnamefont
  {Lederer}}, \bibinfo {author} {\bibfnamefont {Y.}~\bibnamefont {Schattner}},
  \bibinfo {author} {\bibfnamefont {E.}~\bibnamefont {Berg}}, \ and\ \bibinfo
  {author} {\bibfnamefont {S.~A.}\ \bibnamefont {Kivelson}},\ }\bibfield
  {title} {\emph {\bibinfo {title} {{Superconductivity and non-Fermi liquid
  behavior near a nematic quantum critical point}}},\ }\href
  {https://www.pnas.org/doi/10.1073/pnas.1620651114} {\bibfield  {journal}
  {\bibinfo  {journal} {Proceedings of the National Academy of Sciences}\
  }\textbf {\bibinfo {volume} {114}},\ \bibinfo {pages} {4905} (\bibinfo {year}
  {2017})}\BibitemShut {NoStop}%
\bibitem [{\citenamefont {Klein}\ and\ \citenamefont
  {Chubukov}(2018)}]{Klein2018}%
  \BibitemOpen
  \bibfield  {author} {\bibinfo {author} {\bibfnamefont {A.}~\bibnamefont
  {Klein}}\ and\ \bibinfo {author} {\bibfnamefont {A.}~\bibnamefont
  {Chubukov}},\ }\bibfield  {title} {\emph {\bibinfo {title}
  {{Superconductivity near a nematic quantum critical point: Interplay between
  hot and lukewarm regions}}},\ }\href {\doibase 10.1103/PhysRevB.98.220501}
  {\bibfield  {journal} {\bibinfo  {journal} {Phys. Rev. B}\ }\textbf {\bibinfo
  {volume} {98}},\ \bibinfo {pages} {220501} (\bibinfo {year}
  {2018})}\BibitemShut {NoStop}%
\bibitem [{\citenamefont {Valenzuela}\ and\ \citenamefont
  {Vozmediano}(2008)}]{Valenzuela2008}%
  \BibitemOpen
  \bibfield  {author} {\bibinfo {author} {\bibfnamefont {B.}~\bibnamefont
  {Valenzuela}}\ and\ \bibinfo {author} {\bibfnamefont {M.~A.}\ \bibnamefont
  {Vozmediano}},\ }\bibfield  {title} {\emph {\bibinfo {title} {Pomeranchuk
  instability in doped graphene}},\ }\href
  {https://iopscience.iop.org/article/10.1088/1367-2630/10/11/113009}
  {\bibfield  {journal} {\bibinfo  {journal} {New Journal of Physics}\ }\textbf
  {\bibinfo {volume} {10}},\ \bibinfo {pages} {113009} (\bibinfo {year}
  {2008})}\BibitemShut {NoStop}%
\bibitem [{\citenamefont {Kiesel}\ \emph {et~al.}(2013)\citenamefont {Kiesel},
  \citenamefont {Platt},\ and\ \citenamefont {Thomale}}]{Kiesel2013}%
  \BibitemOpen
  \bibfield  {author} {\bibinfo {author} {\bibfnamefont {M.~L.}\ \bibnamefont
  {Kiesel}}, \bibinfo {author} {\bibfnamefont {C.}~\bibnamefont {Platt}}, \
  and\ \bibinfo {author} {\bibfnamefont {R.}~\bibnamefont {Thomale}},\
  }\bibfield  {title} {\emph {\bibinfo {title} {{Unconventional Fermi Surface
  Instabilities in the Kagome Hubbard Model}}},\ }\href
  {https://link.aps.org/doi/10.1103/PhysRevLett.110.126405} {\bibfield
  {journal} {\bibinfo  {journal} {Phys. Rev. Lett.}\ }\textbf {\bibinfo
  {volume} {110}},\ \bibinfo {pages} {126405} (\bibinfo {year}
  {2013})}\BibitemShut {NoStop}%
\bibitem [{\citenamefont {Mayorov}\ \emph {et~al.}(2011)\citenamefont
  {Mayorov}, \citenamefont {Elias}, \citenamefont {Mucha-Kruczynski},
  \citenamefont {Gorbachev}, \citenamefont {Tudorovskiy}, \citenamefont
  {Zhukov}, \citenamefont {Morozov}, \citenamefont {Katsnelson}, \citenamefont
  {ko}, \citenamefont {Geim} \emph {et~al.}}]{Novoselov2011}%
  \BibitemOpen
  \bibfield  {author} {\bibinfo {author} {\bibfnamefont {A.}~\bibnamefont
  {Mayorov}}, \bibinfo {author} {\bibfnamefont {D.}~\bibnamefont {Elias}},
  \bibinfo {author} {\bibfnamefont {M.}~\bibnamefont {Mucha-Kruczynski}},
  \bibinfo {author} {\bibfnamefont {R.}~\bibnamefont {Gorbachev}}, \bibinfo
  {author} {\bibfnamefont {T.}~\bibnamefont {Tudorovskiy}}, \bibinfo {author}
  {\bibfnamefont {A.}~\bibnamefont {Zhukov}}, \bibinfo {author} {\bibfnamefont
  {S.}~\bibnamefont {Morozov}}, \bibinfo {author} {\bibfnamefont
  {M.}~\bibnamefont {Katsnelson}}, \bibinfo {author} {\bibfnamefont {V.~F.}\
  \bibnamefont {ko}}, \bibinfo {author} {\bibfnamefont {A.}~\bibnamefont
  {Geim}},  \emph {et~al.},\ }\bibfield  {title} {\emph {\bibinfo {title}
  {Interaction-driven spectrum reconstruction in bilayer graphene}},\ }\href
  {https://www.science.org/doi/10.1126/science.1208683} {\bibfield  {journal}
  {\bibinfo  {journal} {Science}\ }\textbf {\bibinfo {volume} {333}},\ \bibinfo
  {pages} {860} (\bibinfo {year} {2011})}\BibitemShut {NoStop}%
\bibitem [{\citenamefont {Sun}\ \emph {et~al.}(2019)\citenamefont {Sun},
  \citenamefont {Kittaka}, \citenamefont {Sakakibara}, \citenamefont {Machida},
  \citenamefont {Wang}, \citenamefont {Wen}, \citenamefont {Xing},
  \citenamefont {Shi},\ and\ \citenamefont {Tamegai}}]{Tamegai2019}%
  \BibitemOpen
  \bibfield  {author} {\bibinfo {author} {\bibfnamefont {Y.}~\bibnamefont
  {Sun}}, \bibinfo {author} {\bibfnamefont {S.}~\bibnamefont {Kittaka}},
  \bibinfo {author} {\bibfnamefont {T.}~\bibnamefont {Sakakibara}}, \bibinfo
  {author} {\bibfnamefont {K.}~\bibnamefont {Machida}}, \bibinfo {author}
  {\bibfnamefont {J.}~\bibnamefont {Wang}}, \bibinfo {author} {\bibfnamefont
  {J.}~\bibnamefont {Wen}}, \bibinfo {author} {\bibfnamefont {X.}~\bibnamefont
  {Xing}}, \bibinfo {author} {\bibfnamefont {Z.}~\bibnamefont {Shi}}, \ and\
  \bibinfo {author} {\bibfnamefont {T.}~\bibnamefont {Tamegai}},\ }\bibfield
  {title} {\emph {\bibinfo {title} {{Quasiparticle Evidence for the Nematic
  State above ${T}_{\mathrm{c}}$ in
  ${\mathrm{Sr}}_{x}{\mathrm{Bi}}_{2}{\mathrm{Se}}_{3}$}}},\ }\href
  {https://link.aps.org/doi/10.1103/PhysRevLett.123.027002} {\bibfield
  {journal} {\bibinfo  {journal} {Phys. Rev. Lett.}\ }\textbf {\bibinfo
  {volume} {123}},\ \bibinfo {pages} {027002} (\bibinfo {year}
  {2019})}\BibitemShut {NoStop}%
\bibitem [{\citenamefont {Cho}\ \emph {et~al.}(2020)\citenamefont {Cho},
  \citenamefont {Shen}, \citenamefont {Lyu}, \citenamefont {Atanov},
  \citenamefont {Chen}, \citenamefont {Lee}, \citenamefont {Hor}, \citenamefont
  {Gawryluk}, \citenamefont {Pomjakushina}, \citenamefont {Bartkowiak} \emph
  {et~al.}}]{Cho2020}%
  \BibitemOpen
  \bibfield  {author} {\bibinfo {author} {\bibfnamefont {C.-w.}\ \bibnamefont
  {Cho}}, \bibinfo {author} {\bibfnamefont {J.}~\bibnamefont {Shen}}, \bibinfo
  {author} {\bibfnamefont {J.}~\bibnamefont {Lyu}}, \bibinfo {author}
  {\bibfnamefont {O.}~\bibnamefont {Atanov}}, \bibinfo {author} {\bibfnamefont
  {Q.}~\bibnamefont {Chen}}, \bibinfo {author} {\bibfnamefont {S.~H.}\
  \bibnamefont {Lee}}, \bibinfo {author} {\bibfnamefont {Y.~S.}\ \bibnamefont
  {Hor}}, \bibinfo {author} {\bibfnamefont {D.~J.}\ \bibnamefont {Gawryluk}},
  \bibinfo {author} {\bibfnamefont {E.}~\bibnamefont {Pomjakushina}}, \bibinfo
  {author} {\bibfnamefont {M.}~\bibnamefont {Bartkowiak}},  \emph {et~al.},\
  }\bibfield  {title} {\emph {\bibinfo {title} {{Z$_3$-vestigial nematic order
  due to superconducting fluctuations in the doped topological insulators
  ${\mathrm{Nb}}_{x}{\mathrm{Bi}}_{2}{\mathrm{Se}}_{3}$ and
  ${\mathrm{Cu}}_{x}{\mathrm{Bi}}_{2}{\mathrm{Se}}_{3}$}}},\ }\href
  {https://www.nature.com/articles/s41467-020-16871-9} {\bibfield  {journal}
  {\bibinfo  {journal} {Nature Communications}\ }\textbf {\bibinfo {volume}
  {11}},\ \bibinfo {pages} {3056} (\bibinfo {year} {2020})}\BibitemShut
  {NoStop}%
\bibitem [{\citenamefont {Feldman}\ \emph {et~al.}(2016)\citenamefont
  {Feldman}, \citenamefont {Randeria}, \citenamefont {Gyenis}, \citenamefont
  {Wu}, \citenamefont {Ji}, \citenamefont {Cava}, \citenamefont {MacDonald},\
  and\ \citenamefont {Yazdani}}]{Feldman2016}%
  \BibitemOpen
  \bibfield  {author} {\bibinfo {author} {\bibfnamefont {B.~E.}\ \bibnamefont
  {Feldman}}, \bibinfo {author} {\bibfnamefont {M.~T.}\ \bibnamefont
  {Randeria}}, \bibinfo {author} {\bibfnamefont {A.}~\bibnamefont {Gyenis}},
  \bibinfo {author} {\bibfnamefont {F.}~\bibnamefont {Wu}}, \bibinfo {author}
  {\bibfnamefont {H.}~\bibnamefont {Ji}}, \bibinfo {author} {\bibfnamefont
  {R.~J.}\ \bibnamefont {Cava}}, \bibinfo {author} {\bibfnamefont {A.~H.}\
  \bibnamefont {MacDonald}}, \ and\ \bibinfo {author} {\bibfnamefont
  {A.}~\bibnamefont {Yazdani}},\ }\bibfield  {title} {\emph {\bibinfo {title}
  {{Observation of a nematic quantum Hall liquid on the surface of bismuth}}},\
  }\href {https://www.science.org/doi/10.1126/science.aag1715} {\bibfield
  {journal} {\bibinfo  {journal} {Science}\ }\textbf {\bibinfo {volume}
  {354}},\ \bibinfo {pages} {316} (\bibinfo {year} {2016})}\BibitemShut
  {NoStop}%
\bibitem [{\citenamefont {Little}\ \emph {et~al.}(2020)\citenamefont {Little},
  \citenamefont {Lee}, \citenamefont {John}, \citenamefont {Doyle},
  \citenamefont {Maniv}, \citenamefont {Nair}, \citenamefont {Chen},
  \citenamefont {Rees}, \citenamefont {Venderbos}, \citenamefont {Fernandes}
  \emph {et~al.}}]{Little2020}%
  \BibitemOpen
  \bibfield  {author} {\bibinfo {author} {\bibfnamefont {A.}~\bibnamefont
  {Little}}, \bibinfo {author} {\bibfnamefont {C.}~\bibnamefont {Lee}},
  \bibinfo {author} {\bibfnamefont {C.}~\bibnamefont {John}}, \bibinfo {author}
  {\bibfnamefont {S.}~\bibnamefont {Doyle}}, \bibinfo {author} {\bibfnamefont
  {E.}~\bibnamefont {Maniv}}, \bibinfo {author} {\bibfnamefont {N.~L.}\
  \bibnamefont {Nair}}, \bibinfo {author} {\bibfnamefont {W.}~\bibnamefont
  {Chen}}, \bibinfo {author} {\bibfnamefont {D.}~\bibnamefont {Rees}}, \bibinfo
  {author} {\bibfnamefont {J.~W.}\ \bibnamefont {Venderbos}}, \bibinfo {author}
  {\bibfnamefont {R.~M.}\ \bibnamefont {Fernandes}},  \emph {et~al.},\
  }\bibfield  {title} {\emph {\bibinfo {title} {{Three-state nematicity in the
  triangular lattice antiferromagnet
  ${\mathrm{Fe}}_{1/3}{\mathrm{Nb}}{\mathrm{S}}_{2}$}}},\ }\href
  {https://www.nature.com/articles/s41563-020-0681-0} {\bibfield  {journal}
  {\bibinfo  {journal} {Nature materials}\ }\textbf {\bibinfo {volume} {19}},\
  \bibinfo {pages} {1062} (\bibinfo {year} {2020})}\BibitemShut {NoStop}%
\bibitem [{\citenamefont {Haley}\ \emph {et~al.}(2020)\citenamefont {Haley},
  \citenamefont {Weber}, \citenamefont {Cookmeyer}, \citenamefont {Parker},
  \citenamefont {Maniv}, \citenamefont {Maksimovic}, \citenamefont {John},
  \citenamefont {Doyle}, \citenamefont {Maniv}, \citenamefont {Ramakrishna},
  \citenamefont {Reyes}, \citenamefont {Singleton}, \citenamefont {Moore},
  \citenamefont {Neaton},\ and\ \citenamefont {Analytis}}]{Analytis2020}%
  \BibitemOpen
  \bibfield  {author} {\bibinfo {author} {\bibfnamefont {S.~C.}\ \bibnamefont
  {Haley}}, \bibinfo {author} {\bibfnamefont {S.~F.}\ \bibnamefont {Weber}},
  \bibinfo {author} {\bibfnamefont {T.}~\bibnamefont {Cookmeyer}}, \bibinfo
  {author} {\bibfnamefont {D.~E.}\ \bibnamefont {Parker}}, \bibinfo {author}
  {\bibfnamefont {E.}~\bibnamefont {Maniv}}, \bibinfo {author} {\bibfnamefont
  {N.}~\bibnamefont {Maksimovic}}, \bibinfo {author} {\bibfnamefont
  {C.}~\bibnamefont {John}}, \bibinfo {author} {\bibfnamefont {S.}~\bibnamefont
  {Doyle}}, \bibinfo {author} {\bibfnamefont {A.}~\bibnamefont {Maniv}},
  \bibinfo {author} {\bibfnamefont {S.~K.}\ \bibnamefont {Ramakrishna}},
  \bibinfo {author} {\bibfnamefont {A.~P.}\ \bibnamefont {Reyes}}, \bibinfo
  {author} {\bibfnamefont {J.}~\bibnamefont {Singleton}}, \bibinfo {author}
  {\bibfnamefont {J.~E.}\ \bibnamefont {Moore}}, \bibinfo {author}
  {\bibfnamefont {J.~B.}\ \bibnamefont {Neaton}}, \ and\ \bibinfo {author}
  {\bibfnamefont {J.~G.}\ \bibnamefont {Analytis}},\ }\bibfield  {title} {\emph
  {\bibinfo {title} {Half-magnetization plateau and the origin of threefold
  symmetry breaking in an electrically switchable triangular
  antiferromagnet}},\ }\href {\doibase 10.1103/PhysRevResearch.2.043020}
  {\bibfield  {journal} {\bibinfo  {journal} {Phys. Rev. Res.}\ }\textbf
  {\bibinfo {volume} {2}},\ \bibinfo {pages} {043020} (\bibinfo {year}
  {2020})}\BibitemShut {NoStop}%
\bibitem [{\citenamefont {Ni}\ \emph {et~al.}(2023)\citenamefont {Ni},
  \citenamefont {Antonenko}, \citenamefont {Meese}, \citenamefont {Tian},
  \citenamefont {Huang}, \citenamefont {Haglund}, \citenamefont {Cothrine},
  \citenamefont {Mandrus}, \citenamefont {Fernandes}, \citenamefont {Venderbos}
  \emph {et~al.}}]{LiangWu2023}%
  \BibitemOpen
  \bibfield  {author} {\bibinfo {author} {\bibfnamefont {Z.}~\bibnamefont
  {Ni}}, \bibinfo {author} {\bibfnamefont {D.~S.}\ \bibnamefont {Antonenko}},
  \bibinfo {author} {\bibfnamefont {W.~J.}\ \bibnamefont {Meese}}, \bibinfo
  {author} {\bibfnamefont {Q.}~\bibnamefont {Tian}}, \bibinfo {author}
  {\bibfnamefont {N.}~\bibnamefont {Huang}}, \bibinfo {author} {\bibfnamefont
  {A.~V.}\ \bibnamefont {Haglund}}, \bibinfo {author} {\bibfnamefont
  {M.}~\bibnamefont {Cothrine}}, \bibinfo {author} {\bibfnamefont {D.~G.}\
  \bibnamefont {Mandrus}}, \bibinfo {author} {\bibfnamefont {R.~M.}\
  \bibnamefont {Fernandes}}, \bibinfo {author} {\bibfnamefont {J.~W.}\
  \bibnamefont {Venderbos}},  \emph {et~al.},\ }\bibfield  {title} {\emph
  {\bibinfo {title} {{Signatures of Z$_3$ Vestigial Potts-nematic order in van
  der Waals antiferromagnets}}},\ }\href {https://arxiv.org/abs/2308.07249}
  {\bibfield  {journal} {\bibinfo  {journal} {arXiv:2308.07249}\ } (\bibinfo
  {year} {2023})}\BibitemShut {NoStop}%
\bibitem [{\citenamefont {Hwangbo}\ \emph {et~al.}(2023)\citenamefont
  {Hwangbo}, \citenamefont {Cenker}, \citenamefont {Rosenberg}, \citenamefont
  {Jiang}, \citenamefont {Wen}, \citenamefont {Xiao}, \citenamefont {Chu},\
  and\ \citenamefont {Xu}}]{Hwangbo2023}%
  \BibitemOpen
  \bibfield  {author} {\bibinfo {author} {\bibfnamefont {K.}~\bibnamefont
  {Hwangbo}}, \bibinfo {author} {\bibfnamefont {J.}~\bibnamefont {Cenker}},
  \bibinfo {author} {\bibfnamefont {E.}~\bibnamefont {Rosenberg}}, \bibinfo
  {author} {\bibfnamefont {Q.}~\bibnamefont {Jiang}}, \bibinfo {author}
  {\bibfnamefont {H.}~\bibnamefont {Wen}}, \bibinfo {author} {\bibfnamefont
  {D.}~\bibnamefont {Xiao}}, \bibinfo {author} {\bibfnamefont {J.-H.}\
  \bibnamefont {Chu}}, \ and\ \bibinfo {author} {\bibfnamefont
  {X.}~\bibnamefont {Xu}},\ }\bibfield  {title} {\emph {\bibinfo {title}
  {{Strain Tuning Three-state Potts Nematicity in a Correlated
  Antiferromagnet}}},\ }\href {https://arxiv.org/abs/2308.08734} {\bibfield
  {journal} {\bibinfo  {journal} {arXiv:2308.08734}\ } (\bibinfo {year}
  {2023})}\BibitemShut {NoStop}%
\bibitem [{\citenamefont {Sun}\ \emph {et~al.}(2023)\citenamefont {Sun},
  \citenamefont {Ye}, \citenamefont {Huang}, \citenamefont {Zhou},
  \citenamefont {Huang}, \citenamefont {Li}, \citenamefont {Ye}, \citenamefont
  {Nnokwe}, \citenamefont {Deng}, \citenamefont {Mandrus} \emph
  {et~al.}}]{Sun2023}%
  \BibitemOpen
  \bibfield  {author} {\bibinfo {author} {\bibfnamefont {Z.}~\bibnamefont
  {Sun}}, \bibinfo {author} {\bibfnamefont {G.}~\bibnamefont {Ye}}, \bibinfo
  {author} {\bibfnamefont {M.}~\bibnamefont {Huang}}, \bibinfo {author}
  {\bibfnamefont {C.}~\bibnamefont {Zhou}}, \bibinfo {author} {\bibfnamefont
  {N.}~\bibnamefont {Huang}}, \bibinfo {author} {\bibfnamefont
  {Q.}~\bibnamefont {Li}}, \bibinfo {author} {\bibfnamefont {Z.}~\bibnamefont
  {Ye}}, \bibinfo {author} {\bibfnamefont {C.}~\bibnamefont {Nnokwe}}, \bibinfo
  {author} {\bibfnamefont {H.}~\bibnamefont {Deng}}, \bibinfo {author}
  {\bibfnamefont {D.}~\bibnamefont {Mandrus}},  \emph {et~al.},\ }\bibfield
  {title} {\emph {\bibinfo {title} {{Dimensionality crossover to 2D vestigial
  nematicity from 3D zigzag antiferromagnetism in an XY-type honeycomb van der
  Waals magnet}}},\ }\href {https://arxiv.org/abs/2311.03493} {\bibfield
  {journal} {\bibinfo  {journal} {arXiv:2311.03493}\ } (\bibinfo {year}
  {2023})}\BibitemShut {NoStop}%
\bibitem [{\citenamefont {Tan}\ \emph {et~al.}(2023)\citenamefont {Tan},
  \citenamefont {Occhialini}, \citenamefont {Gao}, \citenamefont {Li},
  \citenamefont {Kitadai}, \citenamefont {Comin},\ and\ \citenamefont
  {Ling}}]{Tan2023}%
  \BibitemOpen
  \bibfield  {author} {\bibinfo {author} {\bibfnamefont {Q.}~\bibnamefont
  {Tan}}, \bibinfo {author} {\bibfnamefont {C.~A.}\ \bibnamefont {Occhialini}},
  \bibinfo {author} {\bibfnamefont {H.}~\bibnamefont {Gao}}, \bibinfo {author}
  {\bibfnamefont {J.}~\bibnamefont {Li}}, \bibinfo {author} {\bibfnamefont
  {H.}~\bibnamefont {Kitadai}}, \bibinfo {author} {\bibfnamefont
  {R.}~\bibnamefont {Comin}}, \ and\ \bibinfo {author} {\bibfnamefont
  {X.}~\bibnamefont {Ling}},\ }\bibfield  {title} {\emph {\bibinfo {title}
  {{Revealing the three-state nematicity in atomically-thin antiferromagnetic
  ${\mathrm{Ni}}{\mathrm{P}}{\mathrm{S}}_{3}$ via magneto-optical effect}}},\
  }\href {https://arxiv.org/abs/2311.12201} {\bibfield  {journal} {\bibinfo
  {journal} {arXiv:2311.12201}\ } (\bibinfo {year} {2023})}\BibitemShut
  {NoStop}%
\bibitem [{\citenamefont {Jin}\ \emph {et~al.}(2021)\citenamefont {Jin},
  \citenamefont {Zhang}, \citenamefont {Guo}, \citenamefont {Chen},
  \citenamefont {Zhou},\ and\ \citenamefont {Li}}]{Jin2021}%
  \BibitemOpen
  \bibfield  {author} {\bibinfo {author} {\bibfnamefont {S.}~\bibnamefont
  {Jin}}, \bibinfo {author} {\bibfnamefont {W.}~\bibnamefont {Zhang}}, \bibinfo
  {author} {\bibfnamefont {X.}~\bibnamefont {Guo}}, \bibinfo {author}
  {\bibfnamefont {X.}~\bibnamefont {Chen}}, \bibinfo {author} {\bibfnamefont
  {X.}~\bibnamefont {Zhou}}, \ and\ \bibinfo {author} {\bibfnamefont
  {X.}~\bibnamefont {Li}},\ }\bibfield  {title} {\emph {\bibinfo {title}
  {{Evidence of Potts-Nematic Superfluidity in a Hexagonal $s{p}^{2}$ Optical
  Lattice}}},\ }\href {\doibase 10.1103/PhysRevLett.126.035301} {\bibfield
  {journal} {\bibinfo  {journal} {Phys. Rev. Lett.}\ }\textbf {\bibinfo
  {volume} {126}},\ \bibinfo {pages} {035301} (\bibinfo {year}
  {2021})}\BibitemShut {NoStop}%
\bibitem [{\citenamefont {Gonz\'alez-Tudela}\ and\ \citenamefont
  {Cirac}(2019)}]{Cirac2019}%
  \BibitemOpen
  \bibfield  {author} {\bibinfo {author} {\bibfnamefont {A.}~\bibnamefont
  {Gonz\'alez-Tudela}}\ and\ \bibinfo {author} {\bibfnamefont {J.~I.}\
  \bibnamefont {Cirac}},\ }\bibfield  {title} {\emph {\bibinfo {title} {Cold
  atoms in twisted-bilayer optical potentials}},\ }\href
  {https://link.aps.org/doi/10.1103/PhysRevA.100.053604} {\bibfield  {journal}
  {\bibinfo  {journal} {Phys. Rev. A}\ }\textbf {\bibinfo {volume} {100}},\
  \bibinfo {pages} {053604} (\bibinfo {year} {2019})}\BibitemShut {NoStop}%
\end{thebibliography}%

\clearpage
\pagebreak

\setcounter{equation}{0}
\setcounter{figure}{0}
\setcounter{table}{0}
\setcounter{page}{1}
\setcounter{section}{0}
\renewcommand{\thesection}{S\Roman{section}}
\renewcommand{\theequation}{S\arabic{equation}}
\renewcommand{\thefigure}{S\arabic{figure}}
\renewcommand{\thetable}{S\Roman{table}}

\onecolumngrid
\begin{center}
	\textbf{\large Supplementary Material: A critical nematic phase with pseudogap-like behavior in twisted bilayers} \\
	\vspace{0.2cm}
\text{Virginia Gali, Matthias Hecker, and Rafael M. Fernandes} \\
\vspace{0.1cm}
\textit{School of Physics and Astronomy, University of Minnesota, Minneapolis, Minnesota 55455, USA} \\

\end{center}
\vspace{1cm}
\twocolumngrid

\section{free-energy expansion of the model hamiltonian}

Here we provide details of the free energy expansion of the microscopic
model introduced in the main text. We are interested in both the free
energy $\mathcal{F}\left[\boldsymbol{\phi}_{\mu}\right]$ of a single
layer, where $\boldsymbol{\phi}_{\mu}=\left(\phi_{\mu,1},\,\phi_{\mu,2}\right)=\left|\boldsymbol{\phi}_{\mu}\right|\left(\cos\alpha_{\mu},\sin\alpha_{\mu}\right)$
is the collective nematic field in layer $\mu$, as well as the free
energy $\mathcal{F}\left[\boldsymbol{\phi}_{+},\,\boldsymbol{\phi}_{-}\right]$
of the twisted bilayer system, where $\boldsymbol{\phi}_{\pm}=\left(\phi_{\pm,1},\,\phi_{\pm,2}\right)=\left|\boldsymbol{\phi}_{\pm}\right|\left(\cos\alpha_{\pm},\sin\alpha_{\pm}\right)$.
For convenience, we first list the resulting free energy expansions
and discuss their implications before delving into the details of
the derivation. For the single layer, we find:
\begin{align}
\mathcal{F}\left[\boldsymbol{\phi}_{\mu}\right] & =r\left|\boldsymbol{\phi}_{\mu}\right|^{2}+u\left|\boldsymbol{\phi}_{\mu}\right|^{4}+\lambda_{3}\left|\boldsymbol{\phi}_{\mu}\right|^{3}\cos\left(3\alpha_{\mu}\right),\label{eq:F_Phi_SL_SM}
\end{align}
which is identical to the $Z_{3}$-Potts model. For the twisted bilayer
system with twist angle $\theta_{\mathrm{tw}}$, on the other hand,
the free energy is given by $\mathcal{F}\left[\boldsymbol{\phi}_{+},\,\boldsymbol{\phi}_{-}\right]=\mathcal{F}_{+}\left[\boldsymbol{\phi}_{+}\right]+\mathcal{F}_{-}\left[\boldsymbol{\phi}_{-}\right]+\mathcal{F}_{\mathrm{m}}\left[\boldsymbol{\phi}_{\pm}\right]$
with:
\begin{align}
\mathcal{F}_{+}\left[\boldsymbol{\phi}_{+}\right] & =\mathcal{F}_{+}^{0}\left[\left|\boldsymbol{\phi}_{+}\right|\right]+\lambda_{6}^{(+)}\left|\boldsymbol{\phi}_{+}\right|^{6}\cos\left(6\alpha_{+}\right)\nonumber \\
 & +\left|\boldsymbol{\phi}_{+}\right|^{3}\left\{ \lambda_{+}^{(3\mathrm{c})}+u_{+}^{(5)}\left|\boldsymbol{\phi}_{+}\right|^{2}\right\} \cos\left(3\theta_{\mathrm{tw}}\right)\cos\left(3\alpha_{+}\right),\label{eq:F_Phip_SM}\\
\mathcal{F}_{-}\left[\boldsymbol{\phi}_{-}\right] & =\mathcal{F}_{-}^{0}\left[\left|\boldsymbol{\phi}_{-}\right|\right]+\lambda_{6}^{(-)}\left|\boldsymbol{\phi}_{-}\right|^{6}\cos\left(6\alpha_{-}\right)\nonumber \\
 & +\left|\boldsymbol{\phi}_{-}\right|^{3}\left\{ \lambda_{+}^{(3\mathrm{s})}+u_{-}^{(5)}\left|\boldsymbol{\phi}_{-}\right|^{2}\right\} \sin\left(3\theta_{\mathrm{tw}}\right)\sin\left(3\alpha_{-}\right),\label{eq:F_Phim_SM}\\
\mathcal{F}_{\mathrm{m}}\left[\boldsymbol{\phi}_{\pm}\right] & =3\lambda_{-}^{(3\mathrm{c})}\cos\left(3\theta_{\mathrm{tw}}\right)\left|\boldsymbol{\phi}_{+}\right|\left|\boldsymbol{\phi}_{-}\right|^{2}\cos\left(2\alpha_{-}+\alpha_{+}\right)\nonumber \\
 & +3\lambda_{-}^{(3\mathrm{s})}\sin\left(3\theta_{\mathrm{tw}}\right)\left|\boldsymbol{\phi}_{+}\right|^{2}\left|\boldsymbol{\phi}_{-}\right|\sin\left(2\alpha_{+}+\alpha_{-}\right)\nonumber \\
 & +u_{3}\left|\boldsymbol{\phi}_{+}\right|^{2}\left|\boldsymbol{\phi}_{-}\right|^{2}\left\{ 2+\cos\left(2\alpha_{+}-2\alpha_{-}\right)\right\} .\label{eq:Fm_Phi_SM}
\end{align}
Here, we defined the term that depends only on the magnitude of the
nematic order parameters as:
\begin{align}
\mathcal{F}_{\pm}^{0}\left[\left|\boldsymbol{\phi}\right|\right] & =r_{\pm}\left|\boldsymbol{\phi}\right|^{2}+u_{\pm}\left|\boldsymbol{\phi}\right|^{4}+u_{\pm}^{(6)}\left|\boldsymbol{\phi}\right|^{6}.\label{eq:F0_Phi_SM}
\end{align}

Importantly, in Eq. (\ref{eq:Fm_Phi_SM}), we note that there is no
bilinear coupling between the two order parameter combinations. However,
in cubic order, they have a linear-quadratic coupling in which one
field acts as an external field to the other. In particular, a non-zero
$\left|\boldsymbol{\phi}_{-}\right|$ induces $\left|\boldsymbol{\phi}_{+}\right|\sim\cos\left(3\theta_{\mathrm{tw}}\right)\left|\boldsymbol{\phi}_{-}\right|^{2}$
whereas a non-zero $\left|\boldsymbol{\phi}_{+}\right|$ induces $\left|\boldsymbol{\phi}_{-}\right|\sim\sin\left(3\theta_{\mathrm{tw}}\right)\left|\boldsymbol{\phi}_{+}\right|^{2}$.
Note that this type of linear-biquadratic coupling involving nematic
order parameters is well understood and also emerges in the contexts
of spin-polarized and valley-polarized nematics \citep{Chichinadze2020,Xu2020,Mandal2023}.
This means that, in the untwisted case ($\theta_{\mathrm{tw}}=0$),
the condensation of the odd-parity nematic order parameter $\left|\boldsymbol{\phi}_{-}\right|$
triggers also standard nematic order. Note that, depending on the
sign of $\lambda_{6}^{(-)}$, $\boldsymbol{\phi}_{-}$ also triggers
either an electric polarization $P_{z}\propto\left|\boldsymbol{\phi}_{-}\right|^{3}\sin\left(3\alpha_{-}\right)$,
which transforms as the $A_{2u}$ irrep of $\mathsf{D_{6h}}$, or
the $l=7$ electric multipole (octacosahectapole) $\Upsilon_{7}\propto\left|\boldsymbol{\phi}_{-}\right|^{3}\cos\left(3\alpha_{-}\right)$,
which transforms as $A_{1u}$. Conversely, in the $30^{\circ}$-twisted
hexagonal bilayer ($\theta_{\mathrm{tw}}=\pi/6$), condensation of
$\left|\boldsymbol{\phi}_{+}\right|$ also induces $\left|\boldsymbol{\phi}_{-}\right|\sim\left|\boldsymbol{\phi}_{+}\right|^{2}$.
Moreover, analogously to $\boldsymbol{\phi}_{-}$ in the untwisted
case, $\boldsymbol{\phi}_{+}$ induces, depending on the sign of $\lambda_{6}^{(+)}$,
either the electric polarization $P_{z}\propto\left|\boldsymbol{\phi}_{+}\right|^{3}\sin\left(3\alpha_{+}\right)$,
which transforms as $B_{2}$ in $\mathsf{D_{6d}}$, or the $l=6$
electric multipole (tetrahexacontapole) $\Upsilon_{6}\propto\left|\boldsymbol{\phi}_{+}\right|^{3}\cos\left(3\alpha_{+}\right)$,
which transforms as $B_{1}$.

\begin{figure}[t]
\includegraphics[width=1\columnwidth]{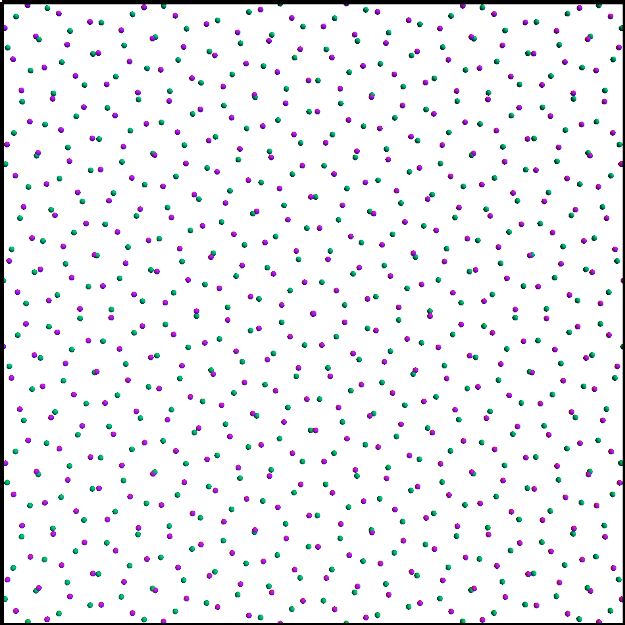} \caption{Quasi-crystalline pattern in real space of the $30^{\circ}$-twisted
bilayer system. Top-layer sites (purple) and bottom-layer sites (green)
can come arbitrarily close to each other, but they are only perfectly
aligned at the origin. \label{fig:twisted_lattice}}
\end{figure}

From Eqs. (\ref{eq:F_Phip_SM}) and (\ref{eq:F_Phim_SM}), we conclude
that unless the twist angle $\theta_{\mathrm{tw}}=n\frac{\pi}{6}$,
with $n\in\mathbb{N}^{0}$, the universality class of the nematic
transition is $Z_{3}$-Potts. Only for $\theta_{\mathrm{tw}}=n\frac{\pi}{6}$
either $\boldsymbol{\phi}_{+}$ (when $n$ is odd) or $\boldsymbol{\phi}_{-}$
(when $n$ is even) acquires the more exotic $Z_{6}$-clock universality.
Note that the linear-quadratic couplings only renormalize the quartic
coefficient of the leading channel upon integrating out the fluctuations
of the uncondensed order parameter. For the two specific twist angles
studied in the main text, the above Landau expansions simplify to
\begin{align}
\theta_{\mathrm{tw}} & =0: & \mathcal{F}_{+}\left[\boldsymbol{\phi}_{+}\right] & =\mathcal{F}_{+}^{0}\left[\left|\boldsymbol{\phi}_{+}\right|\right]+\lambda_{+}^{(3\mathrm{c})}\left|\boldsymbol{\phi}_{+}\right|^{3}\cos\left(3\alpha_{+}\right),\label{eq:F_Phip_SM-1}\\
 &  & \mathcal{F}_{-}\left[\boldsymbol{\phi}_{-}\right] & =\mathcal{F}_{-}^{0}\left[\left|\boldsymbol{\phi}_{-}\right|\right]+\lambda_{6}^{(-)}\left|\boldsymbol{\phi}_{-}\right|^{6}\cos\left(6\alpha_{-}\right),\label{eq:F_Phim_SM-1}
\end{align}
and
\begin{align}
\theta_{\mathrm{tw}} & =\frac{\pi}{6}\!: & \mathcal{F}_{+}\left[\boldsymbol{\phi}_{+}\right] & =\mathcal{F}_{+}^{0}\left[\left|\boldsymbol{\phi}_{+}\right|\right]+\lambda_{6}^{(+)}\left|\boldsymbol{\phi}_{+}\right|^{6}\cos\left(6\alpha_{+}\right),\label{eq:F_Phip_SM-2}\\
 &  & \mathcal{F}_{-}\left[\boldsymbol{\phi}_{-}\right] & =\mathcal{F}_{-}^{0}\left[\left|\boldsymbol{\phi}_{-}\right|\right]+\lambda_{+}^{(3\mathrm{s})}\left|\boldsymbol{\phi}_{-}\right|^{3}\sin\left(3\alpha_{-}\right).\label{eq:F_Phim_SM-2}
\end{align}
To emphasize that the $30^{\circ}$-twisted hexagonal bilayer system
forms a quasi-crystalline pattern with no periodicity, we plot the
corresponding arrangement of the sites of the top and bottom layers
in Fig. \ref{fig:twisted_lattice}.

To derive the Landau coefficients occurring in Eqs. (\ref{eq:F_Phi_SL_SM})-(\ref{eq:F0_Phi_SM}),
we start from the microscopic model Hamiltonian. Using the layer index
$\mu\in\{b,t\}$ for bottom and top layers, the single-layer nematic
Hamiltonian is given by 
\begin{align}
\mathcal{H}_{\mu} & =\sum_{\boldsymbol{k}}\boldsymbol{c}_{\boldsymbol{k},\mu}^{\dagger}\left\{ \left(\varepsilon_{\boldsymbol{k}}+\delta_{0}f_{\boldsymbol{k}}^{A_{1}}\right)-g\left(\boldsymbol{\phi}_{\mu}\cdot\boldsymbol{f}_{\boldsymbol{k}}^{E_{2}}\right)\right\} \sigma^{0}\boldsymbol{c}_{\boldsymbol{k},\mu}^{\phantom{\dagger}},\label{eq:H_mu_SM}
\end{align}
with the electronic operator $\boldsymbol{c}_{\boldsymbol{k},\mu}\equiv\left(c_{\mathbf{k}\uparrow,\mu},c_{\mathbf{k}\downarrow,\mu}\right)$,
the bare dispersion $\varepsilon_{\boldsymbol{k}}=\frac{\boldsymbol{k}^{2}}{2m}-\mu_{0}$
with momentum $\boldsymbol{k}=\left(k_{x},k_{y}\right)=\left|\boldsymbol{k}\right|\left(\cos\theta,\sin\theta\right)$
and chemical potential $\mu_{0}$. The two-component nematic order
parameter in the $\left(d_{x^{2}-y^{2}},d_{xy}\right)$-symmetry channel
is given by $\boldsymbol{\phi}_{\mu}=\left(\phi_{\mu,1},\phi_{\mu,2}\right)^{T}=\left|\boldsymbol{\phi}_{\mu}\right|\boldsymbol{b}_{\alpha_{\mu}}$
where
\begin{align}
\boldsymbol{b}_{\alpha} & =\left(\cos\alpha,\sin\alpha\right)^{T},\label{eq:b_alpha_SM}
\end{align}
and $g$ is the nematic coupling constant. The form factors are represented
with respect to their transformation properties within the $\mathsf{D_{6}}$
point group, i.e. with respect to the corresponding irreducible representation
(irrep). We define 
\begin{align}
f_{\boldsymbol{k}}^{B_{1}} & =\hat{k}_{x}^{3}-3\hat{k}_{x}\hat{k}_{y}^{2}=\cos\left(3\theta\right),\label{eq:f_b1}\\
f_{\boldsymbol{k}}^{B_{2}} & =\hat{k}_{y}^{3}-3\hat{k}_{y}\hat{k}_{x}^{2}=-\sin\left(3\theta\right),\label{eq:f_B2}\\
f_{\boldsymbol{k}}^{A_{1}} & =\big(f_{\boldsymbol{k}}^{B_{1}}\big)^{2}-\big(f_{\boldsymbol{k}}^{B_{2}}\big)^{2}=\cos\left(6\theta\right),\label{eq:f_A1}\\
f_{\boldsymbol{k}}^{A_{2}} & =-2f_{\boldsymbol{k}}^{B_{1}}f_{\boldsymbol{k}}^{B_{2}}=\sin\left(6\theta\right),\label{eq:f_A2}\\
\boldsymbol{f}_{\boldsymbol{k}}^{E_{2}} & =\left(\begin{array}{c}
\hat{k}_{x}^{2}-\hat{k}_{y}^{2}\\
-2\hat{k}_{x}\hat{k}_{y}
\end{array}\right)=\left(\begin{array}{c}
\cos\left(2\theta\right)\\
-\sin\left(2\theta\right)
\end{array}\right),\label{eq:f-E2}
\end{align}
where $\hat{\boldsymbol{k}}=\boldsymbol{k}/\left|\boldsymbol{k}\right|$.
Note that the symmetry elements of the $\mathsf{D_{6}}$ point group,
\begin{align}
\mathsf{D_{6}} & =\left\{ E,C_{6z}^{\pm1},C_{3z}^{\pm1},C_{2z}\right\} \otimes\left\{ E,C_{2x}\right\} ,\label{eq:D6_SM}
\end{align}
are all preserved in the bilayer system regardless of the actual twist
angle $\theta_{\mathrm{tw}}$, cf. Fig. 2(b) of the main text. Correspondingly,
any momentum summation where the argument does not transform trivially
($A_{1}$) has to vanish for symmetry reasons, for example $\sum_{\boldsymbol{k}}g_{\boldsymbol{k}}^{A_{1}}f_{\boldsymbol{k}}^{A_{2}}=0$
where $g_{\boldsymbol{k}}^{A_{1}}$ is an arbitrary function transforming
as $A_{1}$.

As we couple the two layers (\ref{eq:H_mu_SM}) by a tunneling Hamiltonian
\begin{align}
\mathcal{H}_{b-t} & =t\sum_{\boldsymbol{k}}\boldsymbol{c}_{\boldsymbol{k},t}^{\dagger}\sigma^{0}\boldsymbol{c}_{\boldsymbol{k},b}^{\phantom{\dagger}}+\mathrm{h.c.},\label{eq:H_tunnel_SM}
\end{align}
with tunneling amplitude $t$, we also rotate the bottom (top) layer
by an angle $-\theta_{\mathrm{tw}}/2$ ($+\theta_{\mathrm{tw}}/2$).
This rotation affects the form factors according to:
\begin{align}
\boldsymbol{f}_{\mathcal{R}_{z}\left(\pm\theta_{\mathrm{tw}}/2\right)\,\boldsymbol{k}}^{E_{2}} & =\mathcal{R}_{z}\left(\mp\theta_{\mathrm{tw}}\right)\boldsymbol{f}_{\boldsymbol{k}}^{E_{2}},\label{eq:f_E2_trafo_SM}\\
f_{\mathcal{R}_{z}\left(\pm\theta_{\mathrm{tw}}/2\right)\,\boldsymbol{k}}^{A_{1}} & =f_{\boldsymbol{k}}^{A_{1}}\cos\left(3\theta_{\mathrm{tw}}\right)\mp f_{\boldsymbol{k}}^{A_{2}}\sin\left(3\theta_{\mathrm{tw}}\right),\label{eq:fA1_trafo_SM}
\end{align}
with the rotation matrix 
\begin{align}
\mathcal{R}_{z}\left(\gamma\right) & =\left(\begin{smallmatrix}\cos\gamma & -\sin\gamma\\
\sin\gamma & \cos\gamma
\end{smallmatrix}\right).\label{eq:Rz_SM}
\end{align}

As a result, the total Hamiltonian $\mathcal{H}_{\mathrm{tot}}=\mathcal{H}_{b}+\mathcal{H}_{t}+\mathcal{H}_{b-t}$
becomes
\begin{align}
\mathcal{H}_{\mathrm{tot}} & =\sum_{\boldsymbol{k}}\boldsymbol{c}_{\boldsymbol{k}}^{\dagger}\Big\{\left[\varepsilon_{\boldsymbol{k}}+\delta_{0}f_{\boldsymbol{k}}^{A_{1}}\cos\left(3\theta_{\mathrm{tw}}\right)\right]\tau^{0}+\delta_{0}f_{\boldsymbol{k}}^{A_{2}}\sin\left(3\theta_{\mathrm{tw}}\right)\tau^{z}\nonumber \\
 & \quad-g\left(\boldsymbol{\phi}_{+}\cdot\boldsymbol{f}_{\boldsymbol{k}}^{E_{2}}\tau^{0}+\boldsymbol{\phi}_{-}\cdot\boldsymbol{f}_{\boldsymbol{k}}^{E_{2}}\tau^{z}\right)+t\,\tau^{x}\Big\}\sigma^{0}\boldsymbol{c}_{\boldsymbol{k}},\label{eq:Htot_SM}
\end{align}
where $\boldsymbol{c}_{\boldsymbol{k}}=\left(\boldsymbol{c}_{\boldsymbol{k},b},\,\boldsymbol{c}_{\boldsymbol{k},t}\right)$
denotes the four-component fermionic basis and $\tau^{i}$ are Pauli
matrices acting on the layer subspace. Here, we introduced the symmetrized
and anti-symmetrized combinations of the nematic order parameters
\begin{align}
\boldsymbol{\phi}_{\pm} & =\frac{1}{2}\left(\mathcal{R}_{z}^{T}\left(\theta_{\mathrm{tw}}\right)\boldsymbol{\phi}_{b}\pm\mathcal{R}_{z}\left(\theta_{\mathrm{tw}}\right)\boldsymbol{\phi}_{t}\right).\label{eq:phi_pm_SM}
\end{align}
As explained in the main text, only these combinations have a well-defined
transformation behavior under the reflection symmetry operation $\sigma_{h}$,
i.e. $\boldsymbol{\phi}_{\pm}\overset{\sigma_{h}}{\longrightarrow}\pm\boldsymbol{\phi}_{\pm}$
due to the individual transformation $\boldsymbol{\phi}_{b/t}\overset{\sigma_{h}}{\longrightarrow}\mathcal{R}_{z}\left(\pm2\theta_{\mathrm{tw}}\right)\boldsymbol{\phi}_{t/b}$.

For later use, we will need the symmetry-decomposition of powers of
$\big(\hat{\boldsymbol{\phi}}_{\mu}\cdot\boldsymbol{f}_{\boldsymbol{k}}^{E_{2}}\big)^{l}$
where $\hat{\boldsymbol{\phi}}_{\mu}=\boldsymbol{\phi}_{\mu}/\left|\boldsymbol{\phi}_{\mu}\right|$.
To do this, we introduce the higher-order form factors in the $E_{2}$-channel
given by 
\begin{align}
\boldsymbol{f}_{\boldsymbol{k}}^{2,E_{2}} & =\left(\begin{array}{c}
\big(f_{\boldsymbol{k},1}^{E_{2}}\big)^{2}-\big(f_{\boldsymbol{k},2}^{E_{2}}\big)^{2}\\
-2f_{\boldsymbol{k},1}^{E_{2}}f_{\boldsymbol{k},2}^{E_{2}}
\end{array}\right)=\left(\begin{array}{c}
\cos\left(4\theta\right)\\
\sin\left(4\theta\right)
\end{array}\right),\label{eq:f2_E2_SM}\\
\boldsymbol{f}_{\boldsymbol{k}}^{4,E_{2}} & =\left(\begin{array}{c}
\big(f_{\boldsymbol{k},1}^{2,E_{2}}\big)^{2}-\big(f_{\boldsymbol{k},2}^{2,E_{2}}\big)^{2}\\
-2f_{\boldsymbol{k},1}^{2,E_{2}}f_{\boldsymbol{k},2}^{2,E_{2}}
\end{array}\right)=\left(\begin{array}{c}
\cos\left(8\theta\right)\\
-\sin\left(8\theta\right)
\end{array}\right),\label{eq:f4_E2_SM}
\end{align}
which allows us to compute the symmetry decompositions up to fourth
order:

\begin{align}
h_{\mu\nu,\boldsymbol{k}}^{(2)} & \equiv\left(\hat{\boldsymbol{\phi}}_{\mu}\cdot\boldsymbol{f}_{\boldsymbol{k}}^{E_{2}}\right)\left(\hat{\boldsymbol{\phi}}_{\nu}\cdot\boldsymbol{f}_{\boldsymbol{k}}^{E_{2}}\right)\label{eq:h2_SM}\\
 & =\frac{1}{2}\cos\left(\alpha_{\mu}-\alpha_{\nu}\right)+\frac{1}{2}\boldsymbol{b}_{-\alpha_{\mu}-\alpha_{\nu}}\cdot\boldsymbol{f}_{\boldsymbol{k}}^{2,E_{2}},\nonumber \\
h_{\mu\nu\kappa,\boldsymbol{k}}^{(3)} & \equiv\left(\hat{\boldsymbol{\phi}}_{\mu}\cdot\boldsymbol{f}_{\boldsymbol{k}}^{E_{2}}\right)\left(\hat{\boldsymbol{\phi}}_{\nu}\cdot\boldsymbol{f}_{\boldsymbol{k}}^{E_{2}}\right)\left(\hat{\boldsymbol{\phi}}_{\kappa}\cdot\boldsymbol{f}_{\boldsymbol{k}}^{E_{2}}\right)\label{eq:h3_SM}\\
 & =\frac{1}{4}\left(\boldsymbol{b}_{-\alpha_{\mu}+\alpha_{\nu}+\alpha_{\kappa}}+\boldsymbol{b}_{\alpha_{\mu}-\alpha_{\nu}+\alpha_{\kappa}}+\boldsymbol{b}_{\alpha_{\mu}+\alpha_{\nu}-\alpha_{\kappa}}\right)\cdot\boldsymbol{f}_{\boldsymbol{k}}^{E_{2}}\nonumber \\
 & +\frac{1}{4}\cos\left(\alpha_{\mu}+\alpha_{\nu}+\alpha_{\kappa}\right)f_{\boldsymbol{k}}^{A_{1}}-\frac{1}{4}\sin\left(\alpha_{\mu}+\alpha_{\nu}+\alpha_{\kappa}\right)f_{\boldsymbol{k}}^{A_{2}},\nonumber \\
h_{\mu\nu\kappa\lambda,\boldsymbol{k}}^{(4)} & \equiv\left(\hat{\boldsymbol{\phi}}_{\mu}\cdot\boldsymbol{f}_{\boldsymbol{k}}^{E_{2}}\right)\left(\hat{\boldsymbol{\phi}}_{\nu}\cdot\boldsymbol{f}_{\boldsymbol{k}}^{E_{2}}\right)\left(\hat{\boldsymbol{\phi}}_{\kappa}\cdot\boldsymbol{f}_{\boldsymbol{k}}^{E_{2}}\right)\left(\hat{\boldsymbol{\phi}}_{\lambda}\cdot\boldsymbol{f}_{\boldsymbol{k}}^{E_{2}}\right)\label{eq:h4_SM}\\
 & =\frac{1}{8}\boldsymbol{b}_{\alpha_{\mu}+\alpha_{\nu}+\alpha_{\kappa}+\alpha_{\lambda}}\cdot\boldsymbol{f}_{\boldsymbol{k}}^{4,E_{2}}+\frac{1}{8}\Big[\cos\left(\alpha_{\mu}+\alpha_{\nu}-\alpha_{\kappa}-\alpha_{\lambda}\right)\nonumber \\
 & +\cos\left(\alpha_{\mu}-\alpha_{\nu}+\alpha_{\kappa}-\alpha_{\lambda}\right)+\cos\left(\alpha_{\mu}-\alpha_{\nu}-\alpha_{\kappa}+\alpha_{\lambda}\right)\Big]\nonumber \\
 & +\frac{1}{8}\Big(\boldsymbol{b}_{-\alpha_{\mu}-\alpha_{\nu}-\alpha_{\kappa}+\alpha_{\lambda}}+\boldsymbol{b}_{-\alpha_{\mu}-\alpha_{\nu}+\alpha_{\kappa}-\alpha_{\lambda}}\nonumber \\
 & +\boldsymbol{b}_{-\alpha_{\mu}+\alpha_{\nu}-\alpha_{\kappa}-\alpha_{\lambda}}+\boldsymbol{b}_{\alpha_{\mu}-\alpha_{\nu}-\alpha_{\kappa}-\alpha_{\lambda}}\Big)\cdot\boldsymbol{f}_{\boldsymbol{k}}^{2,E_{2}}.\nonumber 
\end{align}

To the fifth and sixth orders, we will only need the diagonal parts
$h_{\mu,\boldsymbol{k}}^{(l)}=\left(\hat{\boldsymbol{\phi}}_{\mu}\cdot\boldsymbol{f}_{\boldsymbol{k}}^{E_{2}}\right)^{l}$,
given by 
\begin{align}
h_{\mu,\boldsymbol{k}}^{(5)} & =\frac{9}{16}\left(\boldsymbol{b}_{\alpha_{\mu}}\cdot\boldsymbol{f}_{\boldsymbol{k}}^{E_{2}}\right)+h_{\boldsymbol{k}}^{\mathrm{tri}}\left[\frac{5}{16}+\frac{1}{8}\left(\boldsymbol{b}_{-2\alpha_{\mu}}\cdot\boldsymbol{f}_{\boldsymbol{k}}^{2,E_{2}}\right)\right],\label{eq:h5_SM}\\
h_{\mu,\boldsymbol{k}}^{(6)} & =\frac{9}{32}+\frac{3}{8}\left(\boldsymbol{b}_{\alpha_{\mu}}\cdot\boldsymbol{f}_{\boldsymbol{k}}^{E_{2}}\right)h_{\boldsymbol{k}}^{\mathrm{tri}}+\frac{9}{32}\boldsymbol{b}_{-2\alpha_{\mu}}\cdot\boldsymbol{f}_{\boldsymbol{k}}^{2,E_{2}}+\frac{\left(h_{\boldsymbol{k}}^{\mathrm{tri}}\right)^{2}}{16},\label{eq:h6_SM}
\end{align}
where we defined $h_{\boldsymbol{k}}^{\mathrm{tri}}=\cos\left(3\alpha_{\mu}\right)f_{\boldsymbol{k}}^{A_{1}}-\sin\left(3\alpha_{\mu}\right)f_{\boldsymbol{k}}^{A_{2}}$.

\subsection{Single layer}

We start by deriving the free energy expansion for a single layer
$\mu$ described by the Hamiltonian (\ref{eq:H_mu_SM}). The corresponding
electronic Green's function is given by:
\begin{equation}
\hat{\mathcal{G}}^{-1}=\hat{\mathcal{G}}_{0}^{-1}-\hat{\mathcal{V}},
\end{equation}
where 
\begin{align}
\hat{\mathcal{G}}_{0,k}^{-1} & =\left(i\omega_{n}-\varepsilon_{\boldsymbol{k}}-\delta_{0}f_{\boldsymbol{k}}^{A_{1}}\right)\sigma^{0}, & \hat{\mathcal{V}}_{\boldsymbol{k}} & =g\left(\boldsymbol{\phi}_{\mu}\cdot\boldsymbol{f}_{\boldsymbol{k}}^{E_{2}}\right)\sigma^{0},\label{G_and_V-1-1}
\end{align}
denote the bare Green's function and the nematic contribution, respectively.
Performing a fermionic Gaussian integration of the corresponding partition
function with respect to the electronic degrees of freedom, we obtain
the effective bosonic action 
\begin{align}
S_{\mathrm{eff}} & \left[\boldsymbol{\phi}_{\mu}\right]=-\mathrm{Tr}\,\ln\left(-\beta\hat{\mathcal{G}}^{-1}\right)+\frac{\boldsymbol{\phi}_{\mu}^{2}}{2U_{\mathrm{nem}}},\label{eq:S_eff}
\end{align}
where the last term arises from the Hubbard-Stratonovich decoupling
of the fermionic interaction term, with $U_{\mathrm{nem}}$ denoting
the interaction projected onto the nematic channel. Expanding the
effective action (\ref{eq:S_eff}) in terms of the (uniform) order
parameter $\boldsymbol{\phi}_{\mu}$ leads to the Landau expansion
of the free energy, which is given by $\mathcal{F}=S_{\mathrm{eff}}T$
with $S_{\mathrm{eff}}$ evaluated at the saddle point. It is convenient
to use the relation 
\begin{align}
-\mathrm{Tr}\,\ln\left(-\beta\hat{\mathcal{G}}^{-1}\right) & =-\mathrm{Tr}\,\ln\left(-\beta\hat{\mathcal{G}}_{0}^{-1}\right)+\sum_{l=1}^{\infty}\mathcal{S}^{(l)},\label{eq:trlog_expansion}
\end{align}
where
\begin{align}
\mathcal{S}^{(l)} & =\frac{1}{l}\sum_{n,\boldsymbol{k}}\mathrm{tr}_{\sigma}\left(\hat{\mathcal{G}}_{0,k}\hat{\mathcal{V}}_{\boldsymbol{k}}\right)^{l}=\frac{2}{l}\sum_{n,\boldsymbol{k}}\frac{g^{l}}{\mathsf{d}_{k}^{l}}\left(\boldsymbol{\phi}_{\mu}\cdot\boldsymbol{f}_{\boldsymbol{k}}^{E_{2}}\right)^{l},\label{eq:S_l}
\end{align}
and 
\begin{align}
\mathsf{d}_{k} & =i\omega_{n}-\varepsilon_{\boldsymbol{k}}-\delta_{0}f_{\boldsymbol{k}}^{A_{1}}.\label{eq:dk_SL_SM}
\end{align}

As explained above, the $\boldsymbol{k}$ integral only evaluates
to a non-zero value for $A_{1}$-contributions in the decompositions
(\ref{eq:h2_SM})-(\ref{eq:h4_SM}). We then obtain the free-energy
expansion
\begin{align}
\mathcal{F}\left[\boldsymbol{\phi}_{\mu}\right] & =r\left|\boldsymbol{\phi}_{\mu}\right|^{2}+u\left|\boldsymbol{\phi}_{\mu}\right|^{4}+\lambda_{3}\left|\boldsymbol{\phi}_{\mu}\right|^{3}\cos\left(3\alpha_{\mu}\right),\label{eq:S_eff-1}
\end{align}
with the coefficients 
\begin{align}
r & =\frac{1}{2U_{\mathrm{nem}}}+\frac{g^{2}T}{2}\sum_{n,\boldsymbol{k}}\frac{1}{\mathsf{d}_{k}^{2}}, & u & =\frac{3g^{4}T}{16}\sum_{n,\boldsymbol{k}}\frac{1}{\mathsf{d}_{k}^{4}},\nonumber \\
\lambda_{3} & =\frac{g^{3}T}{6}\sum_{n,\boldsymbol{k}}\frac{f_{\boldsymbol{k}}^{A_{1}}}{\mathsf{d}_{k}^{3}}.\label{eq:r_SL_SM}
\end{align}

\subsection{Twisted bilayer}

We now proceed to compute the Landau coefficients of the twisted hexagonal
bilayer system given by the total Hamiltonian (\ref{eq:Htot_SM})
for an arbitrary twist angle $\theta_{\mathrm{tw}}$. The fermionic
Green's function $\hat{\mathcal{G}}^{-1}=\hat{\mathcal{G}}_{0}^{-1}-\hat{\mathcal{V}}$
is given by
\begin{align}
\hat{\mathcal{G}}_{0,k}^{-1} & =\left(F_{k}^{A_{1}}\tau^{0}-F_{\boldsymbol{k}}^{A_{2}}\tau^{z}-t\,\tau^{x}\right)\sigma^{0},\label{G_and_V-1-1-1}\\
\hat{\mathcal{V}}_{\boldsymbol{k}} & =g\left[\left(\boldsymbol{\phi}_{+}\cdot\boldsymbol{f}_{\boldsymbol{k}}^{E_{2}}\right)\tau^{0}+\left(\boldsymbol{\phi}_{-}\cdot\boldsymbol{f}_{\boldsymbol{k}}^{E_{2}}\right)\tau^{z}\right]\sigma^{0},
\end{align}
where, for brevity, we defined 
\begin{align}
F_{k}^{A_{1}} & =i\omega_{n}-\varepsilon_{\boldsymbol{k}}-\delta_{0}f_{\boldsymbol{k}}^{A_{1}}\cos\left(3\theta_{\mathrm{tw}}\right), & F_{\boldsymbol{k}}^{A_{2}} & =\delta_{0}f_{\boldsymbol{k}}^{A_{2}}\sin\left(3\theta_{\mathrm{tw}}\right).\label{eq:FA1_FA2_SM}
\end{align}

The nematic order parameter combinations are parameterized by $\boldsymbol{\phi}_{\pm}=\left|\boldsymbol{\phi}_{\pm}\right|\left(\cos\alpha_{\pm},\sin\alpha_{\pm}\right)$
. Analogous to the previous case, the Gaussian integration generates
the effective free energy
\begin{align}
\mathcal{F}_{\mathrm{eff}} & =-T\,\mathrm{Tr}\,\ln\left(-\beta\hat{\mathcal{G}}^{-1}\right)+\frac{\boldsymbol{\phi}_{+}^{2}}{2U_{\mathrm{+}}}+\frac{\boldsymbol{\phi}_{-}^{2}}{2U_{\mathrm{-}}},\label{eq:S_eff-2}
\end{align}
where, to keep the approach general, we introduced different effective
interactions $U_{\pm}$ in each channel, which is allowed by the symmetry
of the problem. To expand the trace-log in terms of $\boldsymbol{\phi}_{+}$
and $\boldsymbol{\phi}_{-}$ we employ Eq. (\ref{eq:trlog_expansion})
and evaluate 
\begin{align}
\mathcal{F}^{(l)} & =\frac{T}{l}\sum_{n,\boldsymbol{k}}\mathrm{tr}_{\sigma,\tau}\left(\hat{\mathcal{G}}_{0,k}\hat{\mathcal{V}}_{\boldsymbol{k}}\right)^{l}.\label{eq:F_l_SM}
\end{align}

Using the symmetry decompositions (\ref{eq:h2_SM})-(\ref{eq:h6_SM}),
and the same logic as above, i.e. that only $A_{1}$ terms contribute
to a non-zero sum, we compute $\mathcal{F}^{(l)}$ order by order.
While the linear term $\mathcal{F}^{(1)}=0$ vanishes, the quadratic
contribution $\tilde{\mathcal{F}}^{(2)}\equiv\frac{1}{2}\left(\frac{\boldsymbol{\phi}_{+}^{2}}{U_{+}}+\frac{\boldsymbol{\phi}_{-}^{2}}{U_{-}}\right)+\mathcal{F}^{(2)}$
becomes 
\begin{align}
\tilde{\mathcal{F}}^{(2)} & =\frac{\boldsymbol{\phi}_{+}^{2}}{2U_{\mathrm{+}}}+\frac{\boldsymbol{\phi}_{-}^{2}}{2U_{\mathrm{-}}}+2g^{2}T\sum_{n,\boldsymbol{k}}\Big\{ I_{k,+}^{(2)}\big(\boldsymbol{\phi}_{+}\cdot\boldsymbol{f}_{\boldsymbol{k}}^{E_{2}}\big)^{2}\nonumber \\
 & \;+I_{k,-}^{(2)}\big(\boldsymbol{\phi}_{-}\cdot\boldsymbol{f}_{\boldsymbol{k}}^{E_{2}}\big)^{2}+4\,\mathsf{d}_{k}^{-2}F_{k}^{A_{1}}F_{\boldsymbol{k}}^{A_{2}}\big(\boldsymbol{\phi}_{+}\cdot\boldsymbol{f}_{\boldsymbol{k}}^{E_{2}}\big)\big(\boldsymbol{\phi}_{-}\cdot\boldsymbol{f}_{\boldsymbol{k}}^{E_{2}}\big)\Big\}\nonumber \\
 & =r_{+}\left|\boldsymbol{\phi}_{+}\right|^{2}+r_{-}\left|\boldsymbol{\phi}_{-}\right|^{2},\label{eq:Stilde_2_SM}
\end{align}
where we defined the quadratic Landau parameter 
\begin{align}
r_{\pm} & =\frac{1}{2U_{\pm}}+g^{2}T\sum_{n,\boldsymbol{k}}I_{k,\pm}^{(2)},\label{eq:r_pm_BL_SM}
\end{align}
with
\begin{align}
I_{k,\pm}^{(2)} & =\left[\big(F_{k}^{A_{1}}\big)^{2}+\big(F_{\boldsymbol{k}}^{A_{2}}\big)^{2}\pm t^{2}\right]\Big/\mathsf{d}_{k}^{2},\label{eq:I2_pm_BL_SM}\\
\mathsf{d}_{k} & =\big(F_{k}^{A_{1}}\big)^{2}-\big(F_{\boldsymbol{k}}^{A_{2}}\big)^{2}-t^{2}.\label{eq:dk_BL_SM}
\end{align}
The reason why the two nematic channels $\boldsymbol{\phi}_{\pm}$
are decoupled on the quadratic level is because the last term in the
curly bracket in (\ref{eq:Stilde_2_SM}) vanishes, since the decompositions
(\ref{eq:h2_SM}) have no $A_{2}$ contribution. 

For the cubic contribution to the free energy, we obtain
\begin{align}
\mathcal{F}^{(3)} & =\lambda_{+}^{(3\mathrm{c})}\mathrm{c}_{3\theta_{\mathrm{tw}}}\left|\boldsymbol{\phi}_{+}\right|^{3}\mathrm{c}_{3\alpha_{+}}+3\lambda_{-}^{(3\mathrm{c})}\mathrm{c}_{3\theta_{\mathrm{tw}}}\left|\boldsymbol{\phi}_{-}\right|^{2}\left|\boldsymbol{\phi}_{+}\right|\mathrm{c}_{\alpha_{+}+2\alpha_{-}}\nonumber \\
 & +\lambda_{+}^{(3\mathrm{s})}\mathrm{s}_{3\theta_{\mathrm{tw}}}\mathrm{s}_{3\alpha_{-}}\left|\boldsymbol{\phi}_{-}\right|^{3}+\lambda_{-}^{(3\mathrm{s})}3\mathrm{s}_{3\theta_{\mathrm{tw}}}\left|\boldsymbol{\phi}_{+}\right|^{2}\left|\boldsymbol{\phi}_{-}\right|\mathrm{s}_{2\alpha_{+}+\alpha_{-}},\label{eq:S_3_SM}
\end{align}
where we defined $\mathrm{c}_{\gamma}\equiv\cos(\gamma)$ and $\mathrm{s}_{\gamma}\equiv\sin(\gamma)$,
as well as the Landau parameters 
\begin{align}
\lambda_{\pm}^{(3\mathrm{c})} & =\frac{g^{3}T}{3}\sum_{n,\boldsymbol{k}}\left(\frac{\left(i\omega_{n}-\varepsilon_{\boldsymbol{k}}\right)f_{\boldsymbol{k}}^{A_{1}}}{\cos\left(3\theta_{\mathrm{tw}}\right)}-\delta_{0}\left(f_{\boldsymbol{k}}^{A_{1}}\right)^{2}\right)I_{k,\pm}^{(3A)},\label{eq:lambda_3c_SM}\\
\lambda_{\pm}^{(3\mathrm{s})} & =-\frac{g^{3}T}{3}\sum_{n,\boldsymbol{k}}\delta_{0}\left(f_{\boldsymbol{k}}^{A_{2}}\right)^{2}I_{k,\pm}^{(3B)},\label{eq:lambda_3s_SM}
\end{align}
and the integrands 
\begin{align}
I_{k,\pm}^{(3A)} & =\left[\big(F_{k}^{A_{1}}\big)^{2}+3\big(F_{\boldsymbol{k}}^{A_{2}}\big)^{2}+\left(1\pm2\right)t^{2}\right]\Big/\mathsf{d}_{k}^{3},\label{eq:I3A_BL_SM}\\
I_{k,\pm}^{(3B)} & =\left[3\big(F_{k}^{A_{1}}\big)^{2}+\big(F_{\boldsymbol{k}}^{A_{2}}\big)^{2}-\left(1\pm2\right)t^{2}\right]\Big/\mathsf{d}_{k}^{3}.\label{eq:I3B_BL_SM}
\end{align}
At first sight, it might appear that $\lambda_{\pm}^{(3\mathrm{c})}$
in Eq. (\ref{eq:lambda_3c_SM}) is singular at $\theta_{\mathrm{tw}}=\pi/6$.
However, this is not the case. To see this, we symmetrize the first
term in (\ref{eq:lambda_3c_SM}) using a $C_{12z}$ operation upon
which $f_{\mathcal{R}_{z}\left(\frac{2\pi}{12}\right)\boldsymbol{k}}^{A_{1},A_{2}}=-f_{\boldsymbol{k}}^{A_{1},A_{2}}$.
Then, we find the result 
\begin{align}
\frac{\left(i\omega_{n}-\varepsilon_{\boldsymbol{k}}\right)f_{\boldsymbol{k}}^{A_{1}}}{\cos\left(3\theta_{\mathrm{tw}}\right)}\left[I_{k,\pm}^{(3A)}-I_{\left(\omega_{n},\mathcal{R}_{z}\left(\frac{\pi}{6}\right)\boldsymbol{k}\right),\pm}^{(3A)}\right]\nonumber \\
 & \!\!\!\!\!\!\!\!\!\!\!\!\!\!\!\!\!\!\!\!\!\!\!\!\!\!\!\!\!\!\!\!\!\!\!\!\!\!\!\!\!\!\!\!\!\!\!\!\!\!\!\!\!\!\!\!\!\!\!\!\!\!\!\!\!\!=\delta_{0}\left(i\omega_{n}-\varepsilon_{\boldsymbol{k}}\right)^{2}\left(f_{\boldsymbol{k}}^{A_{1}}\right)^{2}\tilde{I}_{k,\pm}^{(3A)}.\label{eq:I3A_symmetrizing}
\end{align}

Since $\tilde{I}_{k,\pm}^{(3A)}$ depends on $\cos\left(3\theta_{\mathrm{tw}}\right)$
only in a trivial (analytical) way, this justifies the overall prefactor
$\cos\left(3\theta_{\mathrm{tw}}\right)$ in Eq. (\ref{eq:S_3_SM}).

For the quartic contribution, we find 
\begin{align}
\mathcal{F}^{(4)} & =u_{+}\left|\boldsymbol{\phi}_{+}\right|^{4}+u_{-}\left|\boldsymbol{\phi}_{-}\right|^{4}+u_{3}\left[\boldsymbol{\phi}_{+}^{2}\boldsymbol{\phi}_{-}^{2}+2\left(\boldsymbol{\phi}_{+}\cdot\boldsymbol{\phi}_{-}\right)^{2}\right],\label{eq:S4_DL_SM}
\end{align}
where the Landau parameters are
\begin{align}
u_{\pm} & =\frac{3T}{8}g^{4}\sum_{n,\boldsymbol{k}}I_{k,\pm}^{(4A)}, & u_{3} & =\frac{3T}{4}g^{4}\sum_{n,\boldsymbol{k}}I_{k}^{(4B)},\label{eq:u_pk_u3_SM}
\end{align}
and the integrands are defined as 
\begin{align}
I_{k,\pm}^{(4A)} & =\frac{1}{\mathsf{d}_{k}^{4}}\Big\{\mathsf{d}_{k}^{2}+8\big(F_{k}^{A_{1}}\big)^{2}\big(F_{\boldsymbol{k}}^{A_{2}}\big)^{2}+4t^{2}\left(1\pm1\right)\big(F_{k}^{A_{1}}\big)^{2}\nonumber \\
 & \quad+4t^{2}\left(-1\pm1\right)\big(F_{\boldsymbol{k}}^{A_{2}}\big)^{2}\Big\},\label{eq:I4A_BL_SM}\\
I_{k}^{(4B)} & =\frac{\big(F_{k}^{A_{1}}\big)^{4}+\big(F_{\boldsymbol{k}}^{A_{2}}\big)^{4}+6\big(F_{k}^{A_{1}}\big)^{2}\big(F_{\boldsymbol{k}}^{A_{2}}\big)^{2}-t^{4}-\frac{2t^{2}}{3}\mathsf{d}_{k}}{\mathsf{d}_{k}^{4}}.\label{eq:I4B_BL_SM}
\end{align}

Since the two order parameters are already coupled at the quartic
level (\ref{eq:S4_DL_SM}), for the fifth- and sixth-order terms we
only derive the contributions that depend on one of the order parameters.
The fifth-order terms becomes 
\begin{align}
\mathcal{F}^{(5)} & =u_{+}^{(5)}\mathrm{c}_{3\theta_{\mathrm{tw}}}\left|\boldsymbol{\phi}_{+}\right|^{5}\mathrm{c}_{3\alpha_{+}}+u_{-}^{(5)}\mathrm{s}_{3\theta_{\mathrm{tw}}}\left|\boldsymbol{\phi}_{-}\right|^{5}\mathrm{s}_{3\alpha_{-}},\label{eq:DL_S5_SM}
\end{align}
where we used again the notation $\mathrm{c}_{\gamma}\equiv\cos(\gamma)$
and $\mathrm{s}_{\gamma}\equiv\sin(\gamma)$, and defined the Landau
parameters
\begin{align}
u_{+}^{(5)} & =\frac{T}{2}g^{5}\sum_{n,\boldsymbol{k}}\left(\frac{\left(i\omega_{n}-\varepsilon_{\boldsymbol{k}}\right)f_{\boldsymbol{k}}^{A_{1}}}{\cos\left(3\theta_{\mathrm{tw}}\right)}-\delta_{0}\left(f_{\boldsymbol{k}}^{A_{1}}\right)^{2}\right)I_{k,+}^{(5)},\label{eq:DL_u5p_SM}\\
u_{-}^{(5)} & =-\frac{T}{2}g^{5}\delta_{0}\sum_{n,\boldsymbol{k}}\left(f_{\boldsymbol{k}}^{A_{2}}\right)^{2}I_{k,-}^{(5)}.\label{eq:DL_u5m_SM}
\end{align}
The integrand is given by:
\begin{align}
I_{k,\pm}^{(5)} & =\frac{1}{\mathsf{d}_{k}^{5}}\Big\{\frac{5}{2}\left[\big(F_{k}^{A_{1}}\big)^{2}+\big(F_{\boldsymbol{k}}^{A_{2}}\big)^{2}\pm t^{2}\right]^{2}-\left(1\pm1\right)\big(F_{k}^{A_{1}}\big)^{4}\nonumber \\
 & \quad-\left(1\mp1\right)\big(F_{\boldsymbol{k}}^{A_{2}}\big)^{4}\Big\}.\label{eq:I5_BL_SM}
\end{align}
The $\cos\left(3\theta_{\mathrm{tw}}\right)$ in the denominator in
Eq. (\ref{eq:DL_u5p_SM}) cancels for the same reasons as in relation
(\ref{eq:I3A_symmetrizing}).

Finally, for the sixth-order we find
\begin{align}
\mathcal{F}^{(6)} & =u_{+}^{(6)}\left|\boldsymbol{\phi}_{+}\right|^{6}+u_{-}^{(6)}\left|\boldsymbol{\phi}_{-}\right|^{6}+\lambda_{6}^{(+)}\left|\boldsymbol{\phi}_{+}\right|^{6}\mathrm{c}_{6\alpha_{+}}+\lambda_{6}^{(-)}\left|\boldsymbol{\phi}_{-}\right|^{6}\mathrm{c}_{6\alpha_{-}},\label{eq:F6_BL_SM}
\end{align}
with
\begin{align}
u_{\pm}^{(6)} & =\frac{5g^{6}T}{12}\sum_{n,\boldsymbol{k}}I_{k,\pm}^{(6)},\label{eq:DL_u6_SM}\\
\lambda_{6}^{(\pm)} & =\frac{g^{6}T}{24}\sum_{n,\boldsymbol{k}}\Big[\big(f_{\boldsymbol{k}}^{A_{1}}\big)^{2}-\big(f_{\boldsymbol{k}}^{A_{2}}\big)^{2}\Big]I_{k,\pm}^{(6)},\label{eq:DL_l6_SM}
\end{align}
as well as
\begin{align}
I_{k,\pm}^{(6)} & =\frac{1}{2\mathsf{d}_{k}^{3}}+\frac{\big(F_{\boldsymbol{k}}^{A_{2}}\big)^{2}+\frac{1\pm1}{2}t^{2}}{\mathsf{d}_{k}^{6}}\Big[3\big(F_{k}^{A_{1}}\big)^{2}+\big(F_{\boldsymbol{k}}^{A_{2}}\big)^{2}-\left(1\mp2\right)t^{2}\Big]^{2}.\label{eq:I6_BL_SM}
\end{align}

\section{Spectral function and pseudogap-like behavior in the critical phase}

Here we present details of the evaluation of the one-loop fermionic
self-energy. As shown in the main text, in the critical region where
the fermions couple to the nematic fluctuations, it is given by
\begin{align}
\Sigma\left(\omega,\boldsymbol{k}\right) & =g^{2}A_{0}T\\
 & \sum_{ij}\int_{q}\chi_{\mathrm{nem}}^{ij}\left(\Omega,\boldsymbol{q}\right)f_{\boldsymbol{k}-\frac{\boldsymbol{q}}{2},i}^{E_{2}}f_{\boldsymbol{k}-\frac{\boldsymbol{q}}{2},j}^{E_{2}}\mathcal{G}\left(\omega-\Omega,\boldsymbol{k}-\boldsymbol{q}\right)\nonumber 
\end{align}

Throughout this section, we set both Planck and Boltzmann's constants
to be $\hslash=k_{B}=1$. We focus on the contributions of the fermions
at the Fermi level, with momenta $\boldsymbol{k}_{F}\equiv\left|\boldsymbol{k}_{F}\left(\theta\right)\right|\left(\cos\theta,\sin\theta\right)$,
and for simplicity, we assume a spherical Fermi surface where $\left|\boldsymbol{k}_{F}\left(\theta\right)\right|=k_{F}=\mathrm{const}$.
To keep the discussion general, for the non-interacting fermionic
propagator we use a linearized dispersion and a finite lifetime $\tau_{0}$:
\begin{align}
\mathcal{G}^{-1}\left(\omega,\boldsymbol{k}_{F}-\boldsymbol{q}\right) & =\omega-\boldsymbol{v}_{F}\cdot\boldsymbol{q}+\frac{i}{2\tau_{0}},\label{eq:fermionic_greens_SM}
\end{align}
where $\boldsymbol{v}_{F}=\boldsymbol{k}_{F}/m$ is the Fermi velocity.
We also neglect the $\boldsymbol{q}$ dependence of the vertex function,
which becomes $\boldsymbol{f}_{\boldsymbol{k}-\frac{\boldsymbol{q}}{2}}^{E_{2}}\approx\boldsymbol{f}_{\boldsymbol{k}}^{E_{2}}=\left(\cos2\theta,\,-\sin2\theta\right)$.
Moreover, since we are interested in the thermal transition, we set
the bosonic Matsubara frequency to zero and write the bosonic propagator
as
\begin{align}
\chi_{\mathrm{nem}}^{ij} & \left(0,\boldsymbol{q}\right)=\left(\left|\boldsymbol{q}\right|\big/k_{F}\right)^{\eta-2}\,\delta_{ij}\,\chi_{0},\label{eq:chi_nem_SM}
\end{align}
where $\eta$ is the anomalous exponent characterizing the critical
phase, and the prefactor $\chi_{0}$ has dimensions of $\left(\mathrm{energy}\right)^{-1}$. 

Using all of the above ingredients, the self-energy is given by: 
\begin{equation}
\begin{aligned}\Sigma\left(\omega,\boldsymbol{k}_{F}\right) & =\int\frac{d\boldsymbol{q}}{(2\pi)^{2}}\,\frac{g^{2}\chi_{0}A_{0}T}{\left(\frac{\left|\boldsymbol{q}\right|}{k_{F}}\right)^{2-\eta}\left(\omega-\boldsymbol{v}_{F}\cdot\boldsymbol{q}+\frac{i}{2\tau_{0}}\right)}\mbox{.}\end{aligned}
\end{equation}

We perform a change of variables to dimensionless Cartesian coordinates
$(x,y)$ where $x$ is the momentum $\boldsymbol{q}$ in units of
$k_{F}$ along the direction transverse to the Fermi surface, and
$y$ is the momentum $\boldsymbol{q}$ in units of $k_{F}$ along
the direction longitudinal to the Fermi surface. We obtain 
\begin{equation}
\begin{aligned}\Sigma\left(\omega,\boldsymbol{k}_{F}\right) & =\int_{-\infty}^{\infty}\int_{-\infty}^{\infty}\frac{\kappa T\;dx\,dy}{\left(x^{2}+y^{2}\right)^{1-\frac{\eta}{2}}\left(\frac{\omega}{E_{F}}-2x+\frac{i}{2E_{F}\tau_{0}}\right)}\mbox{,}\end{aligned}
\label{eq:sigma_qpara_qperp_SM}
\end{equation}
where $E_{F}=k_{F}^{2}/2m$ is the Fermi energy and $\kappa$ is a
dimensionless parameter given by $\kappa=\frac{g^{2}k_{F}^{2}A_{0}\chi_{0}}{(2\pi)^{2}E_{F}}$.
Evaluating the integral, we obtain the analytical expression: 
\begin{equation}
\begin{aligned}\Sigma\left(\omega,\boldsymbol{k}_{F}\right) & =\kappa Tc_{\eta}e^{-i\frac{\eta\pi}{2}}\left(\frac{2\omega\tau_{0}+i}{2E_{F}\tau_{0}}\right)^{\eta-1}\mbox{,}\end{aligned}
\end{equation}
where we defined $c_{\eta}$ as 
\begin{equation}
\begin{aligned}c_{\eta} & =\frac{\pi^{\frac{3}{2}}\Gamma\left(\frac{1}{2}-\frac{\eta}{2}\right)}{2^{\eta}\Gamma\left(1-\frac{\eta}{2}\right)}\frac{1}{\sin\left(\frac{\eta\pi}{2}\right)},\end{aligned}
\label{eq:ceta}
\end{equation}
and $\Gamma(x)$ is the Gamma function. The coefficient $c_{\eta}$
varies smoothly in the range $\left[26.6,57.2\right]$ for the allowed
values of $\eta\in\left[1/9,1/4\right]$. The real and imaginary parts
of the spectral function are given by 
\begin{align}
\mathrm{Re}\,\Sigma\left(\omega,\boldsymbol{k}_{F}\right) & =\mathrm{sgn}\left(\omega\right)\,\Sigma_{0}\cos\left(\gamma\right), & \mathrm{Im}\,\Sigma\left(\omega,\boldsymbol{k}_{F}\right) & =\Sigma_{0}\sin\left(\gamma\right)\mbox{,}\label{eq:imresigma}
\end{align}
with magnitude and phase 
\begin{align}
\Sigma_{0} & =\kappa Tc_{\eta}\left(\frac{4\omega^{2}\tau_{0}^{2}+1}{4E_{F}^{2}\tau_{0}^{2}}\right)^{\frac{\eta-1}{2}},\\
\gamma & =\left(\eta-1\right)\arctan\left(\frac{1}{2\left|\omega\right|\tau_{0}}\right)-\frac{\eta\pi}{2}.
\end{align}

We can now compute the spectral function $A\left(\omega,\boldsymbol{k}_{F}\right)=-2\mathrm{Im}\left[\tilde{\mathcal{G}}\left(\omega,\boldsymbol{k}_{F}\right)\right]$,
where $\tilde{\mathcal{G}}^{-1}=\mathcal{G}^{-1}-\Sigma$ is the renormalized
Green's function. The resulting function, 

\begin{align}
A\left(\omega,\boldsymbol{k}_{F}\right) & =\frac{-2\mathrm{Im}\,\Sigma\left(\omega,\boldsymbol{k}_{F}\right)}{\left[\omega-\mathrm{Re}\,\Sigma\left(\omega,\boldsymbol{k}_{F}\right)\right]^{2}+\left[\mathrm{Im}\,\Sigma\left(\omega,\boldsymbol{k}_{F}\right)\right]^{2}},\label{eq:spectral}
\end{align}
is plotted in Fig. 3 of the main text for $T=0.01E_{F}$, $\tau_{0}=50E_{F}^{-1}$,
$\kappa=0.02$ and $\eta\in\left[0.1,0.25\right]$. The pseudogap-like
energy scale set by the distance between the peaks of the spectral
function can be obtained analytically in the limit $\tau_{0}\to\infty$:
\begin{align}
\Sigma_{0} & =\kappa Tc_{\eta}\left(\frac{|\omega|}{E_{F}}\right)^{\eta-1}, & \gamma & =-\frac{\eta\pi}{2}.
\end{align}

Inserting the above expressions into the definition of the spectral
function in Eq. \eqref{eq:spectral} and taking the derivative with
respect to $\omega$, we find the following condition for the maximum
at positive $\omega$, 
\begin{equation}
\begin{aligned}X^{2}-2\kappa a_{\eta}\frac{2\eta-1}{\eta+1}\frac{T}{E_{F}}X+3\kappa^{2}c_{\eta}^{2}\frac{\eta-1}{\eta+1}\frac{T^{2}}{E_{F}^{2}} & =0\end{aligned}
\end{equation}
where we have defined $X\equiv\left(\omega/E_{F}\right)^{2-\eta}$
and $a_{\eta}\equiv c_{\eta}\cos\left(\frac{\eta\pi}{2}\right)$.
The positive root of this polynomial gives the position of the maximum,
$\omega_{\mathrm{max}}$. The value of $\Delta=2\omega_{\mathrm{max}}$
is then given by:
\begin{align}
\Delta & \approx2E_{F}\left[\frac{\kappa T}{E_{F}}\frac{a_{\eta}\left(1-2\eta\right)}{\eta+1}\left(\sqrt{1+\frac{3c_{\eta}^{2}\left(1-\eta^{2}\right)}{a_{\eta}^{2}\left(1-2\eta\right)}}-1\right)\right]^{\frac{1}{2-\eta}}\mbox{.}\label{eq:gap}
\end{align}

The pseudogap-like energy scale $\Delta$ is plotted in Fig. 3(b)
of the main text for different values of the lifetime $\tau_{0}$.
Clearly, the dependence on $\tau_{0}$ is very weak.

\end{document}